\documentclass[twocolumn]{aastex631}
\usepackage{amsmath} 

\begin{document}



\title[UVSAG-I]{UVIT Survey of AGN host Galaxies - I: Star Formation Scenarios}
\shorttitle{UVSAG-I}
\shortauthors{Nandi et al.}

\author[0009-0003-9765-3517]{Payel Nandi}
\affiliation{Joint Astronomy Programme, Department of Physics, Indian Institute of Science, Bangalore 560012, India}
\affiliation{Indian Institute of Astrophysics, Block II, Koramangala, Bangalore 560034, India}
\author[0000-0002-4998-1861]{C. S. Stalin}
\affiliation{Indian Institute of Astrophysics, Block II, Koramangala, Bangalore 560034, India}
\author[0009-0007-7842-9930]{Poulomi Dam}{}
\affiliation{Indian Institute of Astrophysics, Block II, Koramangala, Bangalore 560034, India}
\affiliation{Dipartimento di Fisica e Astronomia, Università di Padova, Vicolo dell’Osservatorio 3, I-35122 Padova, Italy}

\author[0000-0002-4464-8023]{D. J. Saikia}
\affiliation{Inter-University Centre for Astronomy and Astrophysics, Pune 411007, India} 

\correspondingauthor{Payel Nandi}
\email{payel.nandi@iiap.res.in}



\begin{abstract}

Circum-nuclear star formation (SF) is generally seen in galaxies hosting active galactic nuclei (AGN); however, the connection between the AGN activity and SF in them is less well understood. To explore this connection on scales of a few tens of parsec to a few tens of kiloparsec and larger, we carried out an investigation of SF in seven Seyfert type AGN
and one LINER galaxy, using observations with the  Ultra-Violet Imaging Telescope (UVIT) 
on board {\it AstroSat} in the near ultra-Violet (NUV; 2000–3000 \AA) and far ultra-Violet (FUV; 2000–3000 \AA) bands. A total of 1742 star-forming regions were identified, having size scales of 0.010 to 63.642 kpc$^2$. Considering all the galaxies, we found a positive correlation between their total surface density of SF ($\Sigma_{SFR}$) and extinction.  For five galaxies, namely NGC~1365, NGC~4051, NGC~4321, NGC~5033 and
NGC~6814, we found a gradual decrease of both extinction and $\Sigma_{SFR}$ from the centre to the outer regions. 
 Four sources are found to lie in the main sequence (MS) of star-forming galaxies, and the other four are away from MS. We found the ratio of the star formation rate (SFR) in the nuclear region to the total SFR to be positively correlated with the Eddington ratio. This points to the influence of AGN on enhancing the SF characteristics of the hosts. However, the impact is dominant only in the central nuclear region with no significant effect on the larger scales probed in this work.  

\end{abstract}

\keywords{ Active galactic nuclei (16); Seyfert galaxies (1447); Star formation
(1569); Galaxy photometry (611); Ultraviolet astronomy(1736) }


\section{Introduction} \label{sec:intro}

The observational evidences of close correlation between the mass of the 
super-massive black holes that power active galactic nuclei (AGN) via accretion and 
their host galaxy properties \citep{2000ApJ...539L...9F,2000ApJ...539L..13G,
2002MNRAS.331..795M,2003ApJ...589L..21M,2004ApJ...604L..89H} clearly indicate 
that AGN and star formation (SF) are closely linked. This connection is believed
to be via AGN feedback processes, which play an important role 
in the SF characteristics of their hosts from the central nuclear
scales to larger galactic scales \citep{2017NatAs...1E.165H,2019NatAs...3...48S,
2022MNRAS.512.3906R,2023Galax..11...47C,2023ApJ...950...81N,2024MNRAS.527.1612Y}. 
Theoretical studies do invoke feedback to understand galaxy evolution 
(\citealt{2022MNRAS.514.2936W} and references therein).  Also,
simulations do find quenching of SF in galaxies
with AGN feedback \citep{2023arXiv231016086B}. A viable feedback 
mechanism is AGN outflows, which can impact the 
SF characteristic of AGN hosts by either suppressing (negative feedback 
process;\citealt{2017NatAs...1E.165H,2023arXiv230910572H,2023ApJ...959..116N}) or enhancing (positive feedback 
process; \citealt{2017Natur.544..202M,2019MNRAS.485.3409G,2021ApJ...906...38Z}) 
SF.  Also, both processes could co-exist in a galaxy as has been recently 
observed \citep{2013ApJ...774...66Z,2019ApJ...881..147S,2021A&A...645A..21G}.

One possibility to study the impact of AGN on the SF characteristics of their
hosts is to map the star forming regions in galaxies hosting AGN and look for 
correlations if any between the deduced SF and AGN properties.  It is natural to 
expect that the influence of the central AGN on their hosts could decrease from 
the center to the outskirts of the galaxy \citep{2015AJ....150...43T}.
While simulations \citep{2023arXiv231107576B} and observations \citep{2023ApJ...953...26L} 
favour feedback processes to operate in 
the central kilo parsec scale region, its effect on kilo parsec galaxy scale is 
uncertain \citep{2017A&A...601A.143F}.
In the nearby Universe, Seyfert type AGN are 
ideal targets to investigate this connection as the 
resolution offered by ground-based imaging observations will enable
one to probe SF on scales of a few hundreds of parsec to a few tens of 
kilo parsec.  Circumnuclear SF is generally observed
in this category of AGN \citep{1998MNRAS.300..388D,2015MNRAS.451.3173A,
2018MNRAS.477.1086H,2019MNRAS.487.3958D}. They have also  been studied for
SF for more than three decades in different spatial scales
as well as at different wavelengths \citep{2001RMxAA..37....3G,2004MNRAS.355..273C,2007ApJ...671.1388D,
2019MNRAS.482..194B,2022MNRAS.512.3906R,2023ApJ...953L...9Z}. 
Despite years of research, the nature of SF in AGN host galaxies
{\it vis-a-vis} normal galaxies is not settled and highly debated.
For example, in a nearby dwarf Seyfert type AGN, NGC 4395,  \cite{2023ApJ...950...81N} 
found regions of high SF close to the center which  
could be due its AGN activity.  Considering the far-infrared emission to be
due to star formation, \cite{1987ApJ...312..555R} 
found that the far-infrared luminosities of Seyfert galaxies and 
star-burst galaxies are indistinguishable arguing for the bulk of the 
far-infrared emission in Seyfert galaxies to be be due to SF. Alternatively,
\cite{2016MNRAS.458.4512G} found that in a few of the Seyfert galaxies studied
by them, the bulk of the nuclear far-infrared luminosity is contributed
by the dust heated by the AGN. 
\cite{2004MNRAS.355..273C} studied SF 
history extensively in Seyfert 2 galaxies and found that they 
are heterogeneous 
and hosts a mixture of young, intermediate and old stellar populations.

To gain a comprehensive understanding of SF in galaxies, it is 
crucial to employ a range of wavelength bands. This is because galaxies contain 
diverse stellar populations and multi-phase interstellar medium (ISM), making 
it necessary to utilize various tracers for SF across the 
electromagnetic spectrum, from X-ray to radio wavelengths. In this context, 
radio and millimeter-wave bands serve as tracers for SF activity via their continuum and molecular line emission. On the other hand, H$\alpha$ and ultraviolet (UV) 
emissions act as indicators of ionizing radiation emitted by hot stars. Sometimes, 
the continuum UV photons interact with the surrounding gas and dust in the 
ISM, giving rise to infrared (IR) emissions. H$\alpha$ 
emission is particularly informative as it reveals the presence of stellar 
populations with ages ranging from 1 to 10 million years. UV emission, on the 
other hand, primarily traces young, massive star clusters, providing insights 
into stellar populations aged between 10 and 100 million years \citep{2012ARA&A..50..531K}. It is worth 
emphasizing that the presence of dust within a galaxy can diminish the intensity 
of UV light emitted by stars by either scattering or absorbing UV photons. 
Nevertheless, accounting for this extinction effect and applying suitable 
corrections can yield accurate and valuable results.

Until now, most studies of SF in Seyfert galaxies have been 
conducted in the optical, IR, or radio wavelengths using ground-based and 
space-based observatories. While there have been a few studies in the 
UV band using the Galaxy Evolution Explorer (GALEX; \citealt{2005ApJ...619L...1M}), the resolution 
provided by GALEX is often insufficient to resolve SF on 
parsec scales. A limited number of studies have used the Hubble Space Telescope, which 
offers the capability to resolve parsec-scale structures, but it has a restricted 
field of view (FoV), making it observationally expensive to study a large number 
of sources comprehensively.
To address this gap, we have undertaken a systematic investigation of the 
SF properties of Seyfert galaxies using the Ultraviolet Imaging 
Telescope (UVIT). This telescope provides moderate angular resolution, 
better than 1.5$^{\prime\prime}$ \citep{2020AJ....159..158T} and offers 
good FoV coverage, approximately 
$\sim 28\arcmin$ diameter. In Section \ref{sample}, we describe our source selection process. 
Section \ref{obs} outlines our observations and data reduction procedures. In 
Section \ref{annalysis}, we describe the methods we used for our analysis. 
In Section \ref{notes}, we present our research findings on
each of the objects studied in this work, the global
picture of the results of this work is given in Section \ref{global},  
followed by the summary in the final section. 
In this work, we considered a flat $\Lambda$ CDM cosmology with 
$H_0$=70 km$^{-1}$ s$^{-1}$ Mpc$^{-1}$, $\Omega_\Lambda$ = 0.7 and $\Omega_{m}$=0.3.

\section{Our Sample} \label{sample}
Our sample of sources for this study was selected primarily from the catalogue 
of \cite{2006A&A...455..773V}. Our initial criteria involved the identification of 
objects classified as Seyferts in the \cite{2006A&A...455..773V} catalogue with redshift 
$z$ $<$ 0.02. This redshift cut was imposed so that at the resolution of 
the Ultra-Violet Imaging Telescope (UVIT;\citealt{2020AJ....159..158T}), 
the minimum spatial scale that could be probed is about few tens of parsec. We manually 
inspected our sample of Seyfert type AGN in the GALEX database to assess 
their UV emission. Additionally, we imposed a size constraint, requiring that 
the selected objects have an angular size exceeding 2$^\prime$ but lesser than 
12$^\prime$. We also considered some nearby low ionization
nuclear emission line region (LINER) type of AGN for which UVIT observations 
are already available and satisfies the above criteria.  This
resulted in  the identification of 
30 objects. We have acquired observations for few sources in 
our sample. We aim to complete observations of the remaining targets with UVIT 
in the upcoming observing cycles.  Results on one source in our sample, namely 
NGC 4395 is reported in \cite{2023ApJ...950...81N}, while in this paper 
we report the results on another eight sources. The details of 
the sources are given in Table \ref{tab:source}. 

\section{Observations and data reduction}\label{obs}
The observations of our sample of sources were made using UVIT
\citep{2020AJ....159..158T}, one of the payloads on board 
{\it AstroSat} \citep{2006AdSpR..38.2989A}, India's multi-wavelength astronomical 
observatory launched by the Indian Space Research Organization on 
28 September 2015.  UVIT observes simultaneously in the far ultraviolet 
(FUV; 1300$-$1800 \AA) and the near ultraviolet (NUV; 2000$-$3000\AA) over 
a $\sim$28$^{\prime}$ diameter field with a spatial resolution better than 
1.5$^{\prime\prime}$ in multiple filters. The FUV and NUV channels operate in 
the photon counting mode, while the VIS channel (VIS; 3200$-$5500 \AA) operates 
in the integration mode. Images from the VIS channel are used for 
monitoring the drift of the satellite. For the sources in this study, we  
used the data 
available in the archives of the Indian Space Science Data Center 
(ISSDC\footnote{https://astrobrowse.issdc.gov.in/astro\_archive/archive/Search.jsp}).
From ISSDC, we directly took the science ready images for further analysis in 
this work. These science ready images were generated by the UVIT Payload 
Operations Center (UVIT-POC)  using the UVIT-L2 pipeline \citep{2022JApA...43...77G} 
and transferred to ISSDC for archival and dissemination. The observational 
details of the sources used in this study are given in Table \ref{tab:obs}.
Of the eight sources, six sources have observations in both the FUV and NUV 
wavelengths, while two sources have observations only at FUV wavelengths.  The 
details of the filters used are given in Table \ref{tab:filter}. 

\begin{figure*}
    \centering
    \vbox{
    \hbox{
    \includegraphics[scale=0.23]{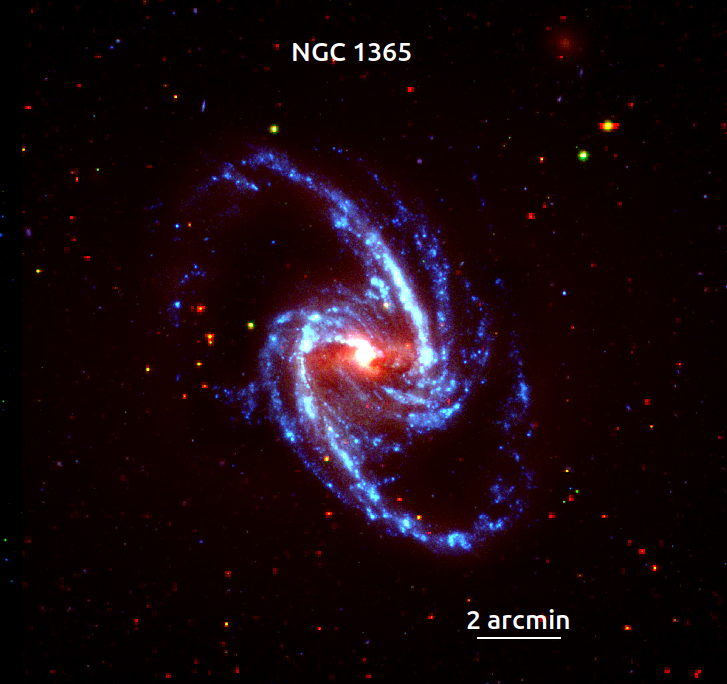}
    \includegraphics[scale=0.23]{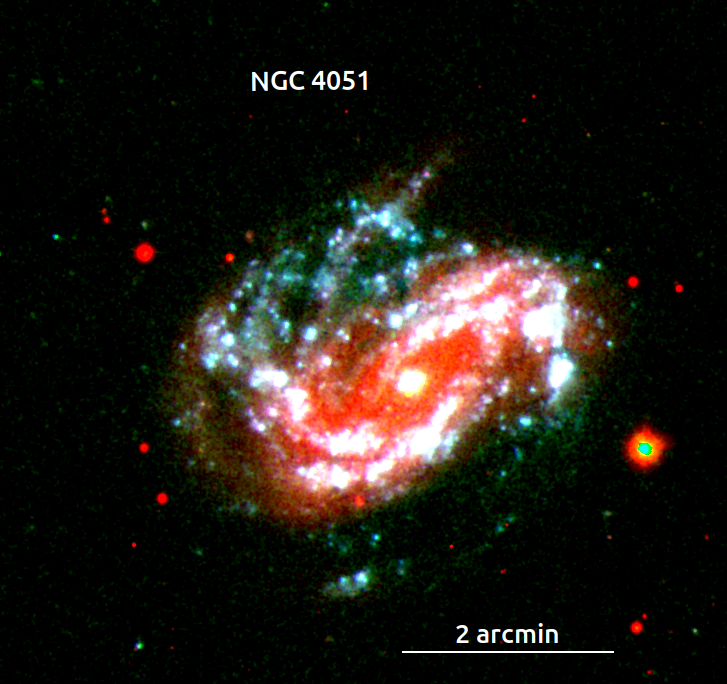}
    \includegraphics[scale=0.23]{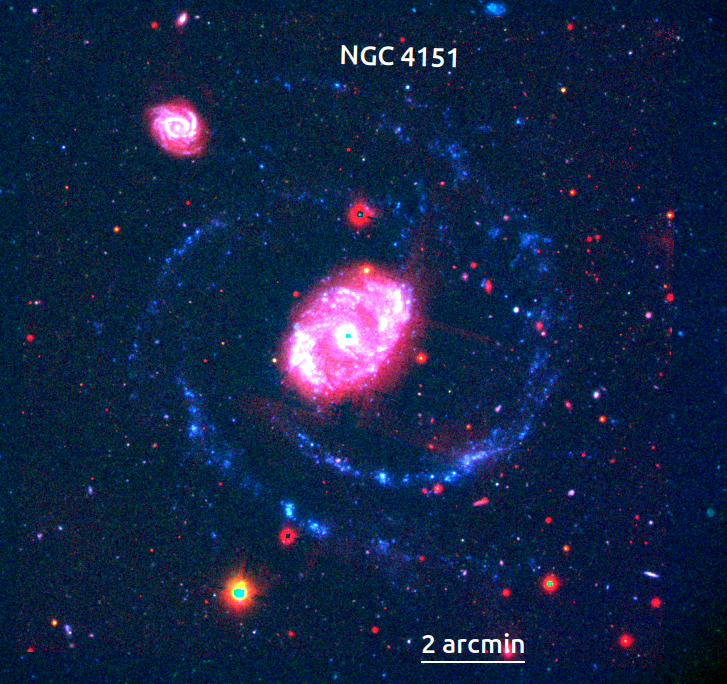}
    }
    \hbox{
    \includegraphics[scale=0.23]{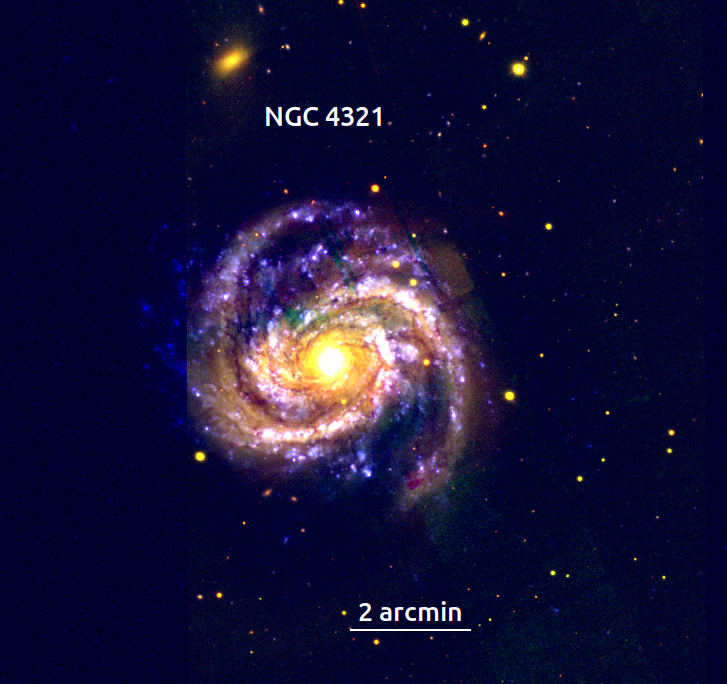}
    \includegraphics[scale=0.23]{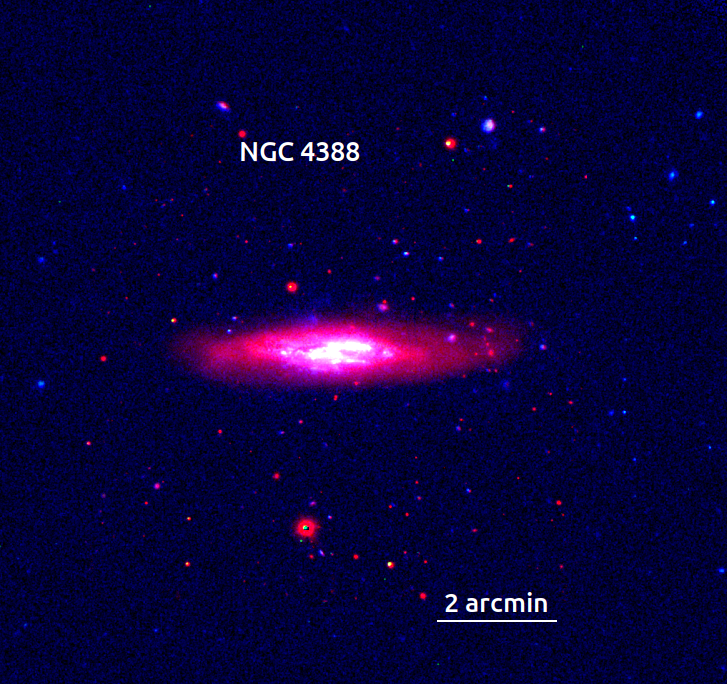}
    \includegraphics[scale=0.23]{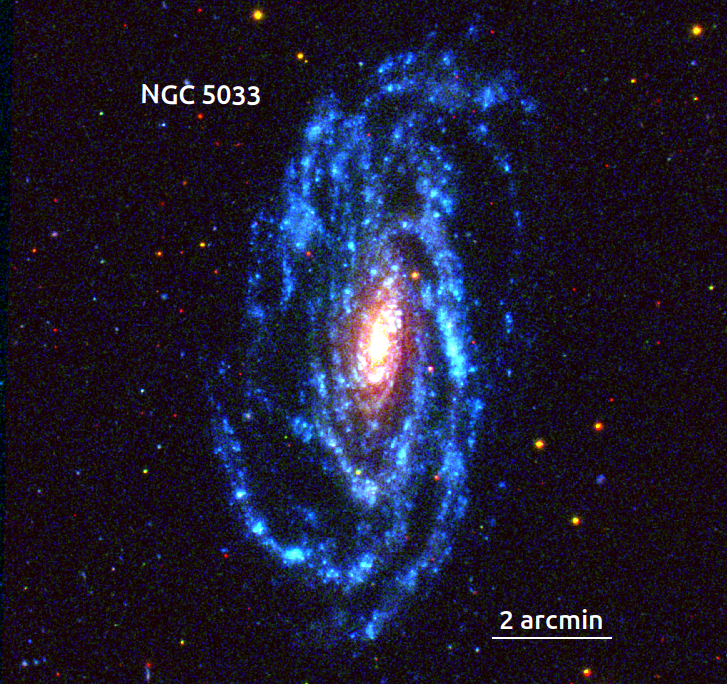}
    }
   \hbox{
    \includegraphics[scale=0.23]{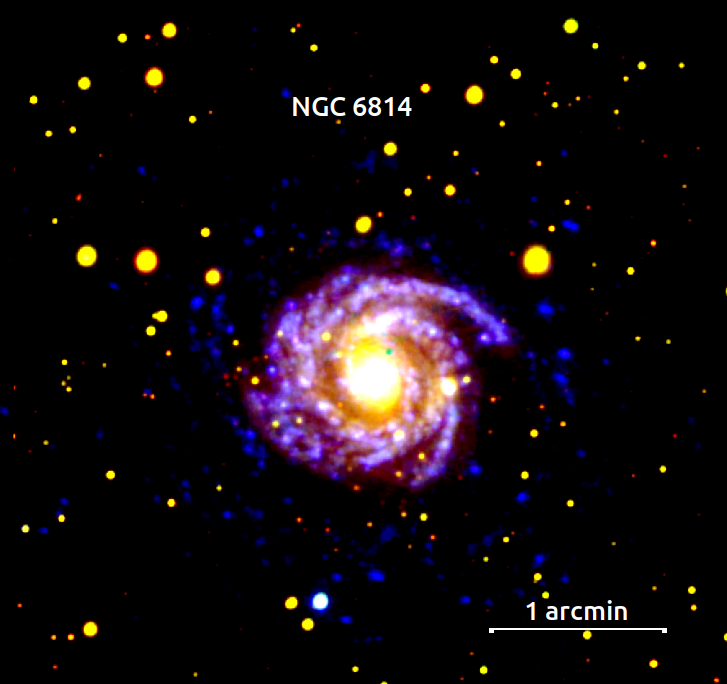}
    \includegraphics[scale=0.23]{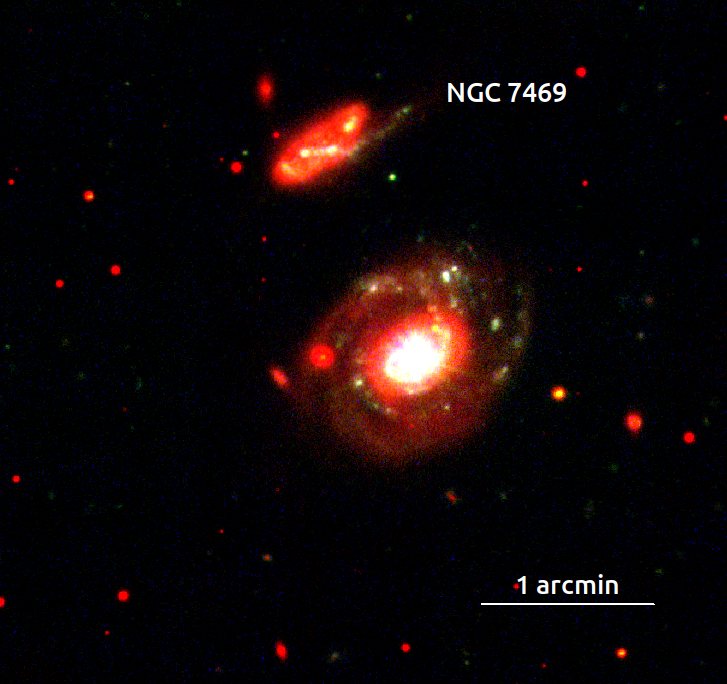}
    }
    }
    \caption{RGB images of the sources. Here, 
red is Pan-STARRS r band (except for  NGC~1365 where it is from the      
Decadal legacy survey), green is UVIT NUV (except for NGC~4321
and NGC~6814 where it is Pan-STARRS  g-band) and blue is UVIT FUV. 
}
\label{fig:rgb}
\end{figure*}

For some sources, the filter-wise combined images taken from ISSDC have exposure 
time smaller than the sum of the individual orbit-wise images. Also, the final 
astrometry has large errors. Therefore, we used the orbit-wise L2 data, aligned 
them and combined them filter wise using the  IRAF\footnote{Image Reduction and 
Analysis Facility}  software package. To ensure precise astrometry of the
combined images, we used the {\it Gaia-EDR3} catalogue \citep{2022arXiv220800211G}, and
carried out astrometry as elaborated in \cite{2023ApJ...950...81N}. 
The measured counts/sec of the sources of interest were converted to 
physical units using the calibrations given in \cite{2020AJ....159..158T}. 
For objects observed multiple times, to generate the final combined images,
we used only those orbit wise images having sufficient signal. The RGB images
of the 8 sources obtained using the final FUV and NUV images from UVIT
along with archival optical images are shown in Fig.\ref{fig:rgb}.

\begin{table*}
\centering
\caption{Details of the sources analysed in this work. Here, RA, DEC, 
redshift (z), morphology, AGN type  and the extinction in V-band A(V) 
are from NED. Here, $^a$ is from \cite{2018ApJ...864...40P}, 
$^b$ is from \cite{2018ApJ...864..146B}, $^c$ is from \cite{2016MNRAS.457.2122G}, 
$^d$ is from \cite{2018A&A...617A..33A}, $^e$ is from \cite{2018ApJ...860...37S}, M$_{\ast}$ is
the stellar mass and R$_{25}$ is the radius of the galaxy to an optical surface
brightness limit of 25 mag/arcsec$^2$ taken from NED.} 
\label{tab:source}
\begin{tabular}{lrrrrrrrrrr}  \hline
Name & RA     & DEC   & $z$ & scale                & Morphology & AGN type  & M$_{\ast}$ & R$_{25}$ & A(V)      \\
     & J2000  & J2000 &     & (pc/$\prime\prime$)  &            &         & (M$_\odot$)    & (arcmin) & (mag)     \\ \hline
    NGC~1365 & 03:33:36.37 & $-$36:08:25.45 & 0.005 & 124.4 & SB(s)b                   & Sy 1.8 & 10.71$\pm$0.10$^{a}$  & 5.61$^\prime$, 3.08$^\prime$, 32$^{\circ}$ & 0.056 \\
    NGC~4051 & 12:03:09.61  &   44:31:52.80 & 0.002 &  41.5 & SAB(rs)bc                & NLSy1  & 10.13$\pm$0.25$^{b}$  & 2.62$^\prime$, 1.94$^\prime$,135$^{\circ}$ & 0.036  \\
    NGC~4151 & 12:10:32.58 &    39:24:20.63 & 0.003 &  62.3 & (R$^{\prime}$SAB(rs)ab   & Sy 1.5 & 10.40$\pm$0.25$^{b}$  & 3.15$^\prime$,2.13$^\prime$,50$^{\circ}$   & 0.074  \\
    NGC~4321 & 12:22:54.83 &    15:49:19.54 & 0.005 & 103.7 & SAB(s)bc                 & LINER  & 10.83$\pm$0.28$^{c}$  & 3.71$^\prime$,3.15$^\prime$,30$^{\circ}$   & 0.072 \\
    NGC~4388 & 12:25:46.75 &    12:39:43.51 & 0.008 & 165.8 & SA(s)b                   & Sy 1.9 &  10.42$^{d}$          & 2.81$^\prime$,0.64$^\prime$,92$^\circ$     & 0.091 \\
    NGC~5033 & 13:13:27.47 &    36:35:38.17 & 0.003 &  62.3 & SA(s)c                   & Sy 1.9 & 11.01$\pm$0.2$^{e}$   & 5.36$^\prime$,2.51$^\prime$,170$^\circ$    & 0.032 \\
    NGC~6814 & 19:42:40.64 & $-$10:19:24.57 & 0.005 & 103.7 & SAB(rs)bc                & Sy 1.5 & 10.34$\pm$0.29$^{b}$  & 1.51$^\prime$,1.41$^\prime$                & 0.509 \\ 
    NGC~7469 & 23:03:15.62 &    08:52:26.39 & 0.016 & 331.0 & (R’)SAB-(rs)a            & Sy 1.2 & 10.88$\pm$0.23$^{b}$  & 0.74$^\prime$,0.54$^\prime$,125$^\circ$    & 0.188\\
\hline
\end{tabular}
\end{table*}

\begin{table*}
\label{tab:obs}
\begin{center}
\vspace{0.5cm} 
\caption{Summary of observations}
\begin{tabular}{lrrrrrrr}
\hline
Name & Observation ID (OBSID) & PI & Date & \multicolumn{2}{c}{Filter} & \multicolumn{2}{c}{Exposure time(sec)} \\
     &                &    &                 & FUV & NUV & FUV & NUV\\
\hline 
NGC~1365 &  A02\_006T01\_9000000776 & Gulab     & 08/11/2016 & F169M  & N279N & 24905 & 37833 \\
         &  A02\_006T01\_9000000802 & Gulab     & 17/11/2016 &        &       &        &       \\
         &  A02\_006T01\_9000000934 & Gulab     & 28/12/2016 &        &       &        &        \\
NGC~4051 &  G05\_248T01\_9000000486 & Stalin    & 11/02/2018 & F172M  & N219M & 26444 & 35336 \\
         &  G08\_071T01\_9000001888 & KPSingh   & 11/02/2018 &        &       &        &        \\
NGC~4151 &  G06\_117T01\_9000001012 & KPSingh   & 08/02/2017 & F154W & N219M  & 67547 & 73548 \\
         &  G06\_117T01\_9000001046 & KPSingh   & 22/02/2017 &       &        &        &        \\
         &  G06\_117T01\_9000001086 & KPSingh   & 17/03/2017 &       &        &        &       \\
         &  G08\_064T01\_9000001814 & KPSingh   & 04/01/2018 &       &        &        &        \\
         &  G08\_064T01\_9000002070 & KPSingh   & 02/05/2018 &       &        &        &        \\
NGC~4321 &  A08\_003T05\_9000003426 & Hutchings & 11/01/2020 & F154W & --     & 6296  &	--\\
NGC~4388 &  A02\_024T01\_9000001044 & Labani    & 21/02/2017 & F154W & N245M  & 13520 & 13621 \\
NGC~5033 &  G06\_087T04\_9000001028 & Stalin    & 14/02/2017 & F148W & N279N  & 2930  & 2948 \\
NGC~6814 &  A05\_037T01\_9000003140 & Pranoti   & 02/09/2019 & F154W & --     & 46782 &	--\\
NGC~7469 &  G08\_071T02\_9000001620 & KPSingh   & 18/10/2017 & F172M & N245M  & 34493 & 50032   \\
\hline
\end{tabular}
\end{center}
\end{table*}

\begin{table}
\label{tab:filter}
\caption{The details of the filters used in this work \citep{2020AJ....159..158T}. Here, $\lambda_{mean}$ and
$\Delta \lambda$ are the mean wavelength and band width in \AA .}
\centering
\begin{tabular}{rrr} \hline
Filter Name & $\lambda_{mean}$(\AA)  & $\Delta\lambda$(\AA) \\ \hline
F148W   &   1481  &  500 \\
F154W   &   1541  &  380 \\
F169M   &   1608  &  290 \\
F172M   &   1717  &  125 \\
N219M   &   2196  &  270 \\
N245M   &   2447  &  280 \\
N279N   &   2792  &   90 \\ \hline
\end{tabular}
\end{table}


\section{Analysis}\label{annalysis}
\begin{figure*}
\vbox{
\hbox{
    \hspace{-0.5cm}\includegraphics[scale=0.32]{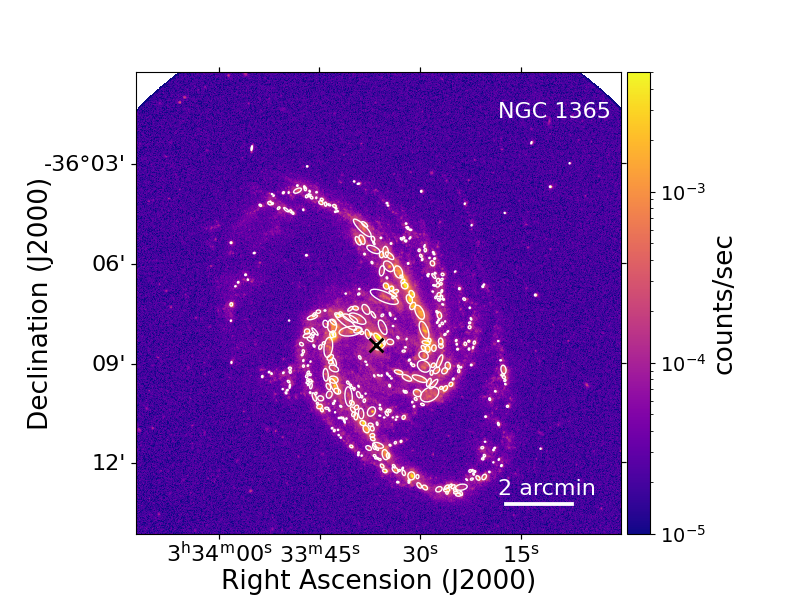}
    \hspace{-0.4cm}\includegraphics[scale=0.32]{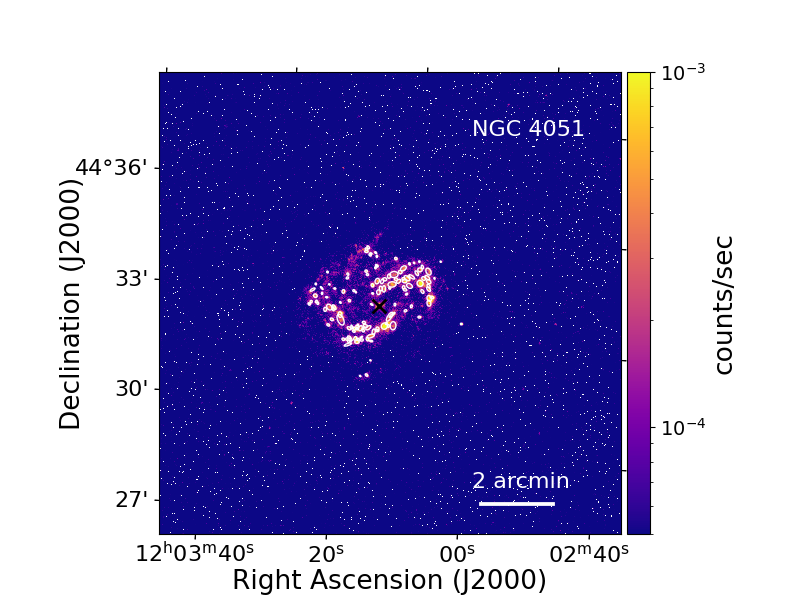}
    \hspace{-0.4cm}\includegraphics[scale=0.32]{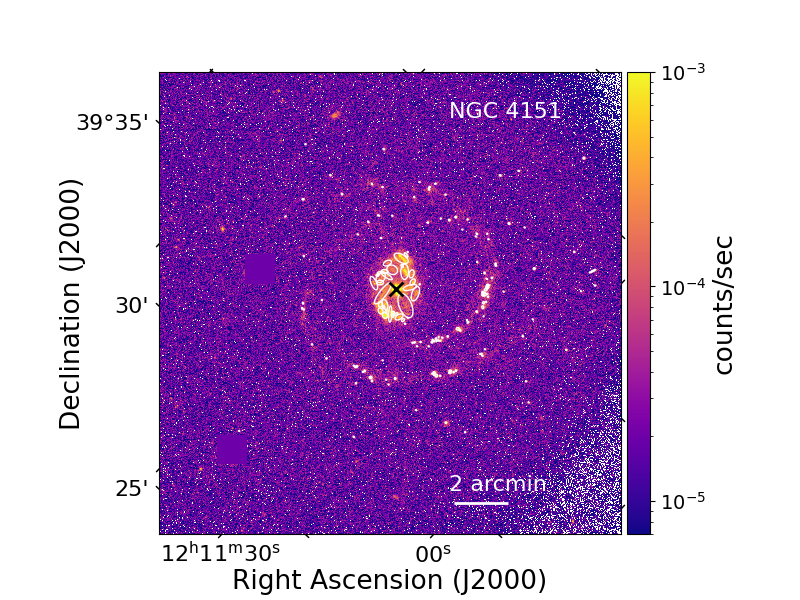}
    }
\hbox{
     \hspace{-0.5cm}\includegraphics[scale=0.32]{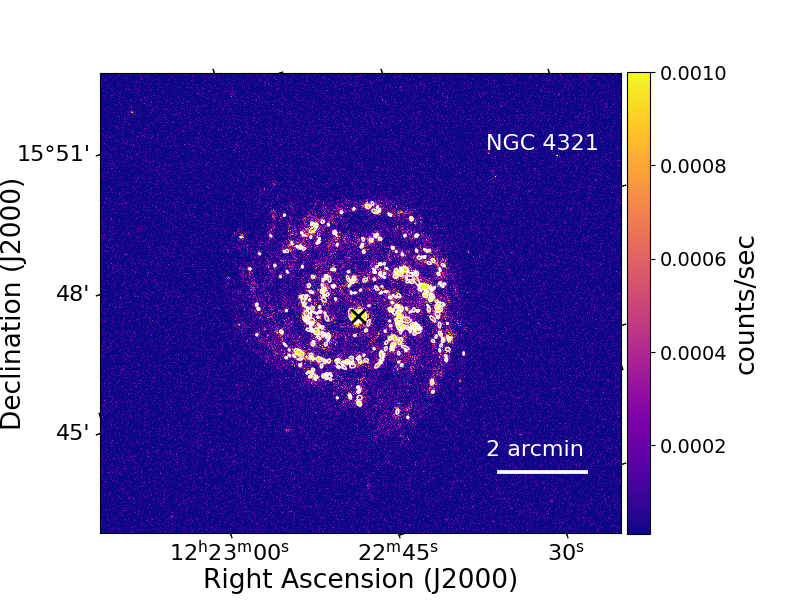}
     \hspace{-0.4cm}\includegraphics[scale=0.32]{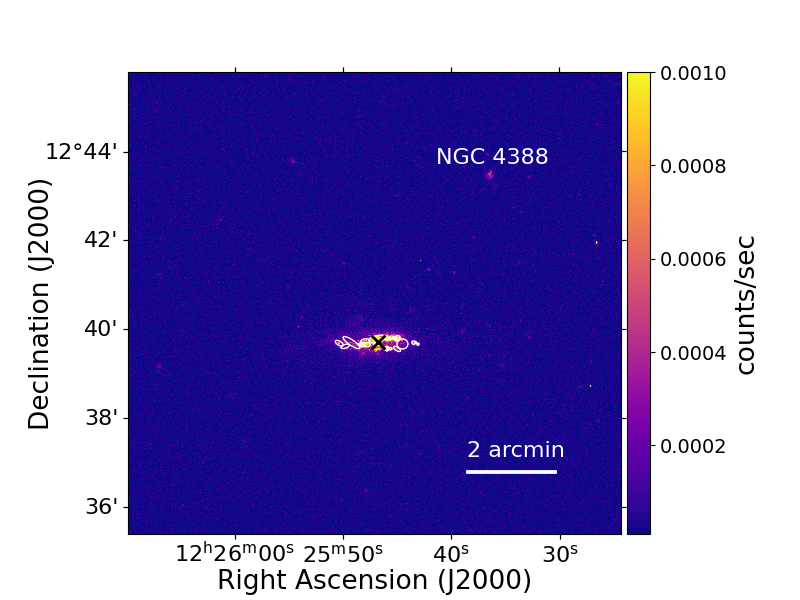}
     \hspace{-0.4cm}\includegraphics[scale=0.32]{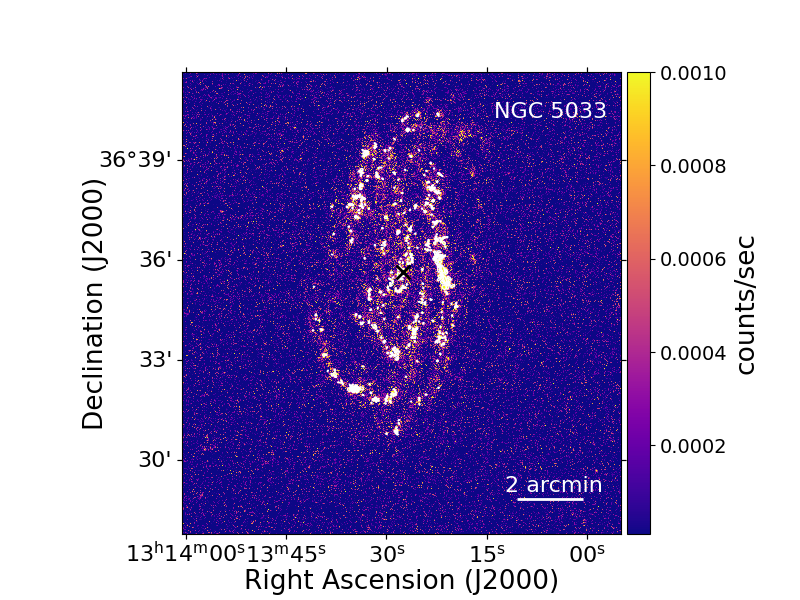}
     }
\hbox{
     \hspace{-0.5cm}\includegraphics[scale=0.32]{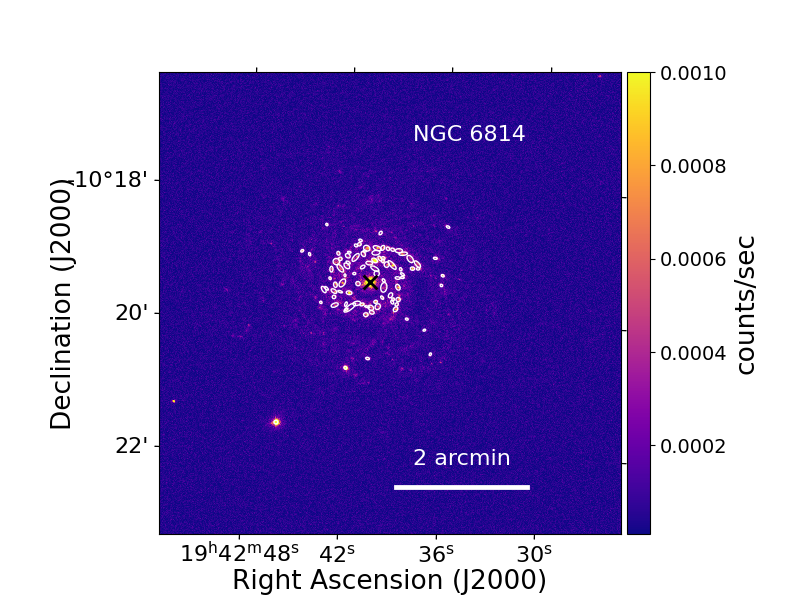}
     \hspace{-0.4cm}\includegraphics[scale=0.32]{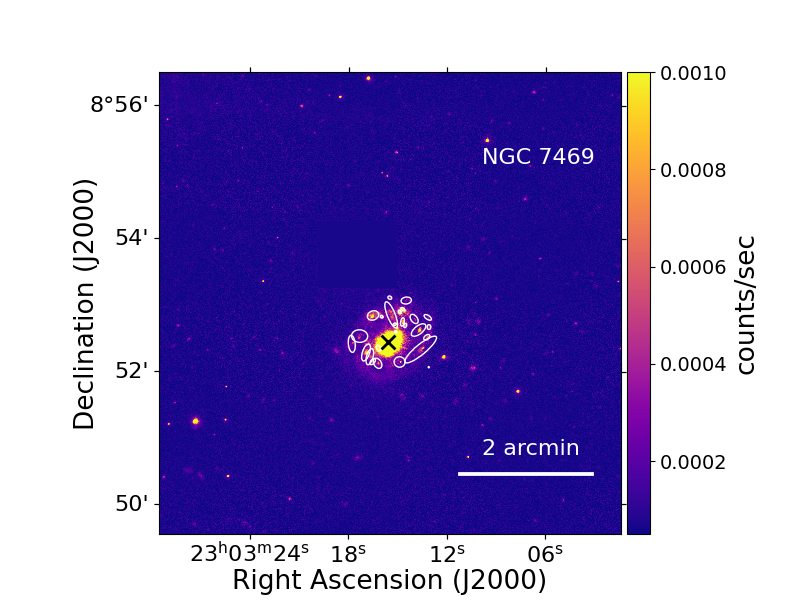}
     }
}
    \caption{FUV images of the objects with the identified star-forming regions marked in white ellipses. The central AGN is marked as a black cross.
     }
    \label{fig:reg}
\end{figure*}

The motivation of this work is to understand the SF characteristics of the host galaxies of AGN. This involves
detection of star forming regions in the galaxies. For this we used the 
{\tt SExtractor} module \citep{1996A&AS..117..393B} in Python. To identify star 
forming regions within each galaxy we adopted a 5$\sigma$ threshold criterion and
followed the details outlined in \cite{2023ApJ...950...81N}. These identified
star forming regions might have contamination from foreground objects. To remove such
foreground objects, we cross-matched the identified regions with the 
{\it Gaia-DR3} \citep{2022arXiv220800211G}
catalogue setting a cutoff in the proper motion of the regions at 10 mas/yr. If 
a detected SF region had a counterpart within 2 arcsecs in the {\it Gaia} catalogue with 
a proper motion exceeding 10 mas/yr, we excluded that SF region from our analysis.
Thus, in total we identified 418 regions in NGC~1365, 131 regions in NGC~4051, 161 regions in NGC~4151, 340 regions in NGC~4321, 20 regions in NGC~4388, 557 regions in NGC~5033, 89 regions in NGC~6814 and 26 regions in NGC~7469. These identified regions for all the galaxies are marked and are shown in Fig. \ref{fig:reg}.

\begin{figure*}
    \centering
    \vbox{
    \hbox{
    \includegraphics[scale=0.30]{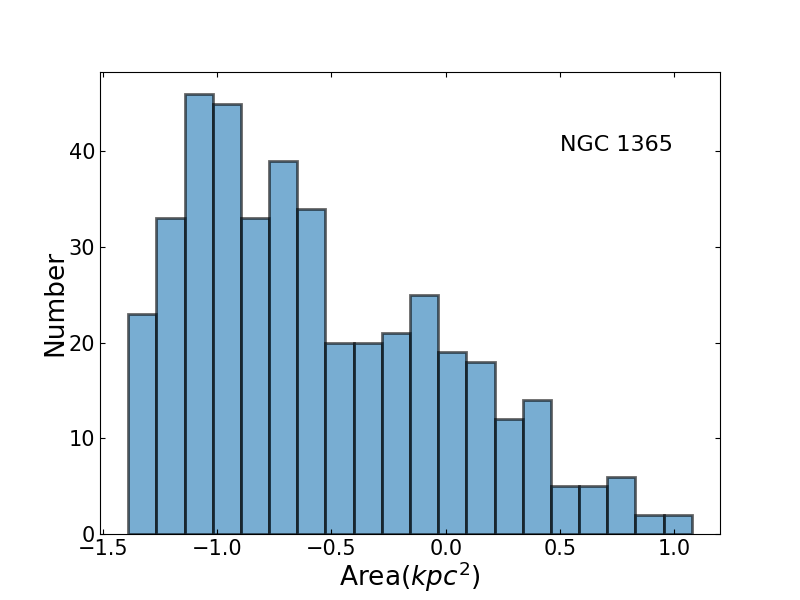}
    \includegraphics[scale=0.30]{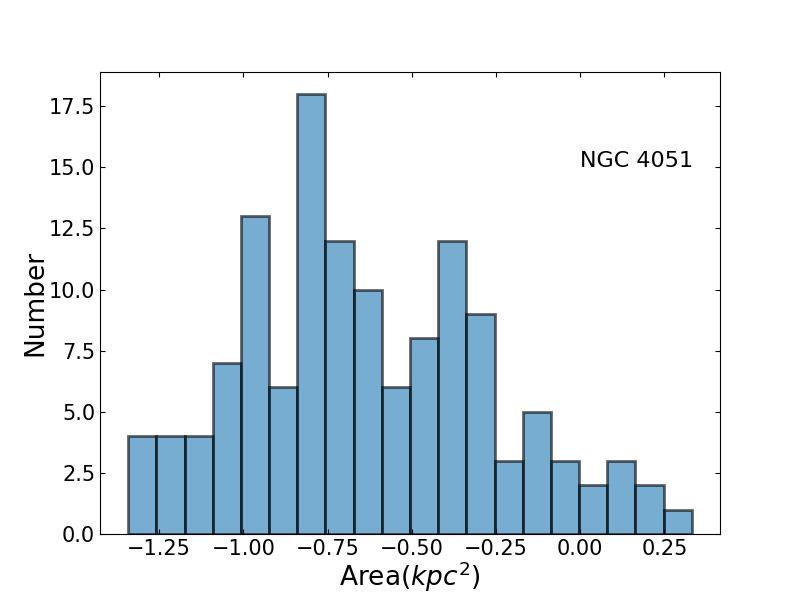}
    \includegraphics[scale=0.30]{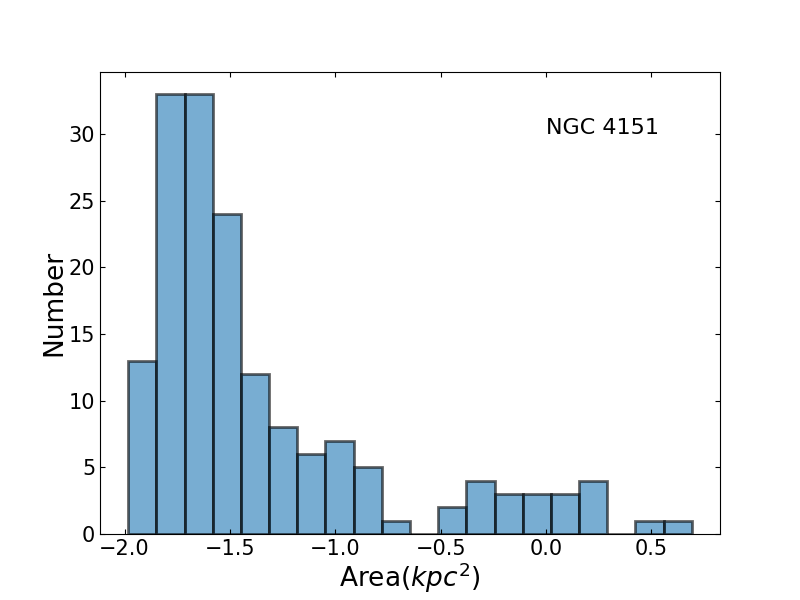}
    }
    \hbox{
    \includegraphics[scale=0.30]{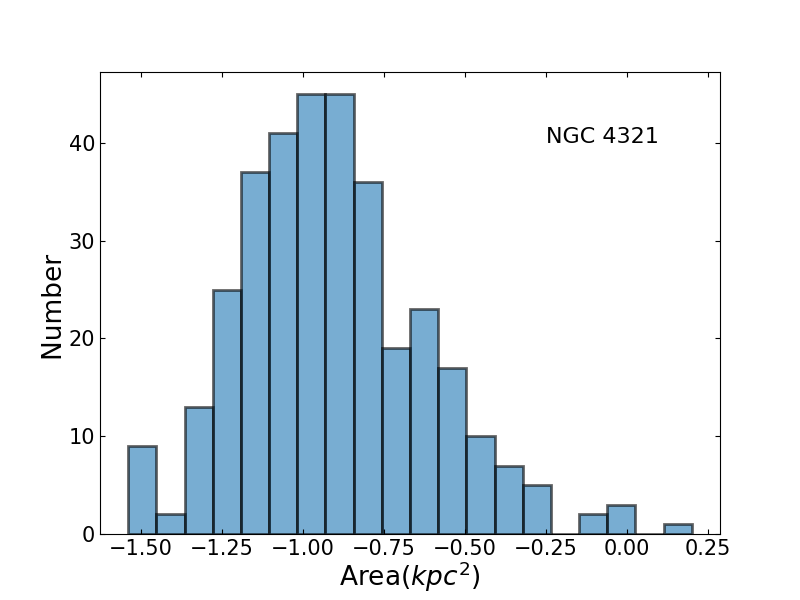}
    \includegraphics[scale=0.30]{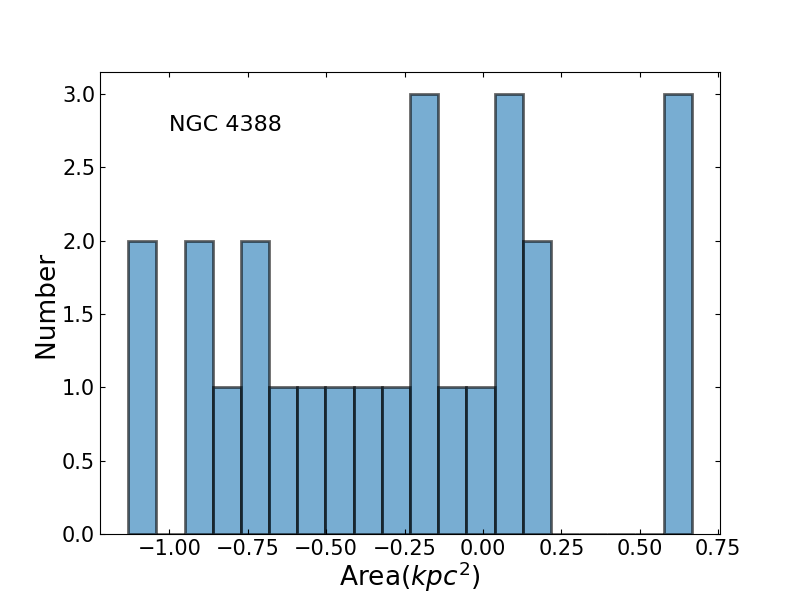}
    \includegraphics[scale=0.30]{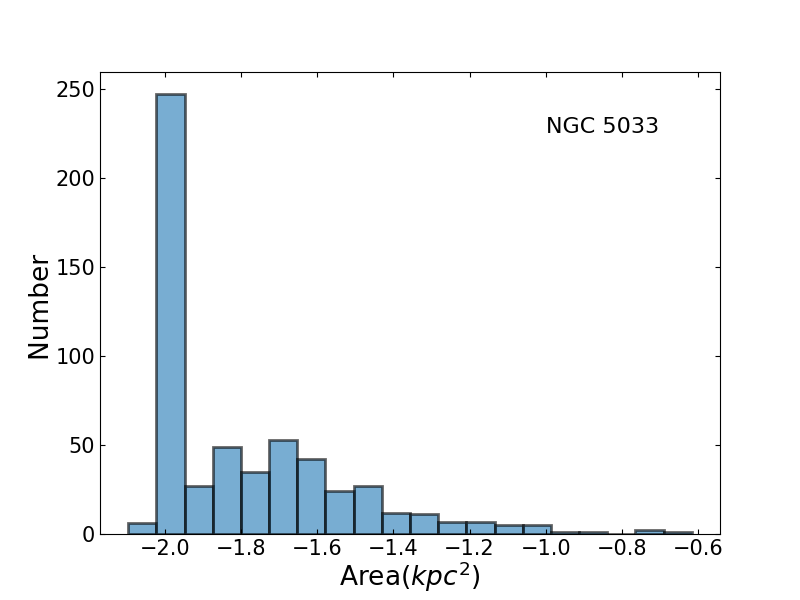}
    }
    \hbox{
    \includegraphics[scale=0.30]{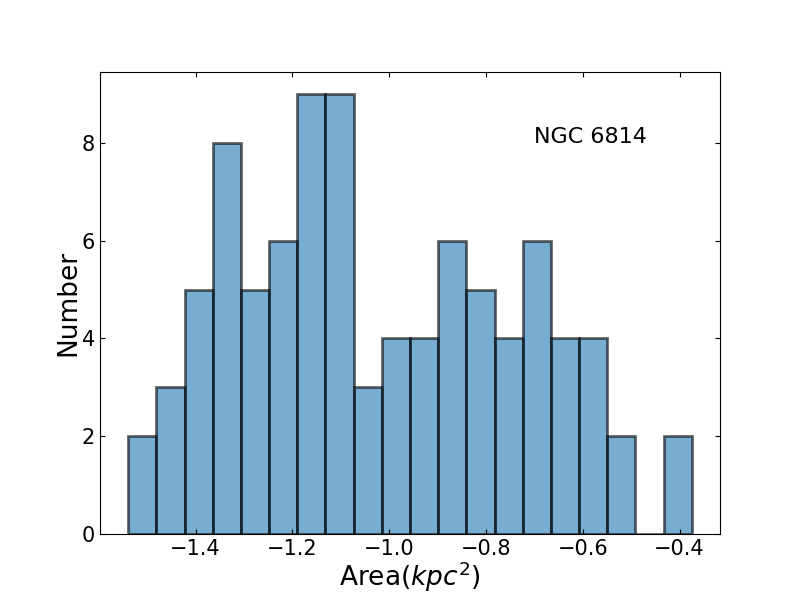}
    \includegraphics[scale=0.30]{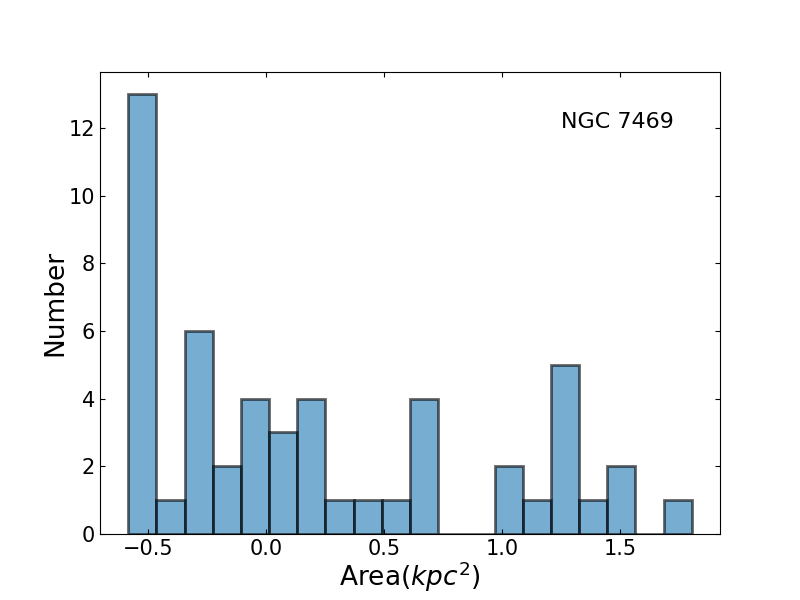}
    }
    }
    \caption{Distributions of the area (log scale) of the identified star-forming regions in 
the sources studied in this work. The names of the sources are given in their respective
panels.}
    \label{fig:area}
\end{figure*}

The sizes of the identified star-forming regions varied, ranging from the point 
spread function (PSF) of the instrument to approximately 4$-$5 times its size. 
To correct for the instrumental resolution, we adopted an assumption of 
elliptical light distribution within the aperture, following the approach 
given in \cite{2023ApJ...950...81N}.  These corrected apertures were subsequently utilized 
for further analysis. The distribution of the area of the identified star forming
regions in each of the galaxies is shown in Fig. \ref{fig:area}.  We determined 
the flux of each of the star forming regions 
via aperture photometry, using {\tt photutils} package \citep{larry_bradley_2020_4044744}. 
We corrected these flux measurements for extinction.  For Milky Way extinction 
correction, we relied on the prescription from \cite{1989ApJ...345..245C} and 
to correct for internal extinction, we used the UV slope ($\beta$)
method \citep{2000ApJ...533..682C}.  For measurements in two UV 
filters, one in FUV and the other in NUV,  $\beta$ was calculated as
\begin{equation}
    \beta =  \frac{m_{FUV} - m_{NUV}} {-2.5 log(\lambda_{FUV}/\lambda_{NUV})} -2.0 
    \label{eq:beta}
\end{equation}

\begin{figure*}
    \centering
    \vbox{
    \hbox{
    \includegraphics[scale=0.30]{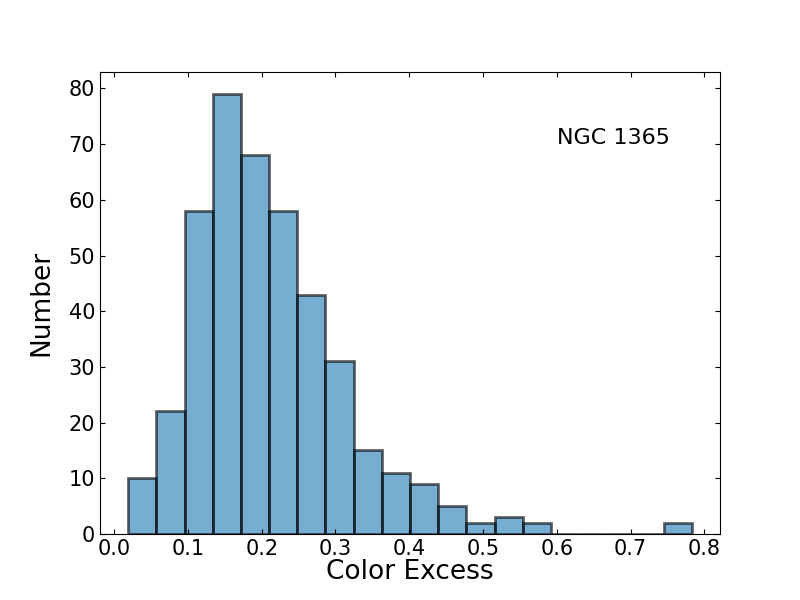}
    \includegraphics[scale=0.30]{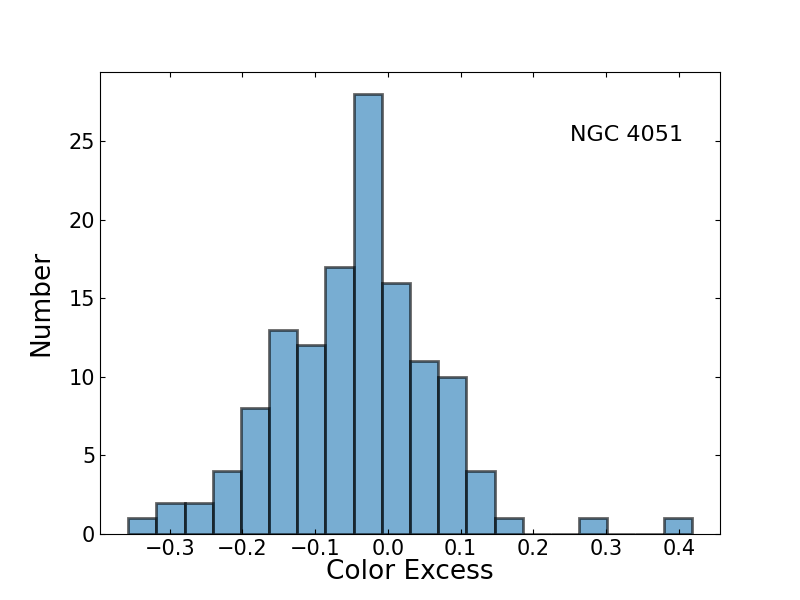}
    \includegraphics[scale=0.30]{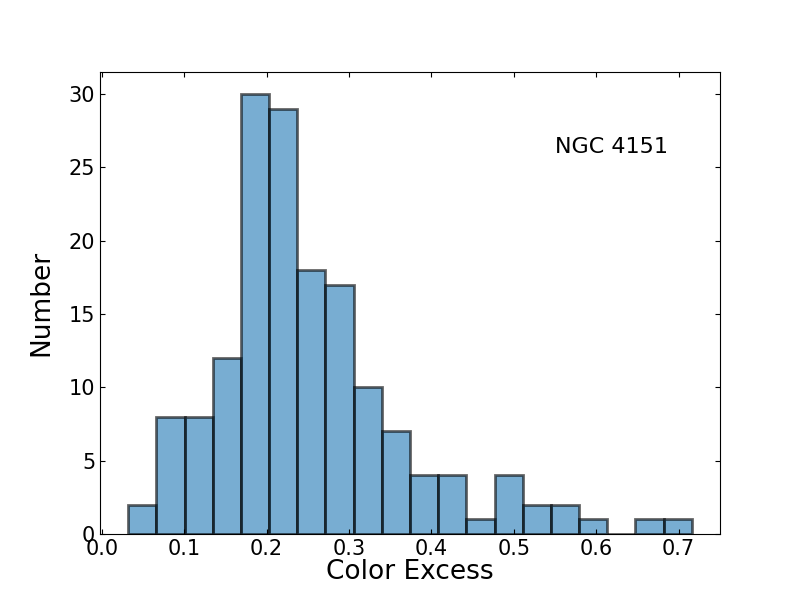}
    }
    \hbox{
    \includegraphics[scale=0.30]{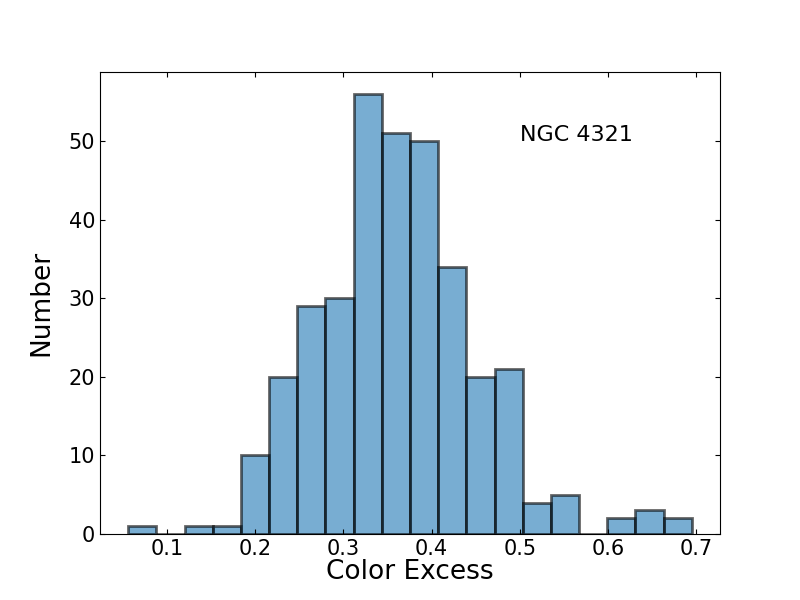}
    \includegraphics[scale=0.30]{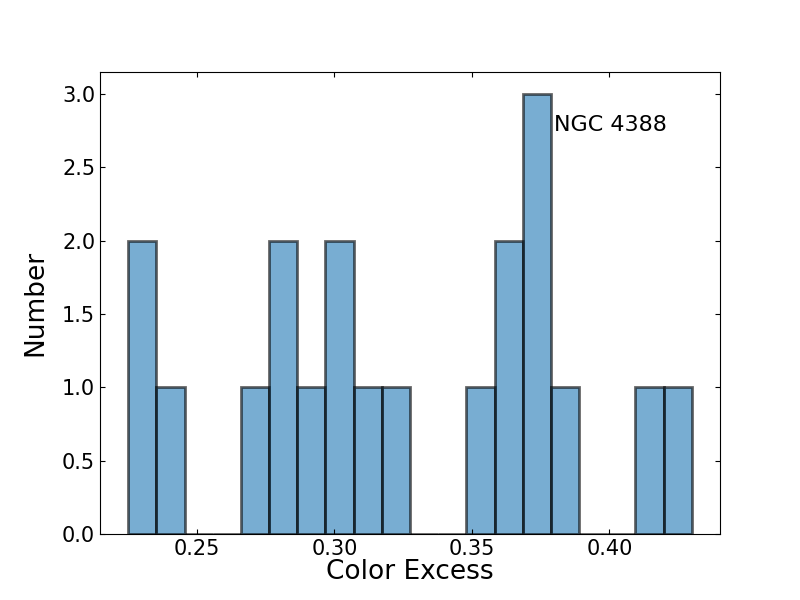}
    \includegraphics[scale=0.30]{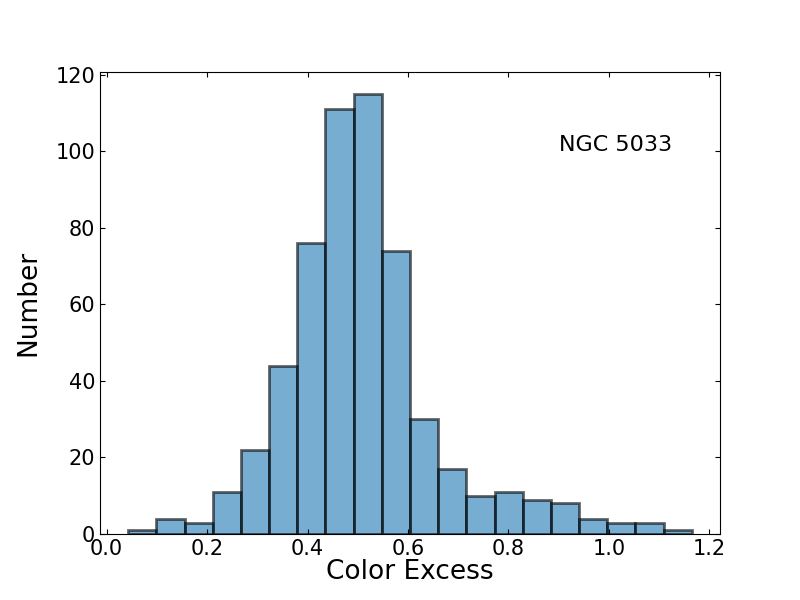}
    }
    \hbox{
    \includegraphics[scale=0.30]{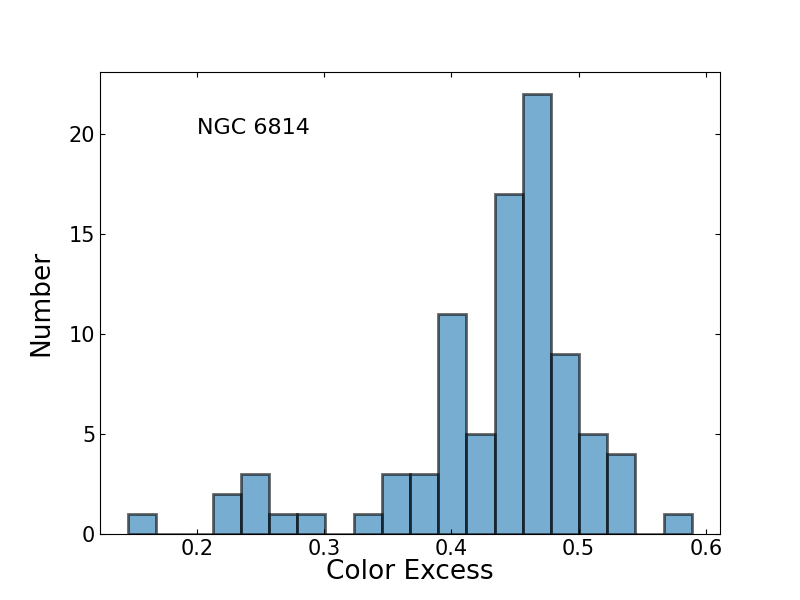}
    \includegraphics[scale=0.30]{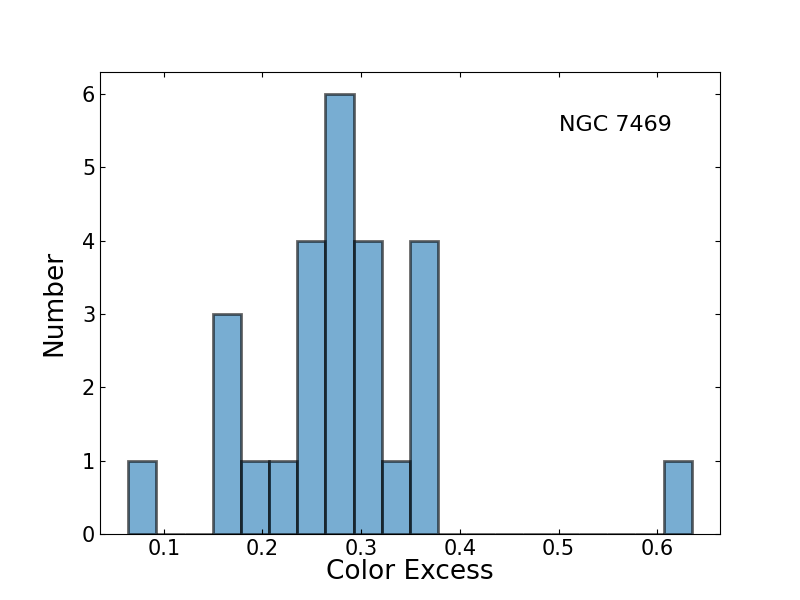}
    }
    }
    \caption{Distributions of E(B$-$V) of the identified star-forming regions in 
the sources studied in this work. The names of the sources are given in their respective
panels.}
    \label{fig:ebv}
\end{figure*}

Here, $m_{FUV}$ and $m_{NUV}$ are the magnitudes in FUV and NUV filters with
wavelengths $\lambda_{FUV}$ and $\lambda_{NUV}$ respectively. The values of
$\beta$ give an idea of the dust obscuration in the 
star forming regions. Using the calculated
$\beta$, we estimated the colour excess, E(B$-$V) using the following relation
\begin{equation}
E(B-V)= (\beta + 2.616)/4.684
\label{eq:ebv}
\end{equation}
The distribution of the E(B$-$V) values for the identified star forming 
regions are given in Fig. \ref{fig:ebv}.
We estimated the internal extinction at any particular wavelength, $\lambda$ as
\begin{equation}
A_{\lambda} = 0.44\times E(B-V)\times k^{\prime}({\lambda})
\label{eq:A_lam}
\end{equation}
where, $k^{\prime}({\lambda}$) is from \cite{2000ApJ...533..682C}. 

\begin{figure*}
    \centering
    \hbox{
    \vbox{
    \hbox{
    \includegraphics[scale=0.21]{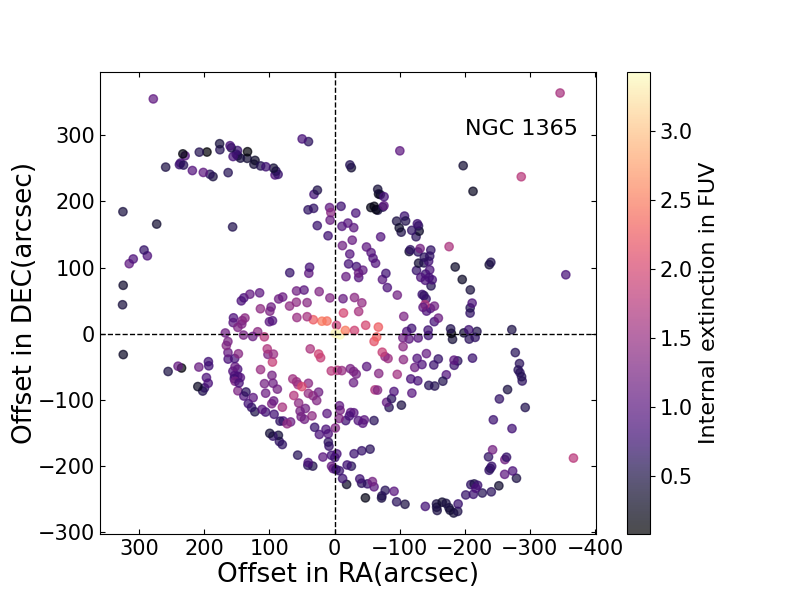}
    \includegraphics[scale=0.21]{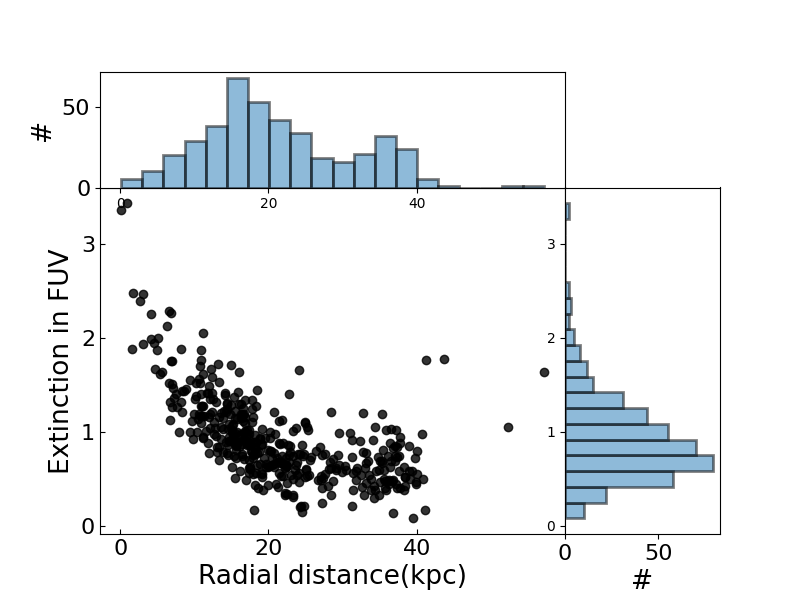}
    }
    \hbox{
    \includegraphics[scale=0.21]{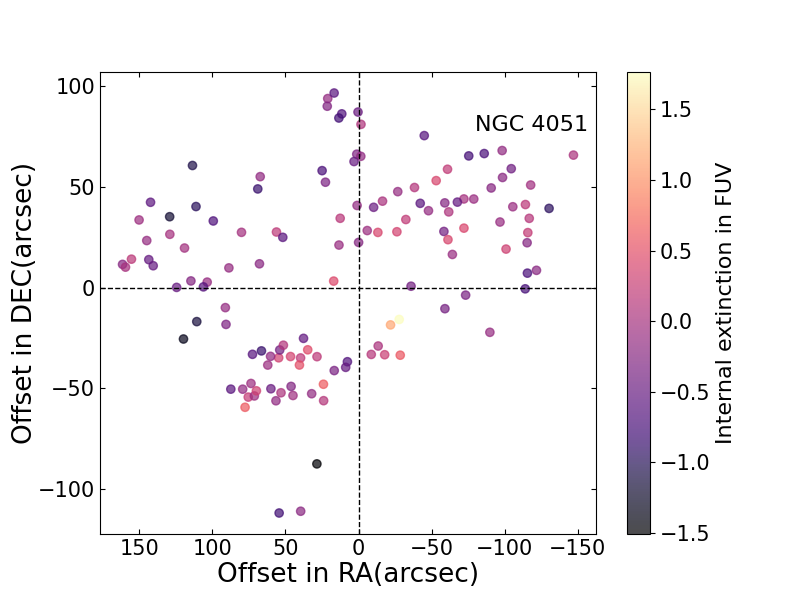}
    \includegraphics[scale=0.21]{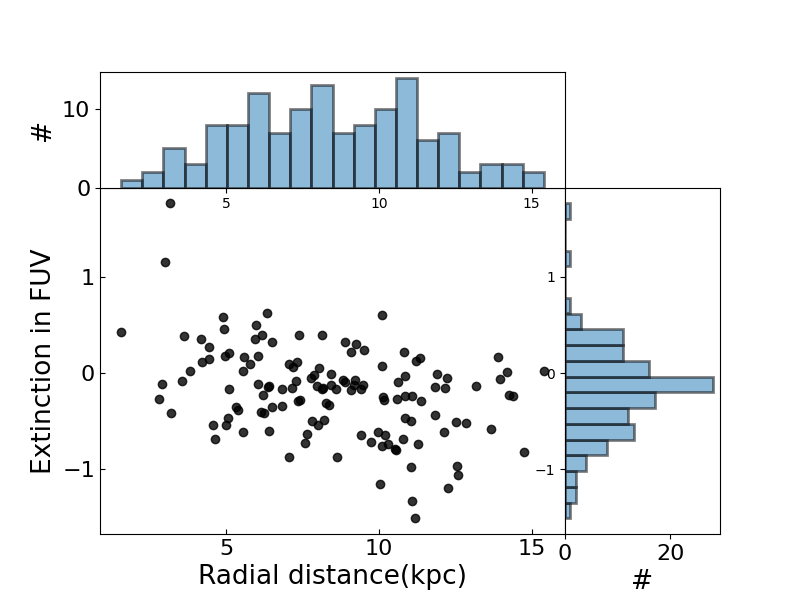}
    }
    \hbox{
    \includegraphics[scale=0.21]{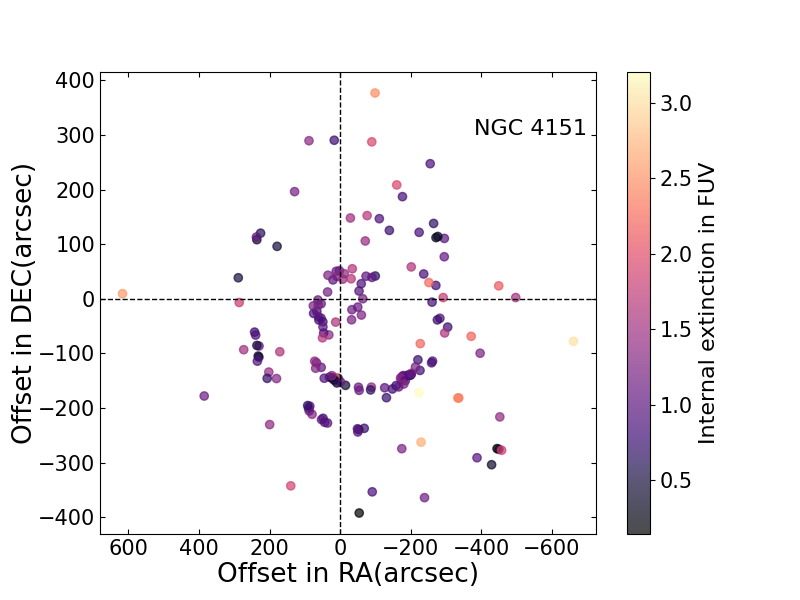}
    \includegraphics[scale=0.21]{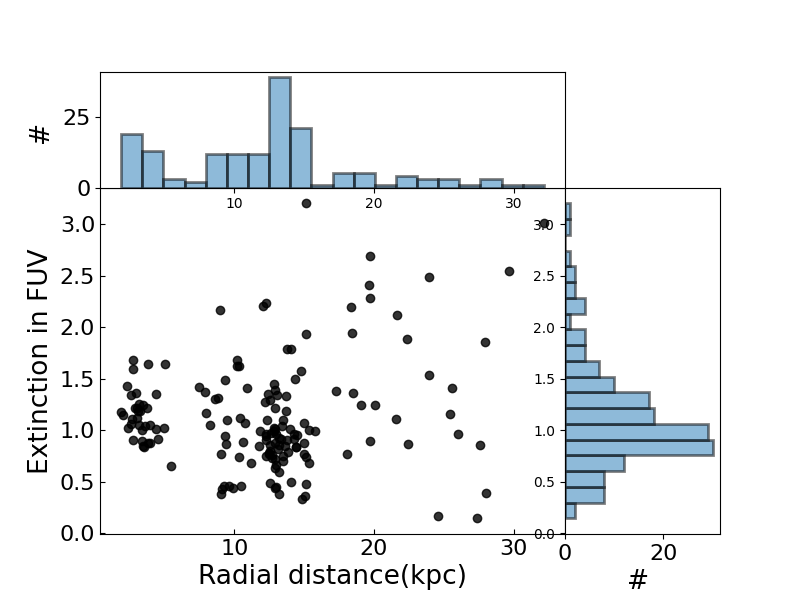}
    }
    \hbox{
    \includegraphics[scale=0.21]{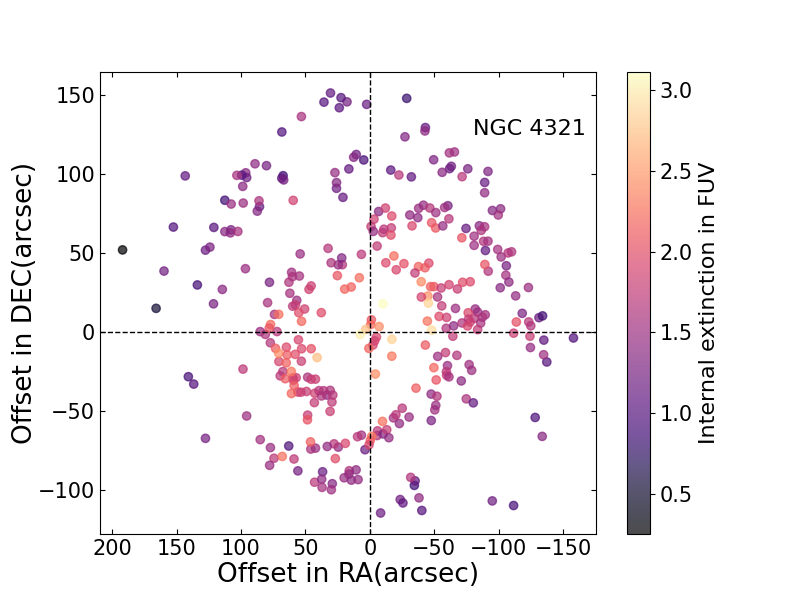}
    \includegraphics[scale=0.21]{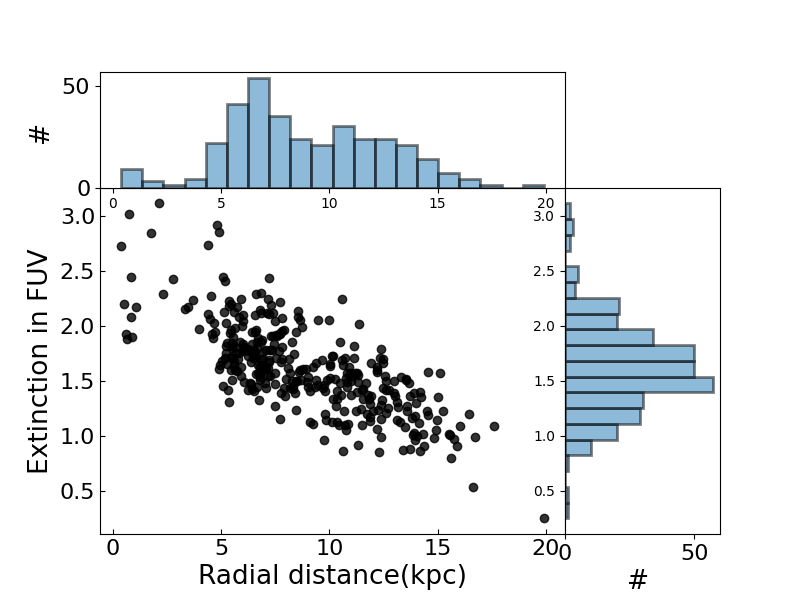}
    }
}
    \vbox{
    \hbox{
    \includegraphics[scale=0.21]{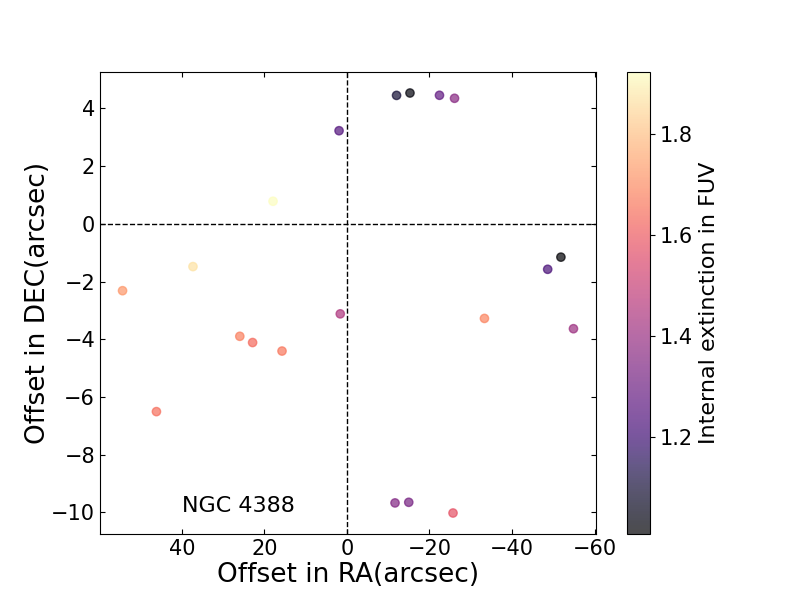}
    \includegraphics[scale=0.21]{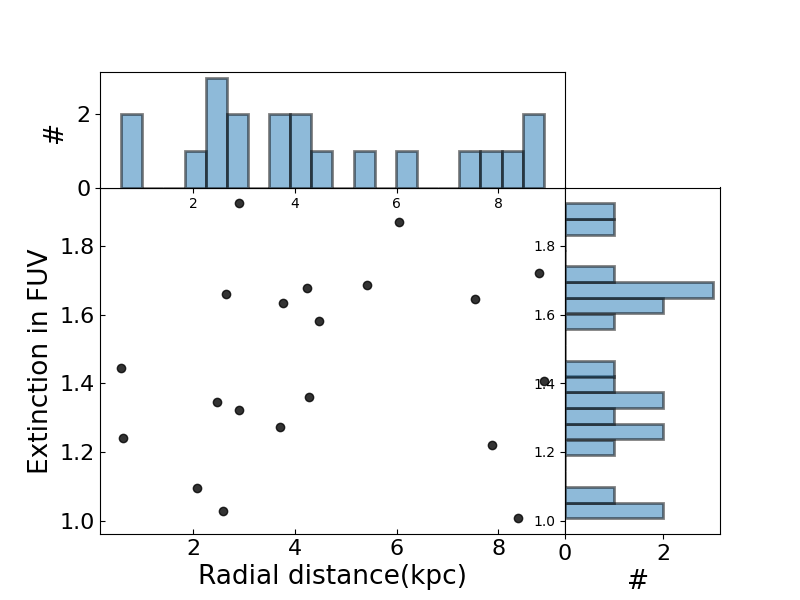}
    }
    \hbox{
    \includegraphics[scale=0.21]{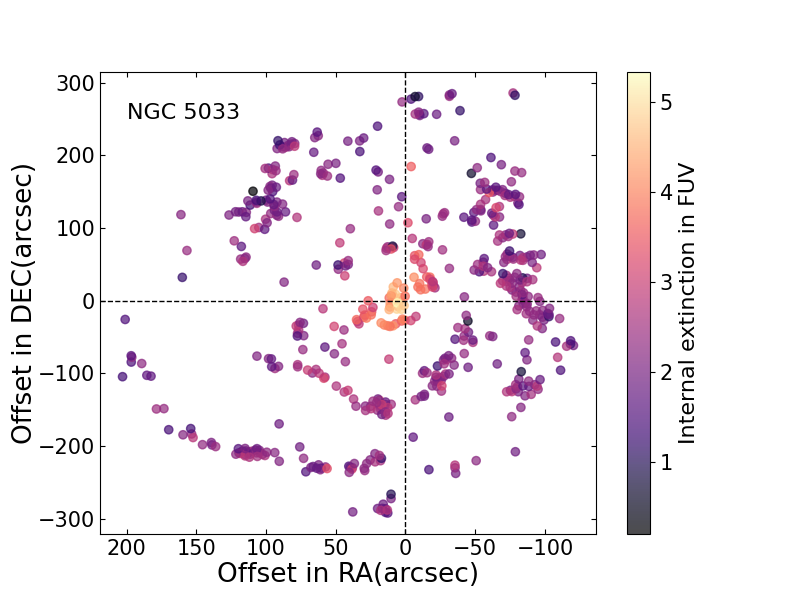}
    \includegraphics[scale=0.21]{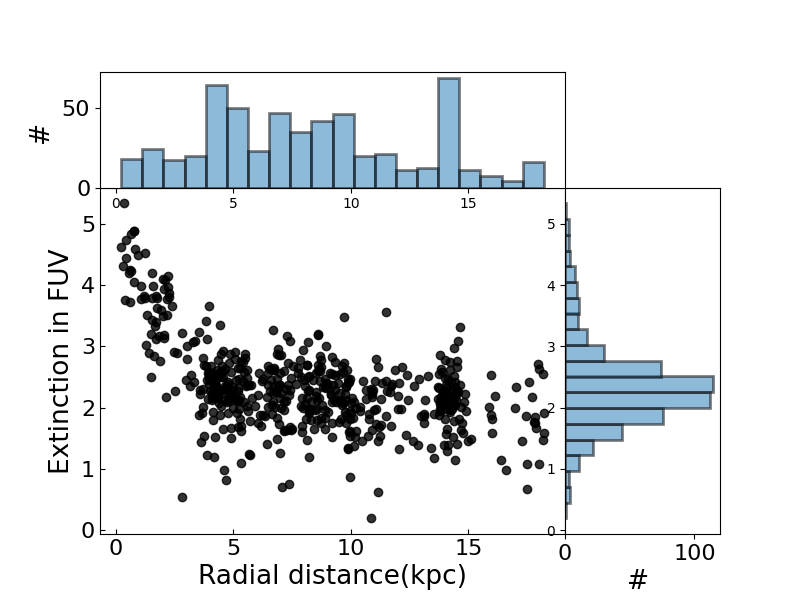}
    }
    \hbox{
    \includegraphics[scale=0.21]{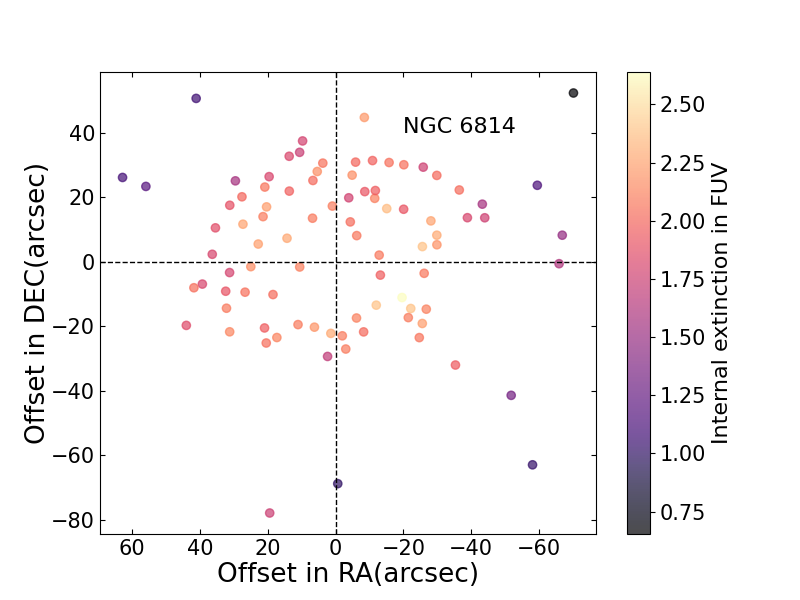}
    \includegraphics[scale=0.21]{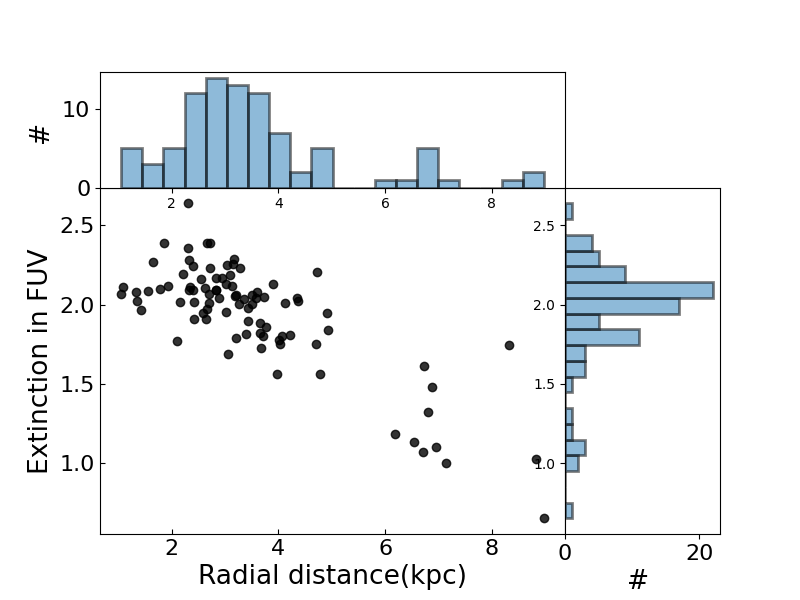}
    }
    \hbox{
    \includegraphics[scale=0.21]{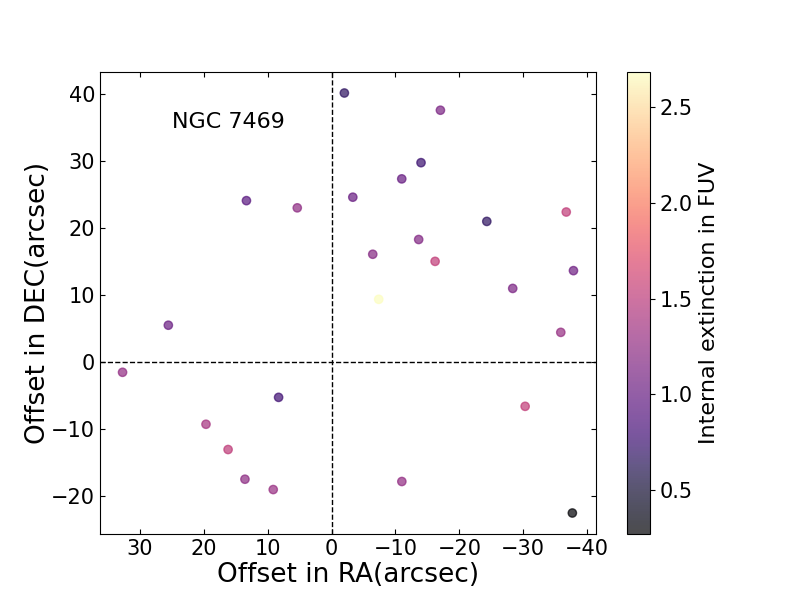}
    \includegraphics[scale=0.21]{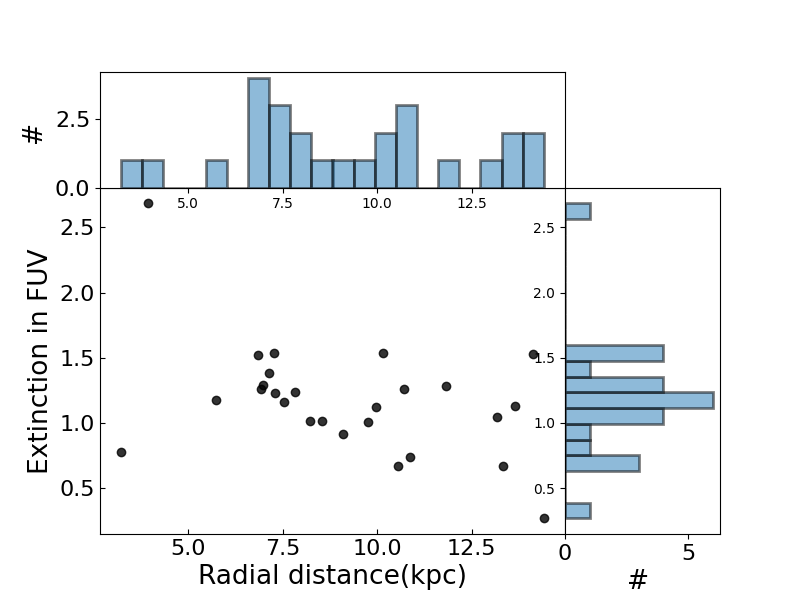}
    }
    }
    }
    \caption{Spatial (left panel) and radial (right panel) variation of internal extinction for each source.}
    \label{fig:extin}
\end{figure*}

In cases where NUV data from UVIT were unavailable, 
we calculated the UV slope using two FUV filters. For regions where neither FUV 
and NUV filters nor two FUV filters were available from UVIT, we utilized the 
GALEX FUV and NUV images. In such instances, we convolved the instrument-corrected 
aperture with the GALEX PSF and employed those apertures 
to determine the extinction correction. The spatial and radial variations of 
internal extinction for our sample of galaxies are shown in Fig. \ref{fig:extin}.

\begin{figure*}
\hbox{
    \vbox{
    \hbox{
     \includegraphics[scale=0.21]{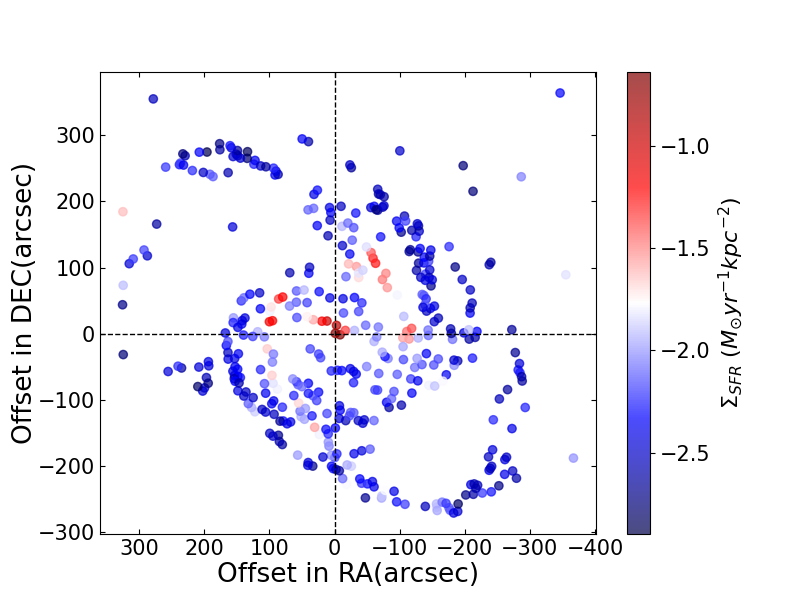}
     \includegraphics[scale=0.21]{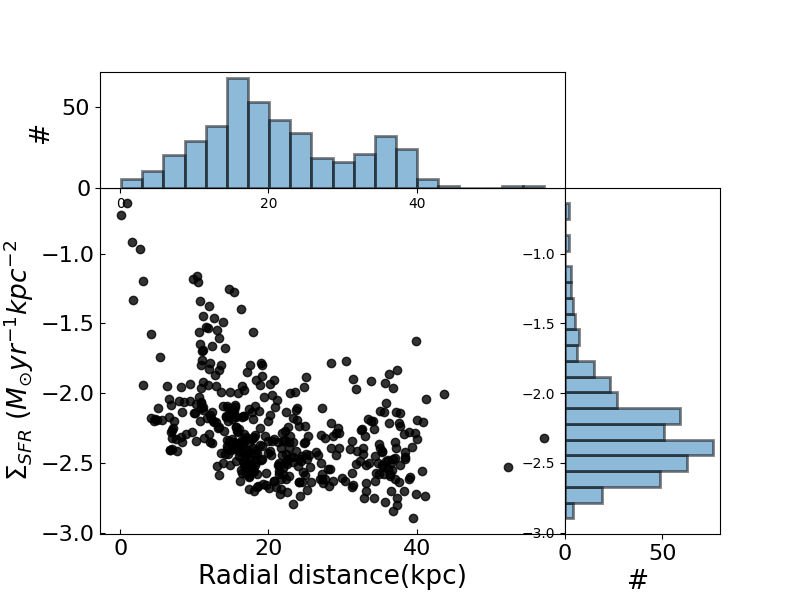}
     }
    \hbox{
     \includegraphics[scale=0.21]{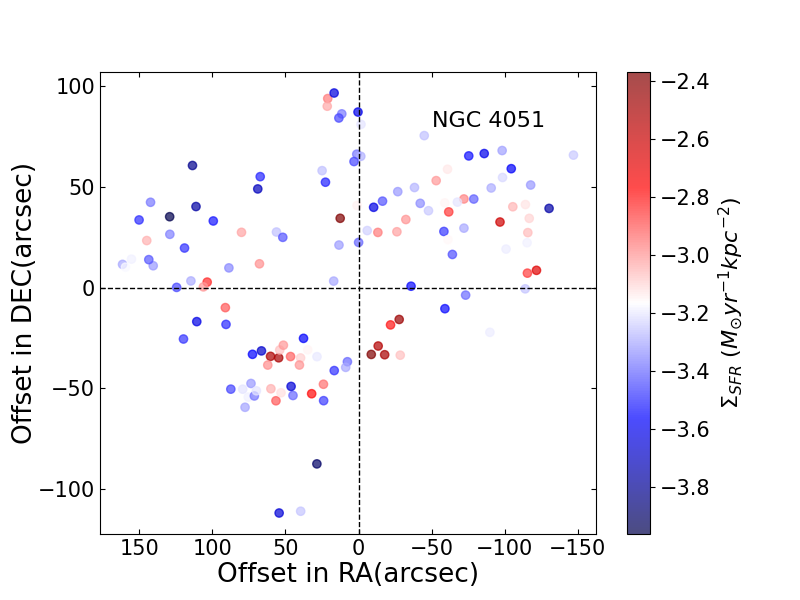}
     \includegraphics[scale=0.21]{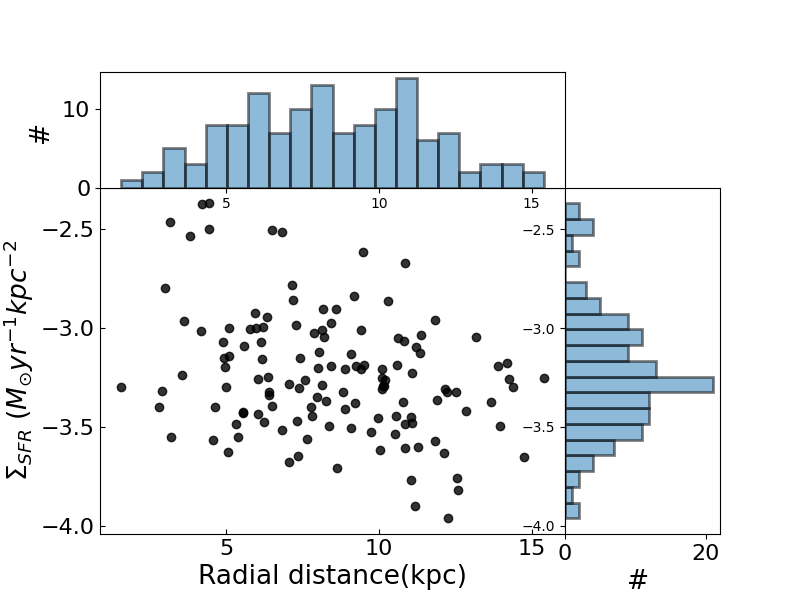}
     }
     \hbox{
     \includegraphics[scale=0.21]{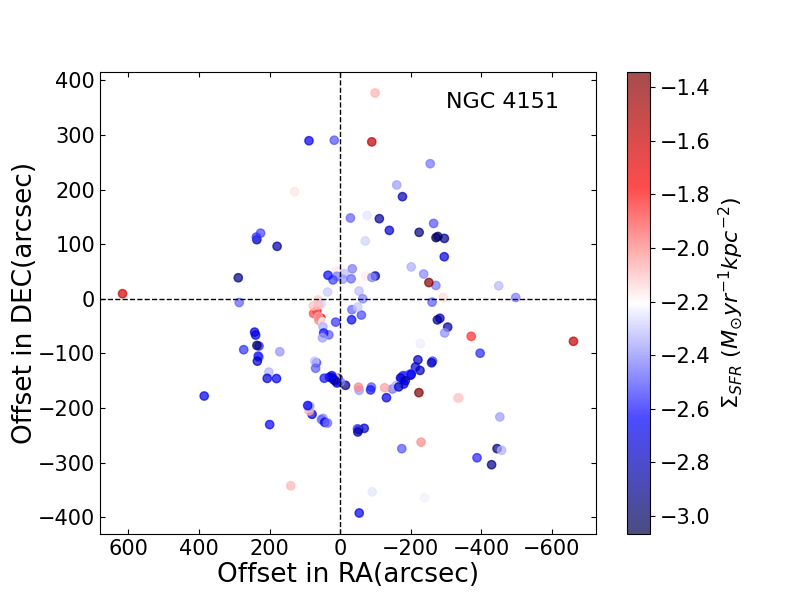}
     \includegraphics[scale=0.21]{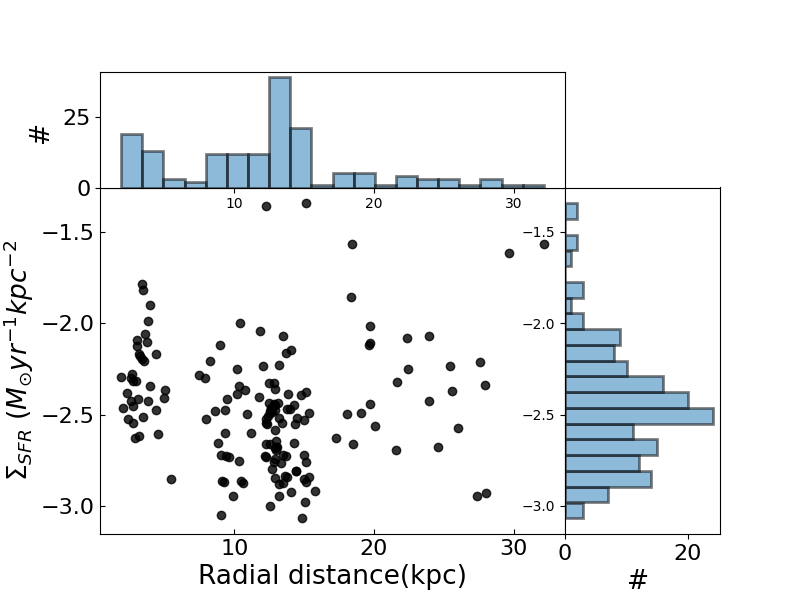}
     }
     \hbox{
     \includegraphics[scale=0.21]{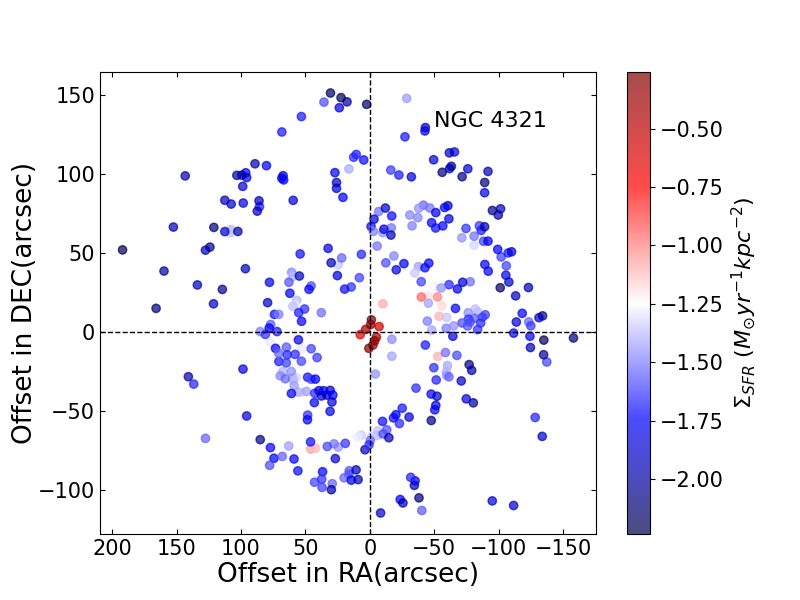}
     \includegraphics[scale=0.21]{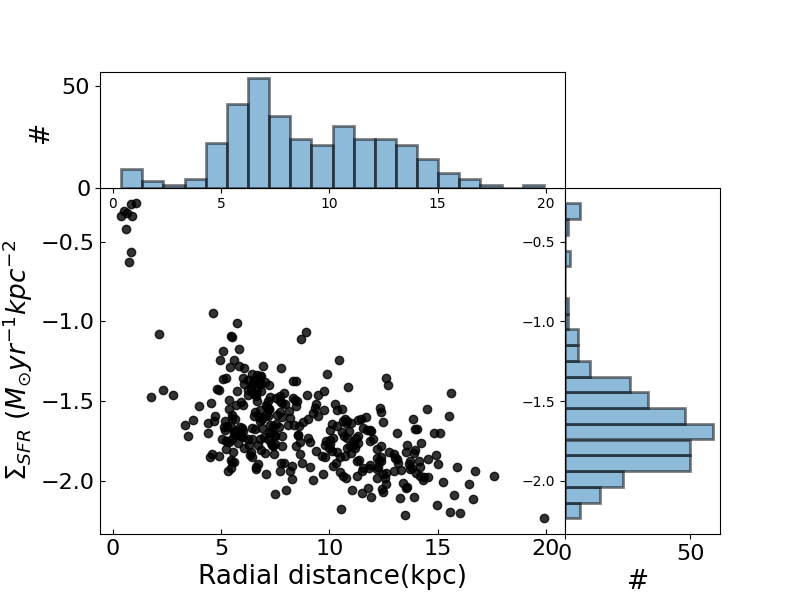}
     }
     }
     
    \vbox{
     \hbox{
     \includegraphics[scale=0.21]{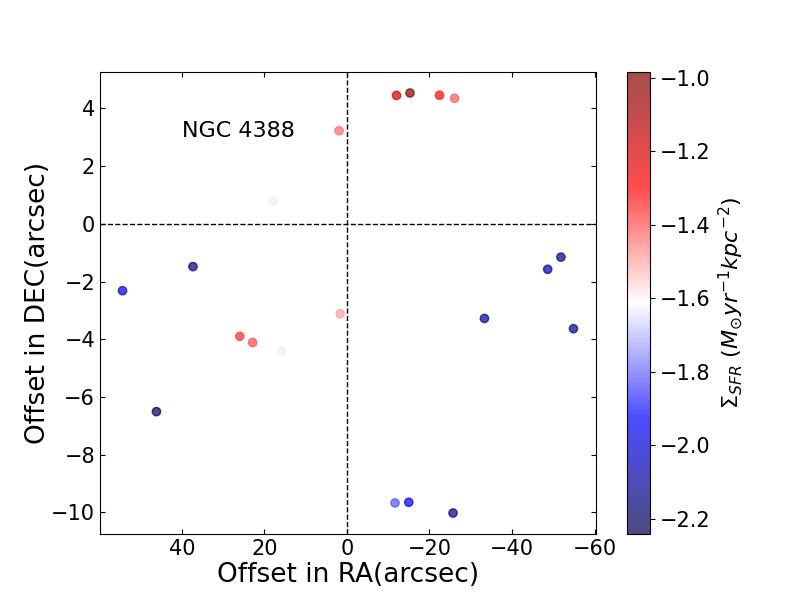}
     \includegraphics[scale=0.21]{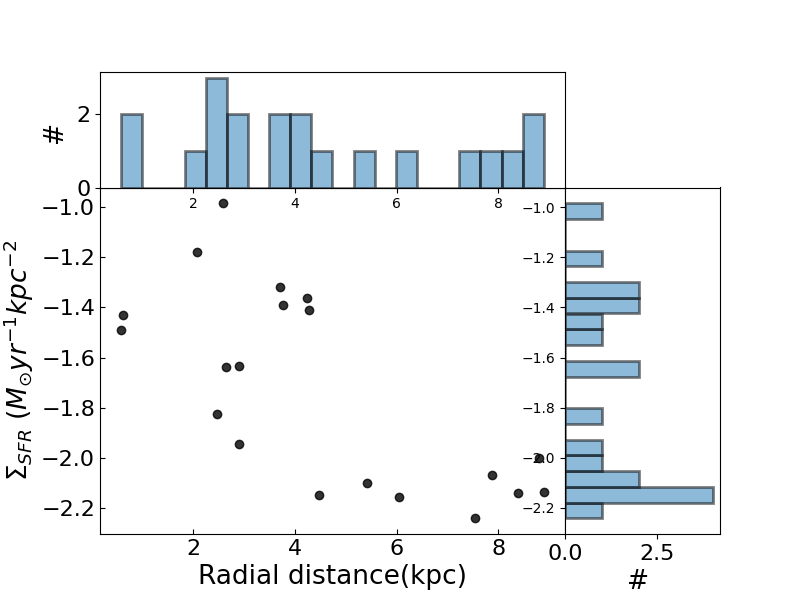}
     }
     \hbox{
     \includegraphics[scale=0.21]{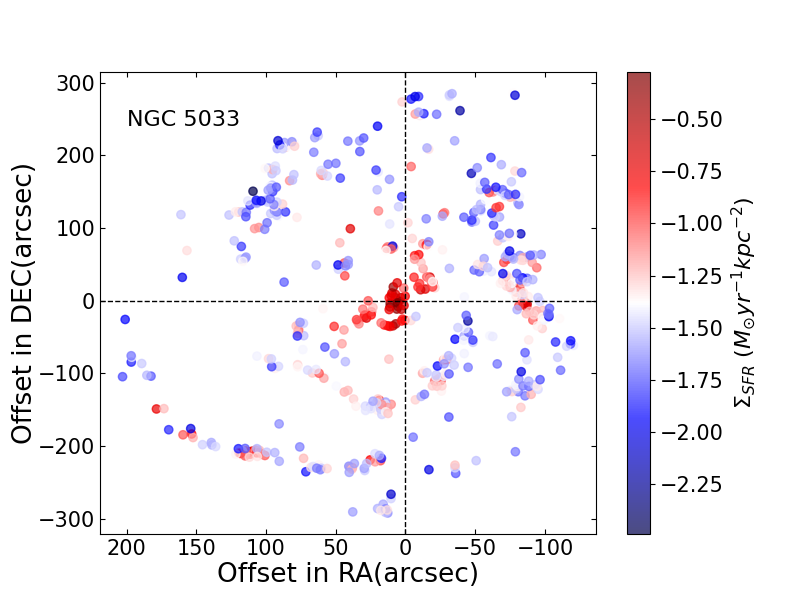}
     \includegraphics[scale=0.21]{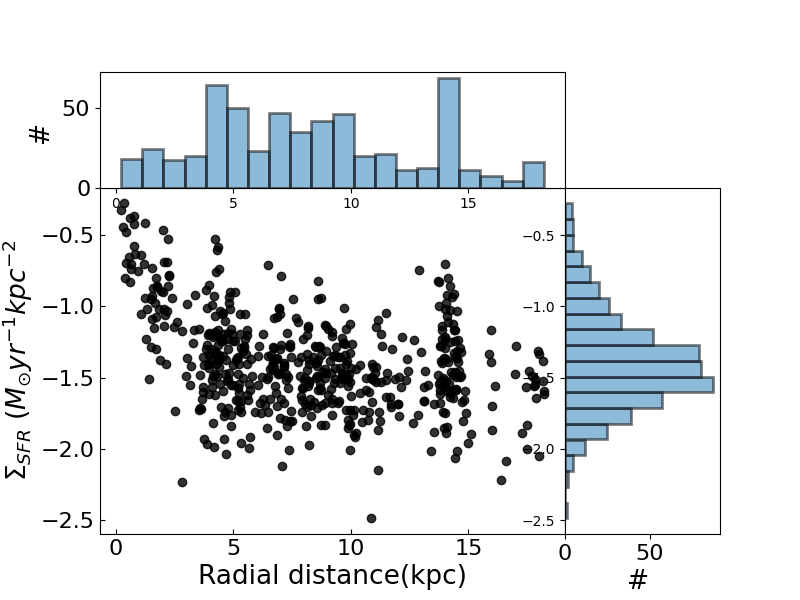}
     }
     \hbox{
     \includegraphics[scale=0.21]{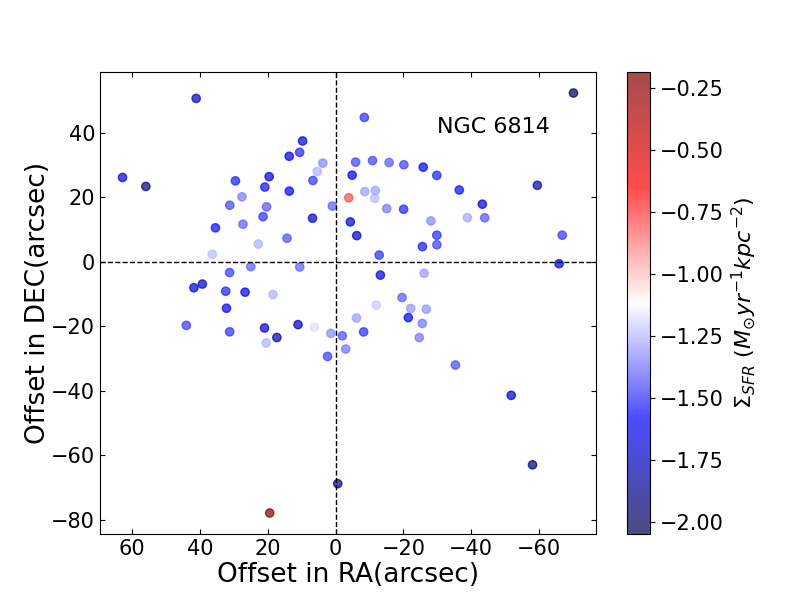}
     \includegraphics[scale=0.21]{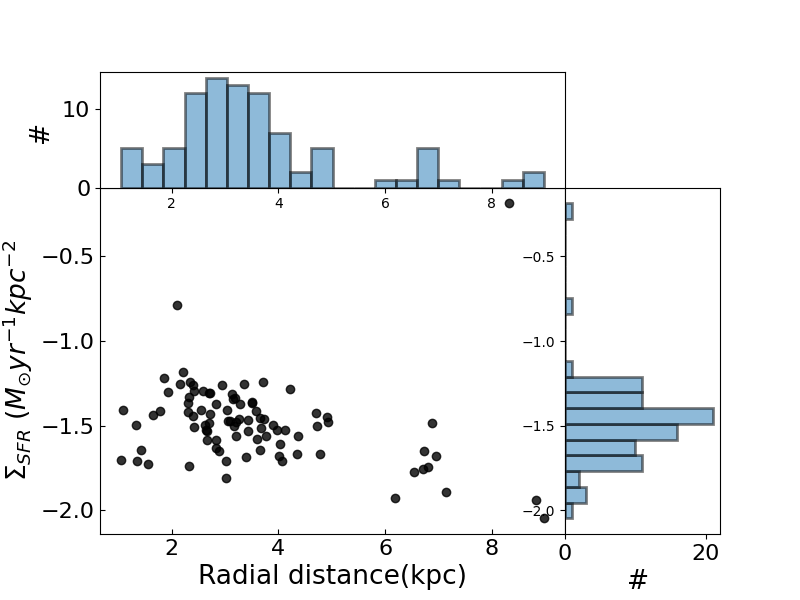}
     }
     \hbox{
     \includegraphics[scale=0.21]{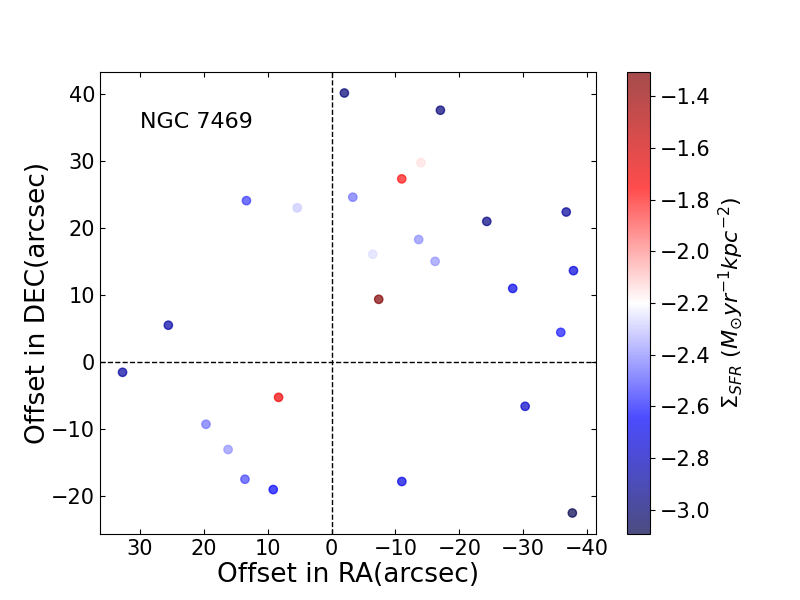}
     \includegraphics[scale=0.21]{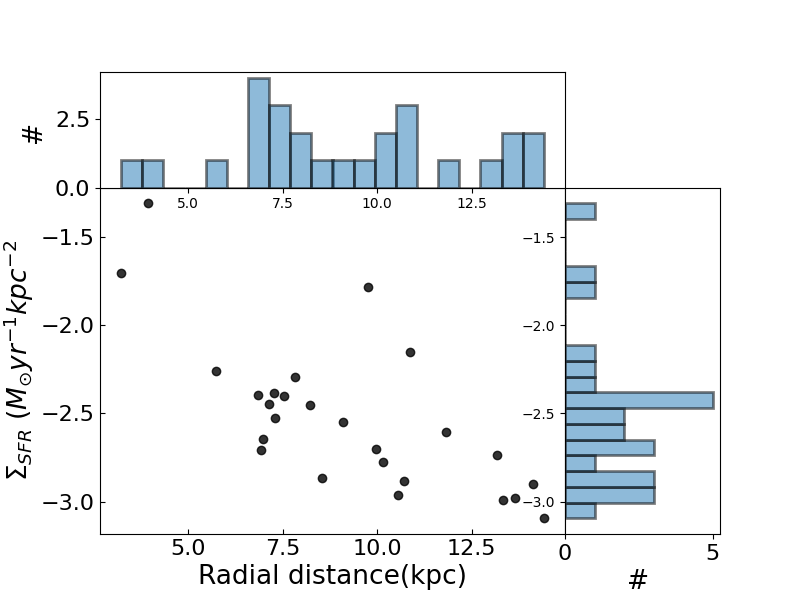}
     }
}
}
    \caption{Spatial (left panel) and radial (right panel) variation of 
$\Sigma_{SFR}$ in log scale.}
    \label{fig:sfr}
\end{figure*}

After getting the extinction value for each SF region, we calculated the intrinsic
luminosities of the star forming regions and used them to estimate the 
star formation rate (SFR) in NUV and FUV wavelengths as follows \citep{2007ApJS..173..267S}.

\begin{subequations}
    \begin{equation}
    log(SFR_{FUV}(M_\odot yr^{-1}))= log[L_{FUV}(W Hz^{-1})] -21.16
    \end{equation}
    \begin{equation}
    log(SFR_{NUV}(M_\odot yr^{-1}))= log[L_{NUV}(W Hz^{-1})] -21.14
    \end{equation}
    \label{eq:sfr}
\end{subequations}
We calculated the surface density of star formation rate ($\Sigma_{SFR}$) by 
taking the ratio of the SFR to the area of the star forming
regions. The spatial and radial variations of $\Sigma_{SFR}$ values
for our sample of sources are shown in Fig. \ref{fig:sfr}.

\section{Notes on Individual sources}\label{notes}

\subsection{NGC~1365}
NGC~1365 at a redshift $z$=0.005 \citep{Vauco91}, is a face-on barred spiral 
galaxy of SB(s)b type \citep{Vauco91} and harbouring a Seyfert 1.8 type 
AGN \citep{Mai95} with a black hole of mass  M$_{BH}$ $\sim$ 10$^{6.5}$ 
M$_\odot$ \citep{Ris09}. NGC 1365 with an X-ray luminosity of L$_{2–10keV}$ = 
10$^{42}$ erg s$^{-1}$ \citep{Ris05} is variable in X-rays, exhibiting 
variability on timescales of hours to years \citep{Bre13}. Notably, 
NGC~1365 is host to a rich population of star clusters and is a prominent galaxy 
within a 30 Mpc distance \citep{Whit23}. At radio frequencies, detection of a ring-like structure in the circumnuclear region with multiple knots of emission was reported by \cite{1994MNRAS.270...46S}. This was followed by a more detailed study which showed the knots of emission to have a non-thermal spectrum, and a suggestion of a possible jet-like structure along a PA$\sim 125^\circ$ \citep{1995A&A...295..585S}. \cite{1999MNRAS.306..479S} confirm the non-thermal nature of the knots and suggest that these are mostly due to multiple supernova remnants (SNRs). They also suggest that the jet-like structure could be due to a nuclear bar which extends to the ring.

We identified 418 star forming regions with the faintest region of 23.56 mag in 
FUV, while the brightest region has a brightness of 13.70 mag in the same FUV 
band. In the NUV, the faintest and the brightest SF regions have brightness of 
23.81 mag and 12.79 mag respectively. These star forming regions 
span a broad range of sizes. Their sizes range from 0.042 $-$ 11.93 kpc$^2$, with 
average and median values of 0.713 kpc$^2$ and 0.215 kpc$^2$ respectively. 

We found the color excess, E(B$-$V) to vary from 0.019 mag to 0.783 mag, with an average of 
0.212 mag and a median of 0.195 mag. In FUV band, the extinction ranged from 0.08 mag to 
3.428 mag, with an average of 0.93 mag and a median of 0.86 mag. Meanwhile, in 
the NUV band, the extinction values range from 0.06 mag to 2.51 mag, with a 
median of 0.63 mag and an average of 0.68 mag. We found high extinction 
in the inner regions compared to the outer regions (see Fig. \ref{fig:extin}).
The $\Sigma_{SFR}$ for the star-forming regions in NGC~1365 varied between 
1.268 $\times$ 10$^{-3}$  $-$  22.96 $\times$ 10$^{-2}$ M$_\odot$ yr$^{-2}$ kpc$^{-2}$ in the 
FUV band, 
with an average$\Sigma_{SFR}$ of 8.91 $\times$ 10$^{-3}$ M$_\odot$ yr$^{-1}$ kpc$^{-2}$. Similarly, in the NUV 
band, we found the  $\Sigma_{SFR}$ values to vary between 
9.72 $\times$ 10$^{-4}$ and 
55.62 $\times$ 10$^{-2}$ M$_\odot$ yr$^{-1}$ kpc$^{-2}$, with a average of 
11.19 $\times$ 10$^{-3}$ M$_\odot$ yr$^{-1}$ kpc$^{-2}$.  We found the 
distribution  of  the $\Sigma_{SFR}$ 
values in both FUV and NUV to be similar. We found the  extinction and $\Sigma_{SFR}$ 
to decrease radially outwards, as can be seen in  Fig. \ref{fig:extin} and 
Fig. \ref{fig:sfr} respectively. The details of the SF regions are given in Table \ref{tab:cat-ngc1365}.

\subsection{NGC~4051}
NGC~4051, an intermediate barred spiral galaxy with a morphological type of 
SAB(rs)bc \citep{Vauco91, Eva96}, is at a redshift of 
$z$ = 0.002 \citep{2006A&A...455..773V}. 
It is a narrow line Seyfert 1 galaxy \citep{Kha74} and powered by a black hole
of mass 6 $\times$ 10$^{5}$ M$_\odot$ \citep{Sei18}. It is luminous
in X-rays wth  L$_{2–10keV}$ = 2.7 $\times$ 10$^{41}$ erg s$^{-1}$ \citep{Pou04}, 
and is  known to be variable across wavelengths and timescales \citep{Pet00, McH04, Jon11, Kum23}. 
High-resolution observations at radio frequencies such as with the VLA A-array at 8.4 GHz show a core and two distinct components on opposite sides, while lower-resolution observations do not resolve well the inner triple but exhibits more diffuse extended emission \citep[e.g.][]{2011MNRAS.412.2641J}. Unlike at X-ray wavelengths, they do not find evidence of significant variability of the core at radio frequencies.
From continuum-subtracted H$\alpha$ images, \cite{Eva96} identified numerous HII knots within its spiral arms.

We detected 131 star-forming regions having FUV brightness between 
23.97 mag (the faintest) and 16.55 mag (the brightest). Similarly, in the NUV 
band, the star forming regions varied in brightness between 24.40 mag at 
the faintest 
end and  16.69 mag at the brightest end. The areas of these 
star forming regions have a wide
range, varying from 0.046 kpc$^2$ to 2.144 kpc$^2$, with the average and 
median value of 0.341 kpc$^2$ and  0.207 kpc$^2$ respectively. 

We found the 
E(B$-$V) values to span a wide range from $-$0.357 mag to 0.418 mag, with an average 
of value of $-$0.046 mag and a median value of $-$0.038 mag. Extinction in the FUV 
band is found to vary  between $-$1.51 mag and 
1.77 mag, while, in the NUV band, we found the extinction values 
to vary  between $-$1.32 mag and 1.55 mag, with mean and median values of 
$-$0.17 mag and $-$0.14 mag, respectively.

We found the $\Sigma_{SFR}$ of the star forming regions to have a wide 
range between 1.09 $\times$ 10$^{-4}$ and 
42.79 $\times$ 10$^{-4}$ M$_\odot$ yr$^{-1}$ kpc$^{-2}$ in the case 
of FUV and between 8.1 $\times$ 10$^{-5}$ and 41.22 $\times$ 10$^{-4}$ 
M$_\odot$ yr$^{-1}$ kpc$^{-2}$ in the case of NUV. The average $\Sigma_{SFR}$ 
in the FUV and NUV bands are 
7.72$\times 10^{-4}$ M$_\odot$ yr$^{-1}$ kpc$^{-2}$ and 6.98$\times 10^{-4}$ M$_\odot$ yr$^{-1}$ kpc$^{-2}$
respectively. We found the $\Sigma_{SFR}$ values to be similar in both FUV and NUV bands.

\subsection{NGC~4151}
NGC~4151 with a well-defined spiral pattern and a central bulge 
\citep{Vauco91, Mun99} is an intermediate face-on spiral galaxy with a morphology of (R')SAB(rs)ab. It hosts a Seyfert 1.5 AGN \citep{2006A&A...455..773V}, 
situated at a redshift of z=0.0033 \citep{Wol13} and powered by a black hole 
of mass 4.57 $\times$ 10$^{7}$ M$_\odot$ \citep{Ben06}. Its X-ray luminosity 
ranges between  (1.3$-$2.1)$\times$10$^{42}$ erg s$^{-1}$ \citep{Wan10} and 
has been studied for variability over a wide range of wavelengths.  It has a  two-sided 
radio jet, and the gas-rich spiral arms, along with the bar, are prominently visible 
in HI images \citep{Bos77}. The outer spiral arms are clearly visible in both FUV and 
NUV images as can be seen in Fig. \ref{fig:rgb} and Fig. \ref{fig:reg}.  These two distinct spiral arms are not prominent in the optical band, which are nicely traced by HI observation \citep{1992MNRAS.259..369P}. The radio structure of NGC~4151 has been studied for decades. It exhibits a two-sided jet with multiple components along a PA$\sim 77^\circ$, with one of the components (C4) coincident with the optical nucleus of the galaxy. C4, which is presumably associated with the supermassive black hole at its centre, was found to vary at radio frequencies, but no significant motion was detected between C4 and the nearby C3 component over a 22-year period \citep{2017MNRAS.472.3842W}.

We detected 161 star forming regions. The faintest and the brightest star-forming regions identified by us in the FUV image of NGC 4151 have a 
magnitude of  23.76 mag and  16.36 mag, respectively. Similarly, in the NUV 
band, the faintest 
and the brightest star-forming regions detected have a brightness value of 
23.91 mag and 16.25 mag respectively. These 
star-forming regions have a wide range of 
area between 0.010 kpc$^2$ to 4.921 kpc$^2$, with an average and median area 
of 0.188 kpc$^2$ and 0.027 kpc$^2$ respectively. We found the color excess, E(B$-$V) to vary between 0.032 mag and 2.446 mag, with an average value of 0.252 mag and a median value 
of 0.227 mag.  Extinction in the FUV band varied between 0.15 mag to 3.21 mag, 
with an average of 1.13  mag and a median of 1.02 mag, while in the NUV band, 
extinction varied from 0.12 mag to 2.66 mag, with a median of 0.94 mag and a 
average of 0.843 mag.

The estimated $\Sigma_{SFR}$ of these SF regions exhibited a wide range, 
with $\Sigma_{SFR}$ values in the FUV band varying from 
8.55 $\times$ 10$^{-4}$ to 0.05 M$_\odot$ yr$^{-1}$ kpc$^{-2}$. The average 
$\Sigma_{SFR}$ in the FUV band is 
48.40 $\times$ 10$^{-4}$ M$_\odot$ yr$^{-1}$ kpc$^{-2}$. In the NUV band, 
$\Sigma_{SFR}$ values ranged between 7.72$ \times$ 10$^{-4}$ and 
0.07 M$_\odot$ yr$^{-1}$ kpc$^{-2}$ 
with an average $\Sigma_{SFR}$ of  57.11 $\times$ 10$^{-4}$ M$_\odot$ yr$^{-1}$ kpc$^{-2}$. The detailed properties of the
440 the star forming regions are given in Table \ref{tab:cat-ngc4151}.

\subsection{NGC~4321}
NGC 4321, classified as a late-type, nearly face-on grand design spiral galaxy 
with a  SAB(s)bc morphology \citep{Vauco91}, is at a  redshift of $z$ = 0.00524 
\citep{All14} and is powered by a black hole of mass (2.5$\pm$0.2)$\times$10$^{7}$ 
M$_\odot$ \citep{Sar02}. It has two bars and a circumnuclear ring \citep{Gar98}. 
It is categorized as HII/LINER and exhibits a relatively high X-ray luminosity, 
exceeding 1.9 $\times$ 10$^{40}$ erg s$^{-1}$ \citep{Gon09}. In the radio, the image made with the Effelsberg telescope with an angular resolution of $\sim$71 arcsec shows 
a bright central region and an extended disk \citep{Urb86}. Higher-resolution observations with the VLA with an angular resolution of 2 arcsec showed the central region to have a roughly circular structure with an angular extent of $\sim$20 arcsec. Radio emission appears to peak at the optical nucleus and also at an enhanced region $\sim$7 arcsec to the east near an optical condensation, besides emission from SN1979c \citep{1981ApJ...243L.151W}. Moreover, strong 
activity has been observed across optical and radio wavelengths, leading to the 
classification of NGC 4321 as a transition galaxy, bridging the gap between 
normal and AGN galaxies \citep{Imm98}.

NGC 4321 has giant molecular associations in various regions, including the bar, 
spiral arms, and circumnuclear ring \citep{Pan17}. It has been observed that 
SF is most active within the circumnuclear ring, with lower activity 
occurring in the inter-arm regions \citep{Pan17}. SF in the 
circumnuclear region or nuclear rings is attributed to bursts triggered by the 
Inner Lindblad resonance (\citealt{Ars89}). Additionally, H$\alpha$ emission 
is prominent on the leading side of the bar and spiral arms, contributing to the 
understanding of star-forming regions in the galaxy \citep{Pan17}. Further studies, 
such as those by \cite{Fer12}, have investigated young star-forming regions in 
the NUV and optical ranges. While these studies confirmed the association of 
H$\alpha$ emission with the nuclear region, they did not find a significant 
correlation between the distance from these regions and the age or dust content.

We identified 340 star-forming regions in NGC 4321. These regions have a wide
range in brightness with values ranging between  22.19 mag and 13.92 mag in 
the FUV band. They have a range of sizes from  
0.029 kpc$^2$ to 1.587 kpc$^2$, with an average area of 0.159 kpc$^2$ and a 
median of 0.116 kpc$^2$. The E(B$-$V) values within these regions 
range between 0.056 mag and 0.695 mag, with an average and median values of 
0.361 mag and 0.356 mag respectively. We found the extinction in the FUV 
band to cover a wide range from 0.25 mag to 3.11 mag, with an average 
and median of 1.62 mag and  1.59 mag, respectively.

The $\Sigma_{SFR}$ within these star formation regions in the FUV band  
have a wide range from 5.82 $\times$ 10$^{-3}$ to 55.17 $\times$ 10$^{-2}$ 
M$_\odot$ yr$^{-1}$ kpc$^{-2}$, with an  average $\Sigma_{SFR}$ of 3.43 $\times$ 10$^{-2}$ 
M$_\odot$ yr$^{-1}$ kpc$^{-2}$.  Both extinction and $\Sigma_{SFR}$ were found 
to show a gradual decline from the center towards the outer regions of the 
galaxy  as can be seen in Fig. \ref{fig:extin} and  Fig. \ref{fig:sfr}. 

\subsection{NGC~4388}
NGC 4388 is a nearly edge-on spiral galaxy, classified as type SA(s)b with a 
major axis position angle of 91$^{\circ}$ \citep{Pat03, Vauco91}. Located at 
a distance of approximately 37.55 Mpc corresponding to a redshift of $z$=0.00842 
\citep{Lu93}, this galaxy exhibits a short bar whose strength depends on the 
eccentricity of stellar or gas orbits \citep{Vei99}. At its center, NGC 4388 
hosts a type 2 Seyfert AGN with a supermassive black hole of mass 
(8.4$\pm$0.2)$\times$10$^{6}$ M$_\odot$ \citep{Tue08, Kuo11}. It is a strong X-ray
source, with an X-ray luminosity of L$_{2-10 keV}$ = 1 $\times$ 10$^{42}$ erg s$^{-1}$ 
\citep{For99}. 
VLA observations have revealed 
an elongated blob with an extent of $\sim$13 arcsec to the north of the galaxy and a collimated structure of $\sim$2.8 towards the south, the observed asymmetry possibly being due to an asymmetric distribution of ISM \citep{Hum91}. More recently \cite{2024ApJ...961..230S} have made a detailed study with the VLA in both total intensity and linear polarization and have suggested that the radio continuum emission is due to AGN winds interacting with the local ISM.

NGC~4388 also exhibits significant dust obscuration, with dust bands extending 
to approximately 1.5 kpc from the center of the disk \citep{Fal98}. Studies have 
shown that the galaxy possesses a truncated HI disc and gas components outside 
the plane, suggesting a possible interaction with the intergalactic 
medium \citep{Sin19}. This interaction is believed to have involved ram 
pressure stripping, resulting in the loss of approximately $85\%$ of the HI gas 
component in NGC 4388, subsequently disrupting star formation in the 
galaxy \citep{Cay90}. Recent investigations have indicated star forming 
regions not coplanar with the disk, contributing to the extended outflows and 
suggesting recent starburst events in the spiral arms and circumnuclear region 
of the galaxy \citep{Dam16}.

We detected 20 star-forming regions in NGC~4388. These regions have FUV brightness
between 22.06 mag and 15.79 mag, while in the NUV the brightness of the 
star forming regions varied between 22.07 mag and 15.72 mag. The areas of these 
regions were found to show  considerable variation, spanning from 0.074 kpc$^2$ 
to 4.636 kpc$^2$, with an average area of 1.268 kpc$^2$ and a median of 0.754 kpc$^2$. 
The E(B$-$V) values span a range from 0.23 mag to 0.43 mag, with an average 
of 0.33 mag and a median of 0.32 mag. Extinction values in the FUV band were 
found to have a wide range from 1.01 mag to 1.92 mag, with a mean and median 
of 1.46 mag and 1.43 mag, respectively, 
while in the NUV band, extinction varied between  0.78 mag and 1.49 mag, with 
a median of 1.11 mag and an average of 1.13 mag.  The $\Sigma_{SFR}$ within these regions in the FUV were found to vary between 5.74$\times$10$^{-3}$ and 103.79$\times$10$^{-3}$ M$_\odot$ yr$^{-1}$ kpc$^{-2}$. 
with an average $\Sigma_{SFR}$ of 27.23$\times$10$^{-3}$ M$_\odot$ yr$^{-1}$ kpc$^{-2}$. 
Similarly, in the NUV band, 
we found $\Sigma_{SFR}$ to range between 7.16$\times$10$^{-3}$ to 108.92$\times$10$^{-3}$ M$_\odot$ 
yr$^{-1}$ kpc$^{-2}$. We found the average $\Sigma_{SFR}$ in the NUV band of 
31.38$\times$10$^{-3}$ M$_\odot$ yr$^{-1}$ kpc$^{-2}$. Properties of the
564 individual star forming regions are given in Table \ref{tab:cat-ngc4388}.

\subsection{NGC~5033}
NGC~5033 is an SA(s)c-type non-barred spiral galaxy at a redshift of 
$z$=0.00291.  It has a central bulge with a mass of approximately 
2 $\times$ 10$^{10}$ M$_\odot$.  It is a Seyfert 1.9 galaxy, and hosts a 
supermassive black hole at its center, with a mass ranging from 
5$-$12 $\times $10$^{6}$ M$_\odot$.  It is luminous in X-rays 
with L$_{2–10 keV}$ = 2.3 $\times$ 10$^{41}$ erg s$^{-1}$ \citep{Ter99}. 
Although high-resolution radio observations show a core-jet morphology, the radio emission from the inner disk is dominated by a starburst with the core-jet structure contributing only 7 percent of the flux density at 1.4 GHz \citep{2007MNRAS.379..275P}. They also report evidence of a `radio spur' due to a hot bubble caused by sequential supernova explosions. The galaxy is variable in X-ray wavelengths \citep{2008A&A...490..995P}. 

The galaxy's flat disk has a high inclination of 67.5$^{\circ}$ and shows signs 
of possible warping, possibly resulting from interactions or mergers. Near 
infrared  and millimeter observations reveal a small nuclear bar, while the 
presence of HII regions following a ring-like pattern is evident in H$\beta$ 
intensity maps. A bright bar of light from the unresolved nucleus has been observed 
in NUV images. Radio 21cm-line studies show similarities between the HI 
distribution and optical characteristics, and it has been found that  
$\Sigma_{SFR}$ is directly related to the total gas 
surface density. Comprehensive CO\textit{(1$-$0)} observations suggest that 
there is no starburst phenomenon in the nuclear region. Instead, star-forming 
regions are present in the disk region, supported by observations from the 
GALEX telescope.  The radio and FIR luminosities of the inner regions indicate 
evidence of a recent short starburst in NGC~5033 \citep{2007MNRAS.379..275P}.

We identified 557 star forming regions with a wide range of brightness. 
The faintest region in the FUV has a brightness of 22.51 mag, while the 
brightest region has a brightness of 15.43 mag. Similarly, in the NUV band, 
the faintest and brightest star forming regions have brightness values of 
22.86 mag and 14.14 mag respectively.  These star forming regions have varied 
sizes, ranging from 0.008 kpc$^2$ to 0.242 kpc$^2$, with an average value of 0.020 kpc$^2$.  We found E(B$-$V) to vary from 
0.04 mag to 1.17 mag, with an average value of 0.51 mag and a median of 
0.49 mag. Extinction values in the FUV were found to lie between 0.19 mag 
and 5.34 mag, with an average of 2.33 mag and a median of 1.58 mag. In the 
NUV band, the extinction is relatively lower, with values between 
0.14 mag and 3.73 mag, with a median of 1.58 mag and an average of 1.63 mag.

The $\Sigma_{SFR}$ of these star forming  regions varied between 
3.25 $\times$ 10$^{-3}$ and 53.02 $\times$ 10$^{-2}$ M$_\odot$ yr$^{-1}$ kpc$^{-2}$ in FUV and between 24.81$\times$10$^{-4}$ and
12.78$\times$10$^{-2}$ M$_\odot$ yr$^{-1}$ kpc$^{-2}$ in NUV. We found 
the $\Sigma_{SFR}$ to gradually decrease radially from the centre to the 
outskirts, as can be seen in Fig. \ref{fig:sfr}.

\subsection{NGC~6814}
NGC~6814 is a grand design spiral galaxy of the SAB(rs)bc type, 
with a face-on orientation \citep{Vauco91}. It is characterized by a compact 
bulge in both near infrared and optical wavelengths, along with a relatively 
weak bar extending approximately ten arcseconds in the North-South direction 
\citep{San04, Sla11}. Situated at a distance of 23.22 Mpc, this galaxy hosts a 
type 1.5 Seyfert AGN at its center \citep{sp05}, with a black hole of mass 
(1.85 $\pm$ 0.35) $\times$ 10$^{7}$ M$_\odot$ \citep{Ben09}. It is variable 
in the optical, UV, and X-ray wavelengths, with the highest variability 
observed in X-rays \citep{Gal21,Tro16}. It is luminous in X-rays with 
L$_{2–10}$ = 2.04 $\times$ 10$^{42}$ erg s$^{-1}$ \citep{Tor18}. At radio frequencies, the central region appears extended in the north-south direction with a possible component towards the west, and no significant jet emission when observed with the VLA A-array,  
 \citep{1984ApJ...285..439U,Xu99}.

Extensive work by \cite{Kna93} has identified numerous HII regions 
within the galaxy, with a comprehensive study of the luminosity functions of 
these regions. Interestingly, the luminosity function exhibits similar 
characteristics in both the outer and inner arms, offering valuable insights 
into the galaxy's SF processes. In addition to HII regions, NGC~6814 
displays ring-like distributions of HII, potentially featuring kinematic 
warping, as suggested by \cite{Lis95}. Moreover, in the UV spectrum, numerous 
star-forming regions are evident along the spiral arms, with these arms 
branching into multiple fragments observable in the GALEX images \citep{San94}.

We identified 89 star forming regions in NGC~6814, with the faintest region 
at 21.21 mag and the brightest at 16.43 mag in the FUV band. These regions 
exhibited a range of sizes, with areas ranging from 0.029 kpc$^2$ to 
0.422 kpc$^2$, with an average area of 0.118 kpc$^2$ and a median area 
of 0.082 kpc$^2$. The E(B$-$V) values in the regions ranged from 0.15 mag 
to 0.59 mag, with an average of 0.43 mag. 
Regarding extinction in the FUV band, it was found to vary between 
0.65 mag to 2.64 mag, with an average of 1.94 mag.

The $\Sigma_{SFR}$ of the star forming  regions  identified in NGC~6814, 
showed variations, with the $\Sigma_{SFR}$ values in the FUV band ranging 
from 8.96 $\times$ 10$^{-3}$ to 65.25 $\times10^{-2}$ M$_\odot$ yr$^{-1}$ kpc$^{-2}$, with an average of 4.19$\times$10$^{-2}$ M$_\odot$ yr$^{-1}$ kpc$^{-2}$. 
We found a gradual radial decrease of extinction and SFR from the centre to the outskirts of the galaxy  (see Fig. \ref{fig:extin} and Fig. \ref{fig:sfr}).

\subsection{NGC~7469}
NGC~7469 is classified as an intermediate, nearly face-on spiral galaxy with 
a morphological type of (R')SAB(rs)a \citep{Vauco91}. This galaxy is located 
at a redshift of $z$=0.01627 \citep{sp05}. It hosts a Seyfert 1.5 type 
AGN \citep{2006A&A...455..773V} with a black hole of mass  1.1$\times$10$^{7}$ M$_\odot$ 
\citep{Pet14,Lu21}.  It is known for its strong X-ray emission, with 
a X-ray luminosity of L$_{2–10keV}$ = 10$^{43}$ erg s$^{-1}$ \citep{Asm15}.

One of the prominent features of this galaxy is its circumnuclear starburst 
ring, situated at a distance of about 1 kiloparsec from its center. This 
starburst ring has been extensively observed across various 
wavelengths \citep{Xu22}. High-resolution radio continuum observations show a nucleus which is the brightest feature, and more extended emission from the circumnuclear starburst with several peaks of emission. The brightest of the compact components is a radio supernova \citep{2001ApJ...553L..19C}.  Additionally, from observations of molecular line emission at millimeter wavelengths \cite{Dav04} 
suggests the presence of either a bar or a pair of spiral arms located 
between these ring like structures. 
\cite{Mar94} suggest that the star forming activities in this galaxy exhibit 
unique characteristics, potentially influenced 
by its H$\alpha$ emission, which may be attributed to the nearby companion 
galaxy IC 5283.

In  NGC~7469, we identified 26 star forming regions. We detected more SF regions in the northern spiral arm which may be due the effect of interaction with nearby galaxy IC~5283 as mentioned by \cite{Mar94}. These regions in FUV 
have brightness ranging between 24.47 mag and  17.67 mag. In the NUV band, the 
faintest star forming region is at 24.63 mag and the brightest is at 
17.65 mag. Also, we found the areas of these star forming regions to vary 
significantly, ranging from 0.295 kpc$^2$ to 63.642 kpc$^2$. On average, the 
star forming regions have an average area of 12.026 kpc$^2$. The E(B$-$V) in these regions range from 0.06 mag to 0.64 mag.
In the FUV band, we found extinction values that range from 0.27 mag to 
2.68 mag, with an average of 1.17 mag. Similarly, in 
the NUV band, we found the extinction to vary from  0.22 mag to 2.21 mag, 
with a median of 0.97 mag.

The $\Sigma_{SFR}$ within these star forming regions exhibits a 
wide range. In FUV, $\Sigma_{SFR}$ varies from  
8.0 $\times$ 10$^{-4}$ to 49.39 $\times$ 10$^{-3}$ M$_\odot$ yr$^{-1}$ kpc$^{-2}$, with an average of 5.72 $\times$ 10$^{-3}$ M$_\odot$ yr$^{-1}$ kpc$^{-2}$. 
In NUV, we found $\Sigma_{SFR}$ to  vary from 7.2 $\times$ 10$^{-4}$ 
to 77.26 $\times$ 10$^{-3}$ M$_\odot$ yr$^{-1}$ kpc$^{-2}$, 
with an average $\Sigma_{SFR}$ of 7.08$\times$10$^{-3}$ M$_\odot$ yr$^{-1}$ kpc$^{-2}$.

\begin{table*}
\tiny
\caption{Summary of the identified star forming regions for all sources. The values of $\Sigma_{SFR}$  are in units of 10$^{-3}$M$_\odot$ yr$^{-1}$ kpc$^{-2}$.}
\label{tab:summary} 
\hspace*{-1cm}
\begin{tabular}{crrrrrrrrrrrrr }
\hline
Name & N & \multicolumn{6}{c}{Mean} & \multicolumn{6}{c}{Median} \\
     &   & area & E(B$-$V) & A$_{FUV}$ & A$_{NUV}$ & $\Sigma_{SFR}^{FUV}$ &$\Sigma_{SFR}^{NUV}$ & area &E(B$-$V) & A$_{FUV}$ & A$_{NUV}$ & $\Sigma_{SFR}^{FUV}$ & $\Sigma_{SFR}^{NUV}$ \\
     &   & (kpc$^2$) & & (mag) & (mag) &  & & (kpc$^2$) & (mag)  & (mag) & (mag) &  &  \\

\hline
NGC~1365 & 418 & 0.713 & 0.212 & 0.93 & 0.68 & 8.90 & 11.19 & 0.215 & 0.195 & 0.86 & 0.63 & 4.65 & 4.83 \\

NGC~4051 & 131 & 0.341 & $-$0.046 &  $-$0.19 & $-$0.17 & 0.77 & 0.70 & 0.207 & $-$0.038 & $-$0.16 & $-$0.14 & 0.54 & 0.48 \\

NGC~4151 & 161 & 0.187 & 0.252 & 1.13 & 0.94 & 4.84 & 5.71 & 0.027 & 0.227 & 1.02 & 0.84 & 3.32 & 3.56 \\

NGC~4321 &	340 & 0.159 & 0.356 & 1.62 & $--$ & 34.37 & $--$ & 0.116 & 0.36 & 1.60 & $--$ & 20.02 & $--$\\

NGC~4388 &	20 & 1.268 & 0.325 & 1.46 & 1.13 & 27.23 & 31.38 & 0.755 & 0.318 & 1.43 & 1.11 & 19.04 & 23.05 \\

NGC~5033 &	557 & 0.021 & 0.509 & 2.33 & 1.63 & 57.73 & 128.34 & 0.013 & 0.494 & 2.26 & 1.58 & 37.75 & 62.33 \\

NGC~6814 &	89 & 0.120 & 0.432 & 1.94 & $--$ & 41.96 & $--$ & 0.083 & 0.452 & 2.03 & $--$ & 32.95 & $--$ \\

NGC~7469 & 26 & 12.026 & 0.277 & 1.17 & 0.96 & 5.71 & 7.08 & 4.973 & 0.277 & 1.17 & 0.97 & 2.66 & 2.88 \\
\hline

\end{tabular}
\end{table*}

\section{Global picture}\label{global}
\begin{figure}
    \centering
    \includegraphics[scale=0.4]{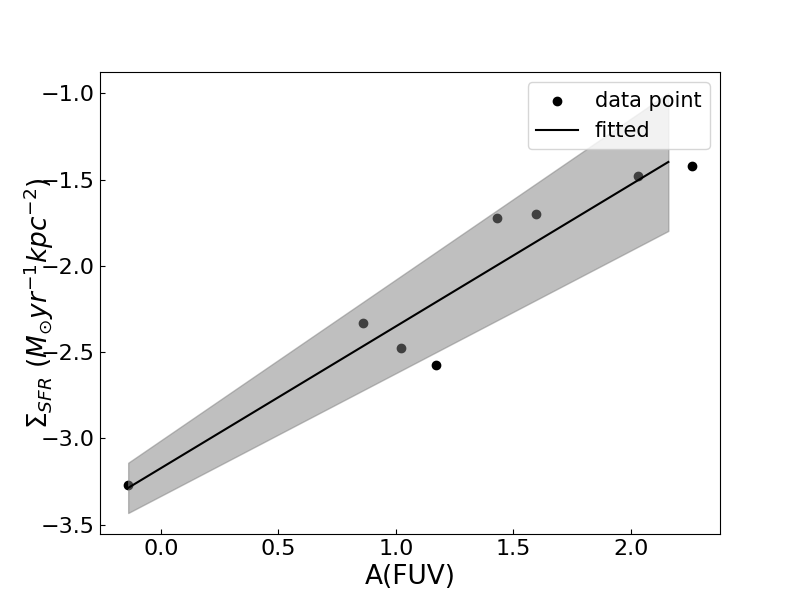}
    \includegraphics[scale=0.4]{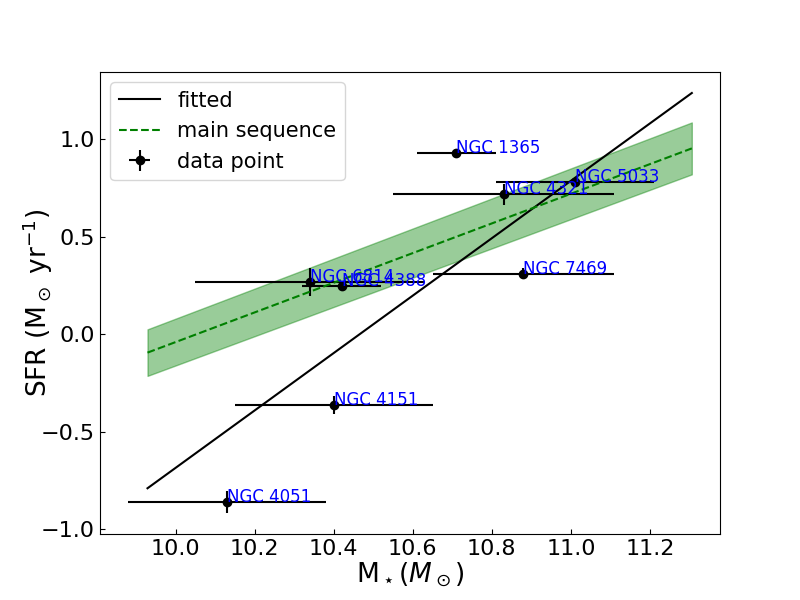}
    \caption{Top panel: Variation of median $\Sigma_{SFR}$ against median extinction in FUV of SF regions for the sources studied in this work. Bottom panel: Variation of SFR with stellar mass. The dotted line is the MS of SF from \citealt{2015ApJ...801L..29R}, and 
the solid line is the best-fitted linear regression line.}  
    \label{fig:glob_ext}
\end{figure}

For the galaxies examined in this study, we observed a positive correlation between the median surface density of star formation and the extinction in the FUV band, with a Pearson correlation coefficient ($r$) of 0.95 and a p-value of 0.0003. This relationship is illustrated in the top panel of Fig. \ref{fig:glob_ext}. The best-fitted line to this correlation is given by:
\begin{equation}
    \log\left(\frac{\Sigma_{SFR}}{M_\odot \text{yr}^{-1} \text{kpc}^{-2}}\right) = (0.82\pm0.11)A(FUV) - (3.17\pm0.16)
\end{equation}

We determined the SFR using Equation \ref{eq:sfr} within the circumnuclear region, employing a 1 kpc radius aperture, and computed the total SFR using the optical R$_{25}$ aperture. In both cases, we corrected for AGN flux contamination using a 1.4$^{\prime\prime}$ radius aperture. Here, R$_{25}$ represents the radius at which the surface brightness reaches 25 mag per square arcsec in the optical B band. The results of these calculations are provided in Table \ref{tab:sfr}. Additionally, detailed properties of each of the identified star-forming regions for all eight galaxies are presented in the Appendix.


We positioned our sources on the stellar mass versus star formation rate plane and compared their locations relative to the main sequence as described by \cite{2015ApJ...801L..29R}. Notably, four of our sources fell within the main sequence, one above it, and three below it. This distribution is illustrated in the bottom panel of Fig. \ref{fig:glob_ext}. The position of the AGN in M$_{\ast}$-SFR below the main sequence has also been noticed by \citep{2007ApJS..173..267S}.

Furthermore, we observed a positive correlation between the star formation rate and the stellar mass. Utilizing a linear regression analysis, we determined a correlation coefficient ($r$) of 0.83 and a p-value of 0.02. The resulting best-fitted line is represented by the equation:

\begin{equation}
    \log\left(\frac{SFR_{total}}{M_\odot \text{yr}^{-1}}\right) = (1.47\pm0.44) \log\left(\frac{M_\star}{M_\odot}\right) - (15.39\pm4.65)
\end{equation}

\begin{figure}
    \centering
    \includegraphics[scale=0.4]{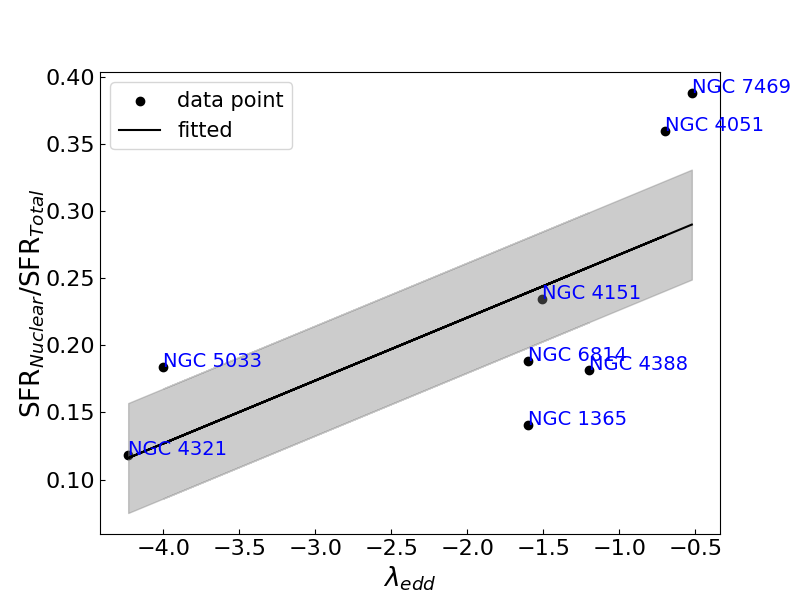}
    \caption{Variation of the ratio of SFR$_{Nuclear}$ to SFR$_{Total}$ with $\lambda_{Edd}$.
The solid line is the linear least squares fit to the data and the shaded region
is 95\% confidence region.}
    \label{fig:edd_sfr}
\end{figure}

We identified a modest positive correlation between the ratio of the nuclear star formation rate (SFR$_{Nuclear}$) to the total SFR (SFR$_{Total}$) and the Eddington ratio, $\lambda_{Edd}$. Here, $\lambda_{Edd}$ represents the ratio of the bolometric luminosity (L$_{Bol}$) to the Eddington luminosity (L$_{Edd}$). To estimate L$_{Bol}$, we utilized the observed L$_{2-10 keV}$ and applied the relation log(L$_{Bol}$) = 0.0378 $\times$ log(L$_{2-10})^2$ - 2.00 $\times$ log(L$_{2-10}$) + 60.5 \citep{2017ApJ...835...74I}, while L$_{Edd}$ was calculated as 1.26 $\times$ 10$^{38}$ (M$_{BH}$/M${_\odot}$) erg s$^{-1}$. This relationship is depicted in Fig. \ref{fig:edd_sfr}. Performing a linear least squares fit to the data points yielded a correlation coefficient ($r$) of 0.67 and a p-value of 0.06. The best fitted line is 
\begin{equation}
    \frac{SFR_{Nuclear}}{SFR_{Total}}= (0.05\pm0.02)log(\lambda_{Edd}) + (0.31\pm0.05)
\end{equation}
This correlation bears resemblance to the findings of \cite{2023A&A...679A.151M} regarding the relationship between SFR and $\lambda_{Edd}$. However, we noted that this correlation is heavily influenced by two points, NGC~7469 and NGC~4051. Thus, further observations with more sources are required to confirm this correlation robustly. Nevertheless, the observed correlation suggests that AGN activity (jet and/or radiation) positively influences the star formation characteristics near the nuclear region, with minimal impact observed at greater distances from the nuclear region. However, disentangling the contribution of AGN jet and radiation is not possible using only the observations reported in this paper. Our observation aligns with both simulations \citep{2023arXiv231107576B} and observational studies \citep{2023ApJ...953...26L}, which support the operation of feedback processes primarily within the central kilo-parsec region in AGN host galaxies.

\begin{table*}
    \centering
    \caption{The properties of the galaxies in FUV. The units of SFR are in M$_{\odot}$ yr$^{-1}$, while that
of $\Sigma_{SFR}$ are in units of 10$^{-3}$ M$_\odot$ yr$^{-1}$ kpc$^{-2}$.}
    \label{tab:sfr}
    \begin{tabular}{rrrrrr}
    \hline
    Name & SFR$_{Total}$ & SFR$_{Nuclear}$  & $\Sigma_{SFR}^{Toal}$  & $\Sigma_{SFR}^{Nuclear}$\\
    \hline
    NGC~1365   & 8.469$\pm$0.128 & 1.192$\pm$0.019 & 2.83$\pm$0.04 & 94.82$\pm$1.62  \\
     NGC~4051  & 0.138$\pm$0.008 & 0.050$\pm$0.003 & 1.40$\pm$0.08 & 15.82$\pm$0.9 \\
     NGC~4151  & 0.434$\pm$0.019 & 0.102$\pm$0.004 & 1.48$\pm$0.07 & 8.12$\pm$0.41 \\
     NGC~4321  & 5.210$\pm$0.281 & 0.617$\pm$0.033 & 3.72$\pm$0.21  & 48.11$\pm$2.60 \\
     NGC~4388  & 0.32$\pm$0.005  & 1.77$\pm$0.02 & 3.22$\pm$0.04 & 25.57$\pm$0.4 \\
     NGC~5033  & 6.058$\pm$0.0997 & 1.116$\pm$0.024 & 10.34$\pm$0.20  & 88.78$\pm$1.91  \\
     NGC~6814  & 1.859$\pm$0.134 & 0.351$\pm$0.025 & 7.25$\pm$0.52 & 27.90$\pm$2.00 \\
     NGC~7469   & 2.042$\pm$0.057 & 0.792$\pm$0.030 & 4.25$\pm$0.14 & 63.00$\pm$2.41  \\
    \hline
    \end{tabular}
\end{table*}

\section{Summary}\label{summary}
We carried out an investigation of the SF characteristics of galaxies hosting AGN using observations carried out in UV. Our sample consists of seven Seyfert and one LINER type AGN. Our approach involved the identification of star-forming regions, carrying out aperture photometry of the identified star-forming regions, 
correcting the derived brightness for both Milky Way and internal extinction,  estimation of SFR and the correlation of SFR with various derived parameters. The findings of this work are summarized below.
\begin{enumerate}
\item We identified many star-forming regions for all eight galaxies hosting AGN. 
For NGC~5033, we detected the maximum number of SF regions. Most of the identified star-forming regions have sizes in excess of 30 parsecs, with areas ranging between 0.010 and 63.642 kpc$^2$. 
\item The extinction corrected SFR of the star-forming regions considering all the sources
studied in this work are found to 
      range between 1.09$\times 10^{-4}$ M$_{\odot}$ yr$^{-1}$ and 22.96$\times 10^{-2}$ M$_{\odot}$ yr$^{-1}$ in FUV, while in NUV it varies between 8.15$\times$ 10$^{-5}$ and 6.634 M$_{\odot}$ yr$^{-1}$. We found the $\Sigma_{SFR}$ in NUV to be larger than that at FUV, though 
      the median values are found to be similar.  
\item We detected two outer spiral arms for NGC~4151 in both FUV and NUV bands. These spiral arms are much extended beyond R25 aperture.
\item For five galaxies,  NGC~1365, NGC~4051, NGC~4321, NGC~5033 and NGC~6814, we found both the internal extinction and the  $\Sigma_{SFR}$ to gradually decrease from the centre towards the outer regions, while this was not significant in the other sources. 
\item We found a positive correlation between the median $\Sigma_{SFR}$ and the median extinction in FUV for the star-forming regions. 
\item Among the observed sources, four are situated within the main sequence (MS) of star-forming galaxies; while the remaining four are positioned away from the MS. 

\item All sources display a positive correlation between SFR and M$_{\ast}$, with a slope that is notably steeper than MS.

\item We found the SFR in the nuclear region dominant over the total SFR. 
\item The ratio of SFR$_{Nuclear}$ to SFR$_{Total}$ is exhibits a weak positive correlation with $\lambda_{Edd}$. This correlation points to the influence of AGN on enhancing the SF characteristics of the hosts and the impact being dominant in the central region with no effect on scales beyond the nuclear region 
probed in this work.
\end{enumerate}

\begin{acknowledgments}
\section*{Acknowledgments}
This manuscript uses the data from the AstroSat mission of the Indian Space Research Organization (ISRO), archived at the Indian Space Science Data Center (ISSDC). This manuscript uses UVIT data processed by the payload 
operations centre at IIA. The UVIT is built in collaboration between IIA, 
IUCAA, TIFR, ISRO, and CSA. We have used the NASA-NED database. PN thanks the Council of Scientific and Industrial Research (CSIR), Government of India, for supporting her research under the CSIR Junior/Senior Research Fellowship program through grant no. $09/079(2867)/2021$-EMR-I. PD wants to thank the Board of Graduate Studies, Indian Institute of Astrophysics, for their support through the Visiting student program.
\end{acknowledgments}

%

\vspace{5mm}
\facilities{UVIT(Astrosat)}


\software{IRAF \citep{1986SPIE..627..733T}, SExtractor \citep{1996A&AS..117..393B}, Photutils \citep{larry_bradley_2020_4044744}}



\appendix

\section{Some extra material}

\begin{table*}[h]
\tiny
    \begin{center}
        \caption{Properties of the identified star forming regions for NGC~1365. 
Here, a,b,$\theta$ are the semi-major axis, semi-minor axis and the position angle 
of the star forming regions, area is the area of the regions in kpc$^2$, E(B$-V$) 
is the colour excess in magnitude, m$_{FUV}$ and m$_{NUV}$ are the magnitudes
if FUV and NUV bands, A$_{FUV}$ and A$_{NUV}$ are the extinction in 
FUV and NUV bands, and $\Sigma_{FUV}$ and $\Sigma_{NUV}$ are the 
surface density of SFR in units of 10$^{-3}$ M$_{\odot}$ yr$^{-1}$ kpc$^{-2}$ in 
FUV and NUV bands respectively.  Only the first 10 entries are given, and the 
table in full is available in the electronic version of the article.} 
        \label{tab:cat-ngc1365}
        \hspace{1cm}
        \begin{tabular}{ccccrccrccrrrr}
        \hline
          RA   & DEC & a & b & $\theta$  & r$_d$ & area & E(B$-$V) & m$_{FUV}$ & m$_{NUV}$ &  A$_{FUV}$ & A$_{NUV}$ & $\Sigma_{FUV}$ & $\Sigma_{NUV}$\\
          \hline
        03:33:26.07 & -36:12:49.39 & 5.58 & 3.45 & 0.009 & 36.26 & 0.936 & 0.102 & 17.97$\pm$0.08 & 18.18$\pm$0.07 & 0.44 & 0.33 & 13.82$\pm$0.46 & 12.0$\pm$0.32\\
03:33:26.09 & -36:12:44.44 & 5.16 & 2.59 & 0.769 & 35.69 & 0.651 & 0.109 & 18.55$\pm$0.1 & 18.74$\pm$0.08 & 0.48 & 0.35 & 11.75$\pm$0.49 & 10.32$\pm$0.35 \\
03:33:26.02 & -36:12:54.31 & 2.65 & 2.13 & 1.504 & 36.84 & 0.275 & 0.112 & 19.57$\pm$0.16 & 19.76$\pm$0.13 & 0.49 & 0.36 & 10.8$\pm$0.7 & 9.54$\pm$0.51 \\
03:33:25.51 & -36:12:41.75 & 3.76 & 1.66 & 0.79 & 35.79 & 0.303 & 0.088 & 19.73$\pm$0.17 & 19.95$\pm$0.14 & 0.38 & 0.28 & 8.5$\pm$0.57 & 7.24$\pm$0.4 \\
03:33:24.70 & -36:12:53.92 & 4.03 & 2.49 & 0.291 & 37.7 & 0.489 & 0.091 & 19.39$\pm$0.14 & 19.61$\pm$0.12 & 0.4 & 0.29 & 7.16$\pm$0.42 & 6.12±0.3 \\
03:33:24.81 & -36:12:50.19 & 3.12 & 2.23  & -0.497 & 37.21 & 0.338 & 0.102 & 19.77$\pm$0.17 & 19.98$\pm$0.14 & 0.45 & 0.33 & 7.3$\pm$0.51 & 6.34$\pm$0.36  \\
03:33:25.11 & -36:12:43.54 & 4.24 & 2.51 & 0.774 & 36.27 & 0.519 & 0.111 & 19.5$\pm$0.16 & 19.68$\pm$0.13 & 0.49 & 0.36 & 6.15$\pm$0.38 & 5.42$\pm$0.28  \\
03:33:24.33 & -36:12:58.13 & 4.89 & 2.59 & -0.051 & 38.42 & 0.614 & 0.097 & 19.84$\pm$0.18 & 20.05$\pm$0.15 & 0.42 & 0.31 & 3.79$\pm$0.27 & 3.27$\pm$0.19 \\ 
03:33:23.91 & -36:12:55.71 & 3.86 & 2.11 & -1.316 & 38.46 & 0.396 & 0.118 & 20.43$\pm$0.24 & 20.61$\pm$0.2 & 0.52 & 0.38 & 3.39$\pm$0.32 & 3.02$\pm$0.24 \\
03:33:27.23 & -36:12:48.16 & 6.02 & 4.37 & 0.492 & 35.39 & 1.278 & 0.183 & 19.32$\pm$0.16 & 19.39$\pm$0.13 & 0.8 & 0.59 & 2.94$\pm$0.18 & 2.89$\pm$0.15 \\
           \hline
        \end{tabular}
    \end{center}
\end{table*}

\begin{table*}[h]
\tiny
    \begin{center}
        \caption{Same as Table \ref{tab:cat-ngc1365}, but for NGC~4051}
        \label{tab:cat-ngc4051}
        \hspace{1cm}
        \begin{tabular}{ccccrccrccrrrr}
        \hline
          RA   & DEC & a & b & $\theta$  & r$_d$ & area & E(B$-$V) & m$_{FUV}$ & m$_{NUV}$ &  A$_{FUV}$ & A$_{NUV}$ & $\Sigma_{FUV}$ & $\Sigma_{NUV}$\\
          \hline
          12:03:09.64 & 44:33:19.98 & 1.87 & 1.17 & 0.809 & 10.86 & 0.106 & -0.11 & 22.32±0.95 & 22.56±0.83 & -0.46 & -0.41 & 0.25±0.09 & 0.21±0.07\\
         12:03:11.02 & 44:33:26.68 & 2.33 & 1.65 & 0.563 & 11.84 & 0.187 & -0.034 & 20.09±0.41 & 20.28±0.36 & -0.14 &  -0.12 & 1.09±0.18 & 0.96±0.14\\
         12:03:11.04 & 44:33:22.87 & 2.81 & 1.68 & 0.024 & 11.38 & 0.23 & -0.068 & 20.04±0.38 & 20.26±0.33 & -0.29 & -0.25 & 0.92±0.14 & 0.8±0.11\\
        12:03:10.37 & 44:33:19.07 & 1.92 & 1.14 & 0.224 & 10.8 & 0.106 & -0.162 & 21.74±0.67 & 22.02±0.59 & -0.68 & -0.6 & 0.42±0.11 & 0.34±0.08\\
        12:03:10.51 & 44:33:17.01 & 2.12 & 1.51 & 0.801 & 10.56 & 0.156 & -0.189 & 21.5±0.58 & 21.8±0.51 & -0.8 & -0.7 & 0.36±0.08 & 0.28±0.06\\
         12:03:10.72 & 44:33:29.44 & 2.69 & 1.48 & 0.945 & 12.13 & 0.193 & -0.144 & 21.73±0.69 & 22.0±0.61 & -0.61 & -0.54 & 0.23±0.06 & 0.19±0.05\\
         12:03:09.50 & 44:33:13.86 & 1.86 & 1.13 & 1.502 & 10.1 & 0.102 & 0.016 & 21.35±0.76 & 21.51±0.67 & 0.07 & 0.06 & 0.62±0.19 & 0.57±0.15\\
          12:03:06.61 & 44:33:08.27 & 2.08 & 1.73 & 0.683 & 10.21 & 0.175 & -0.153 & 20.92±0.48 & 21.19±0.42 & -0.65 -& 0.57 & 0.54±0.1 & 0.44±0.07\\
         12:02:59.81 & 44:32:58.57 & 1.17 & 0.88 & 0.877 & 15.4 & 0.05 & 0.006 & 22.25±1.12 & 22.41±0.98 & 0.03 & 0.02 & 0.56±0.25 & 0.5±0.2\\
         12:03:03.06 & 44:33:00.84 & 1.34 & 1.1 & 1.289 & 12.16 & 0.072 & -0.038 & 21.99±0.93 & 22.18±0.81 & -0.16 & -0.14 & 0.49±0.18  & 0.43±0.14\\
           \hline
        \end{tabular}
    \end{center}
\end{table*}

\begin{table*}[h]
\tiny
    \begin{center}
        \caption{Same as Table \ref{tab:cat-ngc1365}, but for NGC~4151}
        \label{tab:cat-ngc4151}
        \hspace{1cm}
        \begin{tabular}{ccccrccrccrrrr}
        \hline
          RA   & DEC & a & b & $\theta$ &  r$_d$ & area & E(B$-$V) & m$_{FUV}$ & m$_{NUV}$ &  A$_{FUV}$ & A$_{NUV}$ & $\Sigma_{FUV}$ & $\Sigma_{NUV}$\\
          \hline
        12:10:25.98 & 39:30:37.96 & 1.3 & 1.08 & 0.033 & 23.94 & 0.017 & 0.555 & 21.35±0.67 2& 1.01±0.57 & 2.49 & 2.06 & 8.48±2.28 & 12.11±2.78\\
        12:10:15.56 & 39:28:28.52 & 1.85 & 1.32 & 0.824 & 19.69 & 0.03 & 0.199 & 21.66±0.45 & 21.7±0.4 & 0.89 & 0.74 & 3.62±0.66 & 3.69±0.58\\
        12:09:48.51 & 39:23:03.25 & 2.0 & 1.89 & -1.56 & 32.17 & 0.046 & 0.673 & 19.01±0.29 & 18.55±0.25 & 3.01 & 2.5 & 27.07±3.18 & 43.26±4.29\\
        12:09:59.41 & 39:24:23.45 & 1.64 & 1.27 & 0.176 & 23.93 & 0.025 & 0.342 & 21.8±0.6 & 21.69±0.52 & 1.53 & 1.27 & 3.75±0.89 & 4.38±0.9\\
        12:10:26.60 & 39:29:08.51 & 1.36 & 1.09 & 1.256 & 18.4 & 0.018 & 0.435 & 20.01±0.31 & 19.8±0.27 & 1.95 & 1.61 & 27.32±3.44 & 34.8±3.75\\
        12:10:02.62 & 39:24:44.85 & 1.94 & 1.45 & 0.415 & 21.66& 0.034 & 0.4742 & 1.21±0.56 & 20.95±0.48 & 2.12 & 1.76 & 4.77±1.06 & 6.31±1.21\\
        12:10:14.92 & 39:26:39.12 & 1.2 & 1.05 & 0.15 & 15.36 & 0.015 & 0.151 & 22.51±0.62 & 22.59±0.54 & 0.68 & 0.56 & 3.22±0.8 & 3.14±0.68\\
        12:10:12.89 & 39:26:11.72 & 1.85 & 1.16 & -0.093 & 15.78 & 0.026 & 0.222 & 23.0±0.85 & 23.0±0.74 & 1.0 & 0.82 & 1.21±0.41 & 1.26±0.38\\
        12:10:21.88 & 39:27:49.60 & 1.96 & 0.95 & 1.291 & 15.1 & 0.023 & 0.432 & 21.79±0.68 & 21.58±0.58 & 1.94 & 1.6 & 4.23±1.15 & 5.37±1.26 \\
        12:10:20.80 & 39:27:28.18 & 1.38 & 1.18 & -0.499 & 14.41 & 0.02 & 0.187 & 23.03±0.82 & 23.07±0.72 & 0.84 & 0.69 & 1.54±0.51 & 1.55±0.44\\

           \hline
        \end{tabular}
    \end{center}
\end{table*}

\begin{table*}[h]
\tiny
    \begin{center}
        \caption{Same as Table \ref{tab:cat-ngc1365}, but for NGC~4321}
        \label{tab:cat-ngc4321}
        \hspace{1cm}
        \begin{tabular}{ccccrccrccrcrc}
        \hline
          RA   & DEC & a & b & $\theta$  & r$_d$ & area & E(B$-$V) & m$_{FUV}$ & m$_{NUV}$ &  A$_{FUV}$ & A$_{NUV}$ & $\Sigma_{FUV}$ & $\Sigma_{NUV}$\\
          \hline
            12:22:57.31 & 15:51:46.07 & 3.24 & 2.8 & 0.295 & 15.53 & 0.307 & 0.226 & 18.12±0.31 & -- & 1.01 & -- & 25.57±3.21 & -- \\
            12:22:56.97 & 15:51:51.84 & 2.51 & 1.43 & 1.052 & 16.01 & 0.121 & 0.244 & 20.65±0.56 & -- & 1.09 & -- & 6.32±1.42 & -- \\
            12:22:58.48 & 15:51:36.93 & 1.85 & 1.26 & -1.37 & 15.13 & 0.079 & 0.351 & 19.86±0.53 & -- & 1.57  & -- & 19.87±4.2 & -- \\
            12:22:59.49 & 15:51:27.17 & 1.88 & 1.53 & 1.296 & 14.82 & 0.097 & 0.219 & 19.63±0.42  & -- & 0.98 & -- & 20.08±3.39 & -- \\
            12:22:56.51 & 15:51:42.52 & 4.41 & 2.56 & 0.383 & 14.94 & 0.382 & 0.235 & 18.5±0.34 & --  & 1.05 & -- & 14.53±1.98 & -- \\
            12:22:56.12 & 15:51:46.23 & 2.35 & 1.2 & -1.522 & 15.24 & 0.095 & 0.273 & 20.43±0.48 & -- & 1.22 & -- & 9.82±1.9 & -- \\
            12:22:56.42 & 15:51:48.90 & 1.75 & 1.25 & 1.463 & 15.57 & 0.074 & 0.225 & 21.18±0.52 & -- & 1.01 & -- & 6.33±1.32 & --\\
            12:22:55.10 & 15:51:44.64 & 2.84 & 1.4 & 0.688 & 14.97 & 0.134 & 0.255 & 20.42±0.6 & -- & 1.14 & -- & 7.01±1.68 & --\\
            12:23:04.49 & 15:50:59.37 & 1.55 & 1.33 & -1.017 & 17.61 & 0.069 & 0.244 & 20.68±0.69 & --  & 1.09 & -- & 10.66±2.93 & --\\
            12:22:53.03 & 15:51:48.46 & 2.72 & 2.12 & 1.216 & 15.63 & 0.195 & 0.179 & 18.25±0.3 & -- & 0.8 & -- & 35.83±4.24 & --\\
          
           \hline
        \end{tabular}
    \end{center}
\end{table*}

\begin{table*}[h]
\tiny
    \begin{center}
        \caption{Same as Table \ref{tab:cat-ngc1365}, but for NGC~4388}
        \label{tab:cat-ngc4388}
        \hspace{1cm}
        \begin{tabular}{ccccrccrccrrrr}
        \hline
          RA   & DEC & a & b & $\theta$ & r$_d$ & area & E(B$-$V) & m$_{FUV}$ & m$_{NUV}$ &  A$_{FUV}$ & A$_{NUV}$ & $\Sigma_{FUV}$ & $\Sigma_{NUV}$\\
          \hline
          12:25:43.27 & 12:39:42.04 & 0.92 & 0.92 & -0.954 & 8.39 & 0.074 & 0.225 & 22.06±0.68 & 22.07±0.52 & 1.01 & 0.78 & 7.23±1.96 & 7.54±1.55\\
        12:25:43.48 & 12:39:41.62 & 2.42 & 2.04 & -0.846 & 7.87 & 0.428 & 0.272 & 19.97±0.28 & 19.91±0.21 & 1.22 & 0.95 & 8.56±0.97 & 9.48±0.81\\
        12:25:43.07 & 12:39:39.56 & 1.64 & 1.28 & 0.735 & 8.9 & 0.182 & 0.314 & 21.07±0.51 & 20.96±0.38 & 1.41 & 1.09 & 7.31±1.48 & 8.53±1.29\\
        12:25:45.71 & 12:39:47.73 & 1.05 & 0.95 & -0.672 & 2.58 & 0.086 & 0.23 & 19.0±0.17 & 19.0±0.13 & 1.03 & 0.8 & 103.79±7.08 & 108.92±5.65\\
        12:25:45.92 & 12:39:47.65 & 4.78 & 2.05 & -0.224 & 2.08 & 0.848 & 0.244 & 17.01±0.08 & 16.99±0.06  & 1.09 & 0.85 & 66.08±2.04 & 70.63±1.68\\
        12:25:45.23 & 12:39:47.65 & 2.5 & 1.72 & -0.465 & 3.7 & 0.372 & 0.285 & 18.25±0.14 & 18.17±0.1 & 1.27 & 0.99 & 48.23±2.6 & 54.22±2.22\\
        12:25:44.99 & 12:39:47.55 & 3.87 & 3.55 & 0.915 & 4.28 & 1.187 & 0.304 & 17.22±0.09 & 17.11±0.07 & 1.36 & 1.06 & 39.06±1.42 & 44.97±1.24\\
        12:25:46.85 & 12:39:46.42 & 11.45 & 4.69 & 0.223 & 0.62 & 4.636 & 0.277 & 15.79±0.06 & 15.72±0.04 & 1.24 & 0.96 & 37.23±0.82 & 41.47±0.7\\
        12:25:48.46 & 12:39:39.30 & 1.93 & 1.46 & 0.343 & 4.25 & 0.243 & 0.374 & 18.82±0.2 & 18.62±0.15 & 1.68 & 1.3 & 43.59±3.54 & 54.84±3.32\\
        12:25:48.25 & 12:39:39.08 & 2.99 & 2.38 & -1.213 & 3.76 & 0.616 & 0.365 & 17.88±0.13 & 17.69±0.1 & 1.64 & 1.27 & 40.89±2.16 & 50.83±2.01 \\
           \hline
        \end{tabular}
    \end{center}
\end{table*}

\begin{table*}[h]
\tiny
    \begin{center}
        \caption{Same as table \ref{tab:cat-ngc1365}, but for NGC~5033}
        \label{tab:cat-ngc5033}
        \hspace{1cm}
        \begin{tabular}{ccccrccrccrrrr}
        \hline
          RA   & DEC & a & b & $\theta$ &  r$_d$ & area & E(B$-$V) & m$_{FUV}$ & m$_{NUV}$ &  A$_{FUV}$ & A$_{NUV}$ & $\Sigma_{FUV}$ & $\Sigma_{NUV}$\\
          \hline
          13:13:29.99 & 36:30:47.06 & 1.41 & 1.19 & 1.404 & 18.2 & 0.02 & 0.416 & 19.95±0.72 & 19.61±0.6 & 1.91 & 1.33 & 25.6±7.35 & 36.81±8.84\\
        13:13:28.32 & 36:30:45.54 & 0.92 & 0.92 & 0.394 & 18.21 & 0.01 & 0.348 & 20.74±0.95 & 20.52±0.81 & 1.59 & 1.11 & 24.14±9.15 & 30.88±10.02\\
        13:13:28.43 & 36:30:46.39 & 0.92 & 0.92 & 1.229 & 18.16 & 0.01 & 0.321 & 20.52±0.83 & 20.35±0.71 & 1.47 & 1.03 & 29.75±9.84 & 36.38±10.39\\
        13:13:28.80 & 36:30:51.67 & 0.92 & 0.92 & 1.469 & 17.85 & 0.01 & 0.403 & 20.58±0.94 & 20.26±0.79 & 1.85 & 1.29 & 27.99±10.54 & 39.35±12.46\\
        13:13:28.53 & 36:30:57.64 & 1.55 & 0.86 & 0.926 & 17.47 & 0.016 & 0.406 & 20.14±0.77 & 19.81±0.65 & 1.86 & 1.3 & 27.23±8.38 & 38.46±9.94\\
        13:13:28.58 & 36:30:51.86 & 0.92 & 0.92 & 1.295 & 17.83 & 0.01 & 0.381 & 20.48±0.87 & 20.2±0.74 & 1.75 & 1.22 & 30.82±10.77 & 41.77±12.36\\
        13:13:28.43 & 36:30:51.28 & 0.92 & 0.92 & 1.306 & 17.86 & 0.01 & 0.363 & 20.81±0.99 & 20.56±0.85 & 1.66 & 1.16 & 22.77±9.05 & 29.92±10.13\\
        13:13:28.34 & 36:30:49.68 & 0.92 & 0.92 & 1.433 & 17.95 & 0.01 & 0.594 & 19.98±0.93 & 19.31±0.73 & 2.72 & 1.9 & 48.51±17.97 & 94.33±27.39\\
        13:13:28.65 & 36:30:49.22 & 0.92 & 0.92 & -1.35 & 17.99 & 0.01 & 0.575 & 20.03±0.92 & 19.39±0.73 & 2.64 & 1.84 & 46.5±17.14 & 87.66±25.5\\
        13:13:28.16 & 36:31:05.42 & 1.78 & 0.94 & -0.921 & 16.97  & 0.02 & 0.436 & 19.67±0.64 & 19.29±0.54 & 2.0 & 1.4 & 33.27±8.59 & 49.49±10.62\\
        \hline
        \end{tabular}
    \end{center}
\end{table*}

\begin{table*}[h]
\tiny
    \begin{center}
        \caption{Same as Table \ref{tab:cat-ngc1365}, but for NGC~6814}
        \label{tab:cat-ngc6814}
        \hspace{1cm}
        \begin{tabular}{ccccrccrccrrrr}
        \hline
          RA   & DEC & a & b & $\theta$  & r$_d$ & area & E(B$-$V) & m$_{FUV}$ & m$_{NUV}$ &  A$_{FUV}$ & A$_{NUV}$ & $\Sigma_{FUV}$ & $\Sigma_{NUV}$\\
          \hline
          19:42:41.88 & -10:20:43.08 & 1.37 & 1.23 & 0.738 & 8.32 & 0.057 & 0.389 & 16.43±0.21 & -- & 1.74 & -- & 652.5±54.85 & --\\
        19:42:40.54 & -10:20:33.98 & 1.68 & 1.05 & 0.28 & 7.14 & 0.06 & 0.224 & 20.65±0.43 & -- & 1.0 & -- & 12.75±2.2 & --\\
        19:42:36.72 & -10:20:28.14 & 1.3 & 0.91 & -1.042 & 8.82 & 0.04 & 0.229 & 21.21±0.52 & -- & 1.03 & -- & 11.53±2.39 & --\\
        19:42:37.13 & -10:20:06.59 & 1.22 & 0.84 & -0.395 & 6.81 & 0.035 & 0.296 & 20.88±0.5 & -- & 1.32 & -- & 17.91±3.55 & --\\
        19:42:38.23 & -10:19:57.19 & 1.29 & 0.92 & 0.336 & 4.91 & 0.04 & 0.435 & 19.97±0.46 & -- & 1.95 & --  & 35.56±6.61 & --\\
        19:42:41.95 & -10:19:50.36 & 1.67 & 0.8 & -0.444 & 3.35 & 0.045 & 0.455 & 19.36±0.35 & -- & 2.04 & -- & 55.47±7.82 & --\\
        19:42:41.98 & -10:19:45.70 & 2.11 & 1.02 & -1.221 & 3.02 & 0.073 & 0.436 & 19.98±0.4 & -- & 1.96 & -- & 19.46±3.11 & --\\
        19:42:41.74 & -10:19:48.66 & 6.32 & 1.98 & -0.635 & 3.01 & 0.422 & 0.475 & 18.31±0.39 & -- & 2.13 & -- & 15.57±2.45 & --\\
        19:42:43.52 & -10:19:44.90 & 1.15 & 0.99 & 0.815 & 4.93 & 0.038 & 0.411 & 20.09±0.45 & -- & 1.84 & -- & 33.15±5.94 & --\\
        19:42:40.17 & -10:19:42.61 & 1.66 & 1.39 & -0.277 & 1.92 & 0.078 & 0.472 & 18.88±0.35 & --  & 2.12 & -- & 49.8±6.9 & --\\
           \hline
        \end{tabular}
    \end{center}
\end{table*}

\begin{table*}[h]
\tiny
    \begin{center}
        \caption{Same as Table \ref{tab:cat-ngc1365}, but for NGC~7469}
        \label{tab:cat-ngc7469}
        \hspace{1cm}
        \begin{tabular}{ccccrccrccrrrr}
        \hline
          RA   & DEC & a & b & $\theta$ & r$_d$  & area & E(B$-$V) & m$_{FUV}$ & m$_{NUV}$ &  A$_{FUV}$ & A$_{NUV}$ & $\Sigma_{FUV}$ & $\Sigma_{NUV}$\\
          \hline
          23:03:13.11 & 08:52:03.81 & 0.92 & 0.92 & -1.32 & 14.41 & 0.295 & 0.064 & 24.46±2.14 & 24.63±1.66 & 0.27 & 0.22 & 0.8±0.69 & 0.72±0.48\\
        23:03:16.17 & 08:52:21.09 & 4.14 & 2.3 & -1.159 & 3.22 & 3.278 & 0.184 & 18.37±0.18 & 18.41±0.14 & 0.78 & 0.64 & 19.83±1.45 & 19.98±1.15\\
        23:03:15.13 & 08:52:35.74 & 1.68 & 1.49 & -0.205 & 3.93 & 0.859 & 0.635 & 18.83±0.48 & 18.4±0.37 & 2.68 & 2.21 & 49.39±9.51 & 77.26±11.41\\
        23:03:14.89 & 08:52:53.73 & 2.66 & 1.67 & -1.165 & 9.75 & 1.525 & 0.238 & 19.4±0.3 & 19.39±0.23 & 1.0 & 0.83 & 16.49±1.98 & 17.5±1.63\\
        23:03:14.69 & 08:52:56.15 & 3.13 & 1.61 & -0.675 & 10.87 & 1.736 & 0.175 & 20.17±0.38 & 20.22±0.29 & 0.74 & 0.61 & 7.11±1.07 & 7.1±0.83\\
        23:03:16.93 & 08:52:17.07 & 12.04 & 4.8 & 1.246 & 7.13 & 19.877 & 0.327 & 18.27±0.22 & 18.16±0.17 & 1.38 & 1.14 & 3.57±0.31 & 4.13±0.28\\
        23:03:15.19 & 08:52:42.47 & 2.47 & 1.49 & 0.5 & 5.73 & 1.27 & 0.279 & 20.79±0.6 & 20.73±0.46 & 1.18 & 0.97 & 5.51±1.32 & 6.09±1.12\\
        23:03:14.71 & 08:52:44.67 & 5.81 & 2.37 & 1.389 & 7.52 & 4.735 & 0.275 & 19.72±0.37 & 19.66±0.28 & 1.16 & 0.96 & 3.97±0.58 & 4.37±0.5\\
        23:03:15.98 & 08:52:49.40 & 2.07 & 1.58 & -0.847 & 7.83 & 1.126 & 0.292 & 21.0±0.68 & 20.93±0.52 & 1.24 & 1.02 & 5.1±1.38 & 5.71±1.19\\
        23:03:16.51 & 08:52:50.48 & 7.9 & 6.08 & 0.404 & 9.09 & 16.523 & 0.216 & 18.73±0.22 & 18.74±0.17 & 0.91 & 0.75 & 2.82±0.25 & 2.94±0.2\\
           \hline
        \end{tabular}
    \end{center}
\end{table*}




\bibliography{ref}{}

\begin{thebibliography}{}
\expandafter\ifx\csname natexlab\endcsname\relax\def\natexlab#1{#1}\fi
\providecommand{\url}[1]{\href{#1}{#1}}
\providecommand{\dodoi}[1]{doi:~\href{http://doi.org/#1}{\nolinkurl{#1}}}
\providecommand{\doeprint}[1]{\href{http://ascl.net/#1}{\nolinkurl{http://ascl.net/#1}}}
\providecommand{\doarXiv}[1]{\href{https://arxiv.org/abs/#1}{\nolinkurl{https://arxiv.org/abs/#1}}}

\bibitem[{{Agrawal}(2006)}]{2006AdSpR..38.2989A}
{Agrawal}, P.~C. 2006, Advances in Space Research, 38, 2989, \dodoi{10.1016/j.asr.2006.03.038}

\bibitem[{{Allison} {et~al.}(2014){Allison}, {Sadler}, \& {Meekin}}]{All14}
{Allison}, J.~R., {Sadler}, E.~M., \& {Meekin}, A.~M. 2014, \mnras, 440, 696, \dodoi{10.1093/mnras/stu289}

\bibitem[{{{\'A}lvarez-{\'A}lvarez} {et~al.}(2015){{\'A}lvarez-{\'A}lvarez}, {D{\'\i}az}, {Terlevich}, \& {Terlevich}}]{2015MNRAS.451.3173A}
{{\'A}lvarez-{\'A}lvarez}, M., {D{\'\i}az}, A.~I., {Terlevich}, E., \& {Terlevich}, R. 2015, \mnras, 451, 3173, \dodoi{10.1093/mnras/stv1123}

\bibitem[{{Andreani} {et~al.}(2018){Andreani}, {Boselli}, {Ciesla}, {Vio}, {Cortese}, {Buat}, \& {Miyamoto}}]{2018A&A...617A..33A}
{Andreani}, P., {Boselli}, A., {Ciesla}, L., {et~al.} 2018, \aap, 617, A33, \dodoi{10.1051/0004-6361/201832873}

\bibitem[{{Arsenault}(1989)}]{Ars89}
{Arsenault}, R. 1989, \aap, 217, 66

\bibitem[{{Asmus} {et~al.}(2015){Asmus}, {Gandhi}, {H{\"o}nig}, {Smette}, \& {Duschl}}]{Asm15}
{Asmus}, D., {Gandhi}, P., {H{\"o}nig}, S.~F., {Smette}, A., \& {Duschl}, W.~J. 2015, \mnras, 454, 766, \dodoi{10.1093/mnras/stv1950}

\bibitem[{{Bentz} \& {Manne-Nicholas}(2018)}]{2018ApJ...864..146B}
{Bentz}, M.~C., \& {Manne-Nicholas}, E. 2018, \apj, 864, 146, \dodoi{10.3847/1538-4357/aad808}

\bibitem[{{Bentz} {et~al.}(2006){Bentz}, {Denney}, {Cackett}, {Dietrich}, {Fogel}, {Ghosh}, {Horne}, {Kuehn}, {Minezaki}, {Onken}, {Peterson}, {Pogge}, {Pronik}, {Richstone}, {Sergeev}, {Vestergaard}, {Walker}, \& {Yoshii}}]{Ben06}
{Bentz}, M.~C., {Denney}, K.~D., {Cackett}, E.~M., {et~al.} 2006, \apj, 651, 775, \dodoi{10.1086/507417}

\bibitem[{{Bentz} {et~al.}(2009){Bentz}, {Walsh}, {Barth}, {Baliber}, {Bennert}, {Canalizo}, {Filippenko}, {Ganeshalingam}, {Gates}, {Greene}, {Hidas}, {Hiner}, {Lee}, {Li}, {Malkan}, {Minezaki}, {Sakata}, {Serduke}, {Silverman}, {Steele}, {Stern}, {Street}, {Thornton}, {Treu}, {Wang}, {Woo}, \& {Yoshii}}]{Ben09}
{Bentz}, M.~C., {Walsh}, J.~L., {Barth}, A.~J., {et~al.} 2009, \apj, 705, 199, \dodoi{10.1088/0004-637X/705/1/199}

\bibitem[{{Bertin} \& {Arnouts}(1996)}]{1996A&AS..117..393B}
{Bertin}, E., \& {Arnouts}, S. 1996, \aaps, 117, 393, \dodoi{10.1051/aas:1996164}

\bibitem[{{Bing} {et~al.}(2019){Bing}, {Shi}, {Chen}, {S{\'a}nchez}, {Maiolino}, {Riffel}, {Riffel}, {Wylezalek}, {Bizyaev}, {Pan}, \& {Drory}}]{2019MNRAS.482..194B}
{Bing}, L., {Shi}, Y., {Chen}, Y., {et~al.} 2019, \mnras, 482, 194, \dodoi{10.1093/mnras/sty2662}

\bibitem[{{Bollati} {et~al.}(2023){Bollati}, {Lupi}, {Dotti}, \& {Haardt}}]{2023arXiv231107576B}
{Bollati}, F., {Lupi}, A., {Dotti}, M., \& {Haardt}, F. 2023, arXiv e-prints, arXiv:2311.07576, \dodoi{10.48550/arXiv.2311.07576}

\bibitem[{{Bosma} {et~al.}(1977){Bosma}, {Ekers}, \& {Lequeux}}]{Bos77}
{Bosma}, A., {Ekers}, R.~D., \& {Lequeux}, J. 1977, \aap, 57, 97

\bibitem[{Bradley {et~al.}(2020)Bradley, Sip{\H o}cz, Robitaille, Tollerud, Vin{\'{\i}}cius, Deil, Barbary, Wilson, Busko, G{\"u}nther, Cara, Conseil, Bostroem, Droettboom, Bray, Bratholm, Lim, Barentsen, Craig, Pascual, Perren, Greco, Donath, de~Val-Borro, Kerzendorf, Bach, Weaver, D'Eugenio, Souchereau, \& Ferreira}]{larry_bradley_2020_4044744}
Bradley, L., Sip{\H o}cz, B., Robitaille, T., {et~al.} 2020, astropy/photutils: 1.0.0, 1.0.0,  Zenodo, \dodoi{10.5281/zenodo.4044744}

\bibitem[{{Brenneman} {et~al.}(2013){Brenneman}, {Risaliti}, {Elvis}, \& {Nardini}}]{Bre13}
{Brenneman}, L.~W., {Risaliti}, G., {Elvis}, M., \& {Nardini}, E. 2013, \mnras, 429, 2662, \dodoi{10.1093/mnras/sts555}

\bibitem[{{Byrne} {et~al.}(2023){Byrne}, {Faucher-Gigu{\`e}re}, {Wellons}, {Hopkins}, {Angl{\'e}s-Alc{\'a}zar}, {Sultan}, {Wijers}, {Moreno}, \& {Ponnada}}]{2023arXiv231016086B}
{Byrne}, L., {Faucher-Gigu{\`e}re}, C.-A., {Wellons}, S., {et~al.} 2023, arXiv e-prints, arXiv:2310.16086, \dodoi{10.48550/arXiv.2310.16086}

\bibitem[{{Calzetti} {et~al.}(2000){Calzetti}, {Armus}, {Bohlin}, {Kinney}, {Koornneef}, \& {Storchi-Bergmann}}]{2000ApJ...533..682C}
{Calzetti}, D., {Armus}, L., {Bohlin}, R.~C., {et~al.} 2000, \apj, 533, 682, \dodoi{10.1086/308692}

\bibitem[{{Cardelli} {et~al.}(1989){Cardelli}, {Clayton}, \& {Mathis}}]{1989ApJ...345..245C}
{Cardelli}, J.~A., {Clayton}, G.~C., \& {Mathis}, J.~S. 1989, \apj, 345, 245, \dodoi{10.1086/167900}

\bibitem[{{Cayatte} {et~al.}(1990){Cayatte}, {van Gorkom}, {Balkowski}, \& {Kotanyi}}]{Cay90}
{Cayatte}, V., {van Gorkom}, J.~H., {Balkowski}, C., \& {Kotanyi}, C. 1990, \aj, 100, 604, \dodoi{10.1086/115545}

\bibitem[{{Cid Fernandes} {et~al.}(2004){Cid Fernandes}, {Gu}, {Melnick}, {Terlevich}, {Terlevich}, {Kunth}, {Rodrigues Lacerda}, \& {Joguet}}]{2004MNRAS.355..273C}
{Cid Fernandes}, R., {Gu}, Q., {Melnick}, J., {et~al.} 2004, \mnras, 355, 273, \dodoi{10.1111/j.1365-2966.2004.08321.x}

\bibitem[{{Colina} {et~al.}(2001){Colina}, {Alberdi}, {Torrelles}, {Panagia}, \& {Wilson}}]{2001ApJ...553L..19C}
{Colina}, L., {Alberdi}, A., {Torrelles}, J.~M., {Panagia}, N., \& {Wilson}, A.~S. 2001, \apjl, 553, L19, \dodoi{10.1086/320507}

\bibitem[{{Couto} \& {Storchi-Bergmann}(2023)}]{2023Galax..11...47C}
{Couto}, G.~S., \& {Storchi-Bergmann}, T. 2023, Galaxies, 11, 47, \dodoi{10.3390/galaxies11020047}

\bibitem[{{Damas-Segovia} {et~al.}(2016){Damas-Segovia}, {Beck}, {Vollmer}, {Wiegert}, {Krause}, {Irwin}, {We{\.z}gowiec}, {Li}, {Dettmar}, {English}, \& {Wang}}]{Dam16}
{Damas-Segovia}, A., {Beck}, R., {Vollmer}, B., {et~al.} 2016, \apj, 824, 30, \dodoi{10.3847/0004-637X/824/1/30}

\bibitem[{{Davies} {et~al.}(2007){Davies}, {M{\"u}ller S{\'a}nchez}, {Genzel}, {Tacconi}, {Hicks}, {Friedrich}, \& {Sternberg}}]{2007ApJ...671.1388D}
{Davies}, R.~I., {M{\"u}ller S{\'a}nchez}, F., {Genzel}, R., {et~al.} 2007, \apj, 671, 1388, \dodoi{10.1086/523032}

\bibitem[{{Davies} {et~al.}(1998){Davies}, {Sugai}, \& {Ward}}]{1998MNRAS.300..388D}
{Davies}, R.~I., {Sugai}, H., \& {Ward}, M.~J. 1998, \mnras, 300, 388, \dodoi{10.1046/j.1365-8711.1998.01928.x}

\bibitem[{Davies {et~al.}(2004)Davies, Tacconi, \& Genzel}]{Dav04}
Davies, R.~I., Tacconi, L.~J., \& Genzel, R. 2004, The Astrophysical Journal, 602, 148, \dodoi{10.1086/380995}

\bibitem[{{de Vaucouleurs} {et~al.}(1991){de Vaucouleurs}, {de Vaucouleurs}, {Corwin}, {Buta}, {Paturel}, \& {Fouque}}]{Vauco91}
{de Vaucouleurs}, G., {de Vaucouleurs}, A., {Corwin}, Herold~G., J., {et~al.} 1991, {Third Reference Catalogue of Bright Galaxies}

\bibitem[{{Diniz} {et~al.}(2019){Diniz}, {Riffel}, {Storchi-Bergmann}, \& {Riffel}}]{2019MNRAS.487.3958D}
{Diniz}, M.~R., {Riffel}, R.~A., {Storchi-Bergmann}, T., \& {Riffel}, R. 2019, \mnras, 487, 3958, \dodoi{10.1093/mnras/stz1329}

\bibitem[{{Evans} {et~al.}(1996){Evans}, {Koratkar}, {Storchi-Bergmann}, {Kirkpatrick}, {Heckman}, \& {Wilson}}]{Eva96}
{Evans}, I.~N., {Koratkar}, A.~P., {Storchi-Bergmann}, T., {et~al.} 1996, \apjs, 105, 93, \dodoi{10.1086/192308}

\bibitem[{{Falcke} {et~al.}(1998){Falcke}, {Wilson}, \& {Simpson}}]{Fal98}
{Falcke}, H., {Wilson}, A.~S., \& {Simpson}, C. 1998, \apj, 502, 199, \dodoi{10.1086/305886}

\bibitem[{{Ferrarese} \& {Merritt}(2000)}]{2000ApJ...539L...9F}
{Ferrarese}, L., \& {Merritt}, D. 2000, \apjl, 539, L9, \dodoi{10.1086/312838}

\bibitem[{{Ferreras} {et~al.}(2012){Ferreras}, {Cropper}, {Kawata}, {Page}, \& {Hoversten}}]{Fer12}
{Ferreras}, I., {Cropper}, M., {Kawata}, D., {Page}, M., \& {Hoversten}, E.~A. 2012, \mnras, 424, 1636, \dodoi{10.1111/j.1365-2966.2012.21017.x}

\bibitem[{{Fiore} {et~al.}(2017){Fiore}, {Feruglio}, {Shankar}, {Bischetti}, {Bongiorno}, {Brusa}, {Carniani}, {Cicone}, {Duras}, {Lamastra}, {Mainieri}, {Marconi}, {Menci}, {Maiolino}, {Piconcelli}, {Vietri}, \& {Zappacosta}}]{2017A&A...601A.143F}
{Fiore}, F., {Feruglio}, C., {Shankar}, F., {et~al.} 2017, \aap, 601, A143, \dodoi{10.1051/0004-6361/201629478}

\bibitem[{{Forster} {et~al.}(1999){Forster}, {Leighly}, \& {Kay}}]{For99}
{Forster}, K., {Leighly}, K.~M., \& {Kay}, L.~E. 1999, \apj, 523, 521, \dodoi{10.1086/307761}

\bibitem[{{Gaia Collaboration} {et~al.}(2022){Gaia Collaboration}, {Vallenari}, {Brown}, {Prusti}, {de Bruijne}, {Arenou}, {Babusiaux}, {Biermann}, {Creevey}, {Ducourant}, {Evans}, {Eyer}, {Guerra}, {Hutton}, {Jordi}, {Klioner}, {Lammers}, {Lindegren}, {Luri}, {Mignard}, {Panem}, {Pourbaix}, {Randich}, {Sartoretti}, {Soubiran}, {Tanga}, {Walton}, {Bailer-Jones}, {Bastian}, {Drimmel}, {Jansen}, {Katz}, {Lattanzi}, {van Leeuwen}, {Bakker}, {Cacciari}, {Casta{\~n}eda}, {De Angeli}, {Fabricius}, {Fouesneau}, {Fr{\'e}mat}, {Galluccio}, {Guerrier}, {Heiter}, {Masana}, {Messineo}, {Mowlavi}, {Nicolas}, {Nienartowicz}, {Pailler}, {Panuzzo}, {Riclet}, {Roux}, {Seabroke}, {Sordo{\o}rcit}, {Th{\'e}venin}, {Gracia-Abril}, {Portell}, {Teyssier}, {Altmann}, {Andrae}, {Audard}, {Bellas-Velidis}, {Benson}, {Berthier}, {Blomme}, {Burgess}, {Busonero}, {Busso}, {C{\'a}novas}, {Carry}, {Cellino}, {Cheek}, {Clementini}, {Damerdji}, {Davidson}, {de Teodoro}, {Nu{\~n}ez Campos}, {Delchambre}, {Dell'Oro}, {Esquej},
  {Fern{\'a}ndez-Hern{\'a}ndez}, {Fraile}, {Garabato}, {Garc{\'\i}a-Lario}, {Gosset}, {Haigron}, {Halbwachs}, {Hambly}, {Harrison}, {Hern{\'a}ndez}, {Hestroffer}, {Hodgkin}, {Holl}, {Jan{\ss}en}, {Jevardat de Fombelle}, {Jordan}, {Krone-Martins}, {Lanzafame}, {L{\"o}ffler}, {Marchal}, {Marrese}, {Moitinho}, {Muinonen}, {Osborne}, {Pancino}, {Pauwels}, {Recio-Blanco}, {Reyl{\'e}}, {Riello}, {Rimoldini}, {Roegiers}, {Rybizki}, {Sarro}, {Siopis}, {Smith}, {Sozzetti}, {Utrilla}, {van Leeuwen}, {Abbas}, {{\'A}brah{\'a}m}, {Abreu Aramburu}, {Aerts}, {Aguado}, {Ajaj}, {Aldea-Montero}, {Altavilla}, {{\'A}lvarez}, {Alves}, {Anders}, {Anderson}, {Anglada Varela}, {Antoja}, {Baines}, {Baker}, {Balaguer-N{\'u}{\~n}ez}, {Balbinot}, {Balog}, {Barache}, {Barbato}, {Barros}, {Barstow}, {Bartolom{\'e}}, {Bassilana}, {Bauchet}, {Becciani}, {Bellazzini}, {Berihuete}, {Bernet}, {Bertone}, {Bianchi}, {Binnenfeld}, {Blanco-Cuaresma}, {Blazere}, {Boch}, {Bombrun}, {Bossini}, {Bouquillon}, {Bragaglia}, {Bramante}, {Breedt},
  {Bressan}, {Brouillet}, {Brugaletta}, {Bucciarelli}, {Burlacu}, {Butkevich}, {Buzzi}, {Caffau}, {Cancelliere}, {Cantat-Gaudin}, {Carballo}, {Carlucci}, {Carnerero}, {Carrasco}, {Casamiquela}, {Castellani}, {Castro-Ginard}, {Chaoul}, {Charlot}, {Chemin}, {Chiaramida}, {Chiavassa}, {Chornay}, {Comoretto}, {Contursi}, {Cooper}, {Cornez}, {Cowell}, {Crifo}, {Cropper}, {Crosta}, {Crowley}, {Dafonte}, {Dapergolas}, {David}, {David}, {de Laverny}, {De Luise}, {De March}, {De Ridder}, {de Souza}, {de Torres}, {del Peloso}, {del Pozo}, {Delbo}, {Delgado}, {Delisle}, {Demouchy}, {Dharmawardena}, {Di Matteo}, {Diakite}, {Diener}, {Distefano}, {Dolding}, {Edvardsson}, {Enke}, {Fabre}, {Fabrizio}, {Faigler}, {Fedorets}, {Fernique}, {Fienga}, {Figueras}, {Fournier}, {Fouron}, {Fragkoudi}, {Gai}, {Garcia-Gutierrez}, {Garcia-Reinaldos}, {Garc{\'\i}a-Torres}, {Garofalo}, {Gavel}, {Gavras}, {Gerlach}, {Geyer}, {Giacobbe}, {Gilmore}, {Girona}, {Giuffrida}, {Gomel}, {Gomez}, {Gonz{\'a}lez-N{\'u}{\~n}ez},
  {Gonz{\'a}lez-Santamar{\'\i}a}, {Gonz{\'a}lez-Vidal}, {Granvik}, {Guillout}, {Guiraud}, {Guti{\'e}rrez-S{\'a}nchez}, {Guy}, {Hatzidimitriou}, {Hauser}, {Haywood}, {Helmer}, {Helmi}, {Sarmiento}, {Hidalgo}, {Hilger}, {H{\l}adczuk}, {Hobbs}, {Holland}, {Huckle}, {Jardine}, {Jasniewicz}, {Jean-Antoine Piccolo}, {Jim{\'e}nez-Arranz}, {Jorissen}, {Juaristi Campillo}, {Julbe}, {Karbevska}, {Kervella}, {Khanna}, {Kontizas}, {Kordopatis}, {Korn}, {K{\'o}sp{\'a}l}, {Kostrzewa-Rutkowska}, {Kruszy{\'n}ska}, {Kun}, {Laizeau}, {Lambert}, {Lanza}, {Lasne}, {Le Campion}, {Lebreton}, {Lebzelter}, {Leccia}, {Leclerc}, {Lecoeur-Taibi}, {Liao}, {Licata}, {Lindstr{\o}m}, {Lister}, {Livanou}, {Lobel}, {Lorca}, {Loup}, {Madrero Pardo}, {Magdaleno Romeo}, {Managau}, {Mann}, {Manteiga}, {Marchant}, {Marconi}, {Marcos}, {Marcos Santos}, {Mar{\'\i}n Pina}, {Marinoni}, {Marocco}, {Marshall}, {Polo}, {Mart{\'\i}n-Fleitas}, {Marton}, {Mary}, {Masip}, {Massari}, {Mastrobuono-Battisti}, {Mazeh}, {McMillan}, {Messina}, {Michalik},
  {Millar}, {Mints}, {Molina}, {Molinaro}, {Moln{\'a}r}, {Monari}, {Mongui{\'o}}, {Montegriffo}, {Montero}, {Mor}, {Mora}, {Morbidelli}, {Morel}, {Morris}, {Muraveva}, {Murphy}, {Musella}, {Nagy}, {Noval}, {Oca{\~n}a}, {Ogden}, {Ordenovic}, {Osinde}, {Pagani}, {Pagano}, {Palaversa}, {Palicio}, {Pallas-Quintela}, {Panahi}, {Payne-Wardenaar}, {Pe{\~n}alosa Esteller}, {Penttil{\"a}}, {Pichon}, {Piersimoni}, {Pineau}, {Plachy}, {Plum}, {Poggio}, {Pr{\v{s}}a}, {Pulone}, {Racero}, {Ragaini}, {Rainer}, {Raiteri}, {Rambaux}, {Ramos}, {Ramos-Lerate}, {Re Fiorentin}, {Regibo}, {Richards}, {Rios Diaz}, {Ripepi}, {Riva}, {Rix}, {Rixon}, {Robichon}, {Robin}, {Robin}, {Roelens}, {Rogues}, {Rohrbasser}, {Romero-G{\'o}mez}, {Rowell}, {Royer}, {Ruz Mieres}, {Rybicki}, {Sadowski}, {S{\'a}ez N{\'u}{\~n}ez}, {Sagrist{\`a} Sell{\'e}s}, {Sahlmann}, {Salguero}, {Samaras}, {Sanchez Gimenez}, {Sanna}, {Santove{\~n}a}, {Sarasso}, {Schultheis}, {Sciacca}, {Segol}, {Segovia}, {S{\'e}gransan}, {Semeux}, {Shahaf}, {Siddiqui}, {Siebert},
  {Siltala}, {Silvelo}, {Slezak}, {Slezak}, {Smart}, {Snaith}, {Solano}, {Solitro}, {Souami}, {Souchay}, {Spagna}, {Spina}, {Spoto}, {Steele}, {Steidelm{\"u}ller}, {Stephenson}, {S{\"u}veges}, {Surdej}, {Szabados}, {Szegedi-Elek}, {Taris}, {Taylo}, {Teixeira}, {Tolomei}, {Tonello}, {Torra}, {Torra}, {Torralba Elipe}, {Trabucchi}, {Tsounis}, {Turon}, {Ulla}, {Unger}, {Vaillant}, {van Dillen}, {van Reeven}, {Vanel}, {Vecchiato}, {Viala}, {Vicente}, {Voutsinas}, {Weiler}, {Wevers}, {Wyrzykowski}, {Yoldas}, {Yvard}, {Zhao}, {Zorec}, {Zucker}, \& {Zwitter}}]{2022arXiv220800211G}
{Gaia Collaboration}, {Vallenari}, A., {Brown}, A.~G.~A., {et~al.} 2022, arXiv e-prints, arXiv:2208.00211.
\newblock \doarXiv{2208.00211}

\bibitem[{{Gallagher} {et~al.}(2019){Gallagher}, {Maiolino}, {Belfiore}, {Drory}, {Riffel}, \& {Riffel}}]{2019MNRAS.485.3409G}
{Gallagher}, R., {Maiolino}, R., {Belfiore}, F., {et~al.} 2019, \mnras, 485, 3409, \dodoi{10.1093/mnras/stz564}

\bibitem[{{Gallo} {et~al.}(2021){Gallo}, {Gonzalez}, \& {Miller}}]{Gal21}
{Gallo}, L.~C., {Gonzalez}, A.~G., \& {Miller}, J.~M. 2021, \apjl, 908, L33, \dodoi{10.3847/2041-8213/abdcb5}

\bibitem[{{Garc{\'\i}a-Bernete} {et~al.}(2021){Garc{\'\i}a-Bernete}, {Alonso-Herrero}, {Garc{\'\i}a-Burillo}, {Pereira-Santaella}, {Garc{\'\i}a-Lorenzo}, {Carrera}, {Rigopoulou}, {Ramos Almeida}, {Villar Mart{\'\i}n}, {Gonz{\'a}lez-Mart{\'\i}n}, {Hicks}, {Labiano}, {Ricci}, \& {Mateos}}]{2021A&A...645A..21G}
{Garc{\'\i}a-Bernete}, I., {Alonso-Herrero}, A., {Garc{\'\i}a-Burillo}, S., {et~al.} 2021, \aap, 645, A21, \dodoi{10.1051/0004-6361/202038256}

\bibitem[{{Garcia-Burillo} {et~al.}(1998){Garcia-Burillo}, {Sempere}, {Combes}, \& {Neri}}]{Gar98}
{Garcia-Burillo}, S., {Sempere}, M.~J., {Combes}, F., \& {Neri}, R. 1998, \aap, 333, 864.
\newblock \doarXiv{astro-ph/9803006}

\bibitem[{{Garc{\'\i}a-Gonz{\'a}lez} {et~al.}(2016){Garc{\'\i}a-Gonz{\'a}lez}, {Alonso-Herrero}, {Hern{\'a}n-Caballero}, {Pereira-Santaella}, {Ramos-Almeida}, {Acosta-Pulido}, {D{\'\i}az-Santos}, {Esquej}, {Gonz{\'a}lez-Mart{\'\i}n}, {Ichikawa}, {L{\'o}pez-Rodr{\'\i}guez}, {Povic}, {Roche}, \& {S{\'a}nchez-Portal}}]{2016MNRAS.458.4512G}
{Garc{\'\i}a-Gonz{\'a}lez}, J., {Alonso-Herrero}, A., {Hern{\'a}n-Caballero}, A., {et~al.} 2016, \mnras, 458, 4512, \dodoi{10.1093/mnras/stw626}

\bibitem[{{Gebhardt} {et~al.}(2000){Gebhardt}, {Bender}, {Bower}, {Dressler}, {Faber}, {Filippenko}, {Green}, {Grillmair}, {Ho}, {Kormendy}, {Lauer}, {Magorrian}, {Pinkney}, {Richstone}, \& {Tremaine}}]{2000ApJ...539L..13G}
{Gebhardt}, K., {Bender}, R., {Bower}, G., {et~al.} 2000, \apjl, 539, L13, \dodoi{10.1086/312840}

\bibitem[{{Georgiev} {et~al.}(2016){Georgiev}, {B{\"o}ker}, {Leigh}, {L{\"u}tzgendorf}, \& {Neumayer}}]{2016MNRAS.457.2122G}
{Georgiev}, I.~Y., {B{\"o}ker}, T., {Leigh}, N., {L{\"u}tzgendorf}, N., \& {Neumayer}, N. 2016, \mnras, 457, 2122, \dodoi{10.1093/mnras/stw093}

\bibitem[{{Ghosh} {et~al.}(2022){Ghosh}, {Tandon}, {Singh}, {Shelat}, {Tahlani}, {Singh}, {Srinivasan}, {Joseph}, {Devaraj}, {George}, {Mohan}, {Postma}, \& {Stalin}}]{2022JApA...43...77G}
{Ghosh}, S.~K., {Tandon}, S.~N., {Singh}, S.~K., {et~al.} 2022, Journal of Astrophysics and Astronomy, 43, 77, \dodoi{10.1007/s12036-022-09842-7}

\bibitem[{{Gonz{\'a}lez-Mart{\'\i}n} {et~al.}(2009){Gonz{\'a}lez-Mart{\'\i}n}, {Masegosa}, {M{\'a}rquez}, {Guainazzi}, \& {Jim{\'e}nez-Bail{\'o}n}}]{Gon09}
{Gonz{\'a}lez-Mart{\'\i}n}, O., {Masegosa}, J., {M{\'a}rquez}, I., {Guainazzi}, M., \& {Jim{\'e}nez-Bail{\'o}n}, E. 2009, \aap, 506, 1107, \dodoi{10.1051/0004-6361/200912288}

\bibitem[{{Gu} {et~al.}(2001){Gu}, {Dultzin-Hacyan}, \& {de Diego}}]{2001RMxAA..37....3G}
{Gu}, Q., {Dultzin-Hacyan}, D., \& {de Diego}, J.~A. 2001, \rmxaa, 37, 3, \dodoi{10.48550/arXiv.astro-ph/0011419}

\bibitem[{{H{\"a}ring} \& {Rix}(2004)}]{2004ApJ...604L..89H}
{H{\"a}ring}, N., \& {Rix}, H.-W. 2004, \apjl, 604, L89, \dodoi{10.1086/383567}

\bibitem[{{Harrison}(2017)}]{2017NatAs...1E.165H}
{Harrison}, C.~M. 2017, Nature Astronomy, 1, 0165, \dodoi{10.1038/s41550-017-0165}

\bibitem[{{Hennig} {et~al.}(2018){Hennig}, {Riffel}, {Dors}, {Riffel}, {Storchi-Bergmann}, \& {Colina}}]{2018MNRAS.477.1086H}
{Hennig}, M.~G., {Riffel}, R.~A., {Dors}, O.~L., {et~al.} 2018, \mnras, 477, 1086, \dodoi{10.1093/mnras/sty547}

\bibitem[{{Hervella Seoane} {et~al.}(2023){Hervella Seoane}, {Ramos Almeida}, {Acosta Pulido}, {Speranza}, {Tadhunter}, \& {Bessiere}}]{2023arXiv230910572H}
{Hervella Seoane}, K., {Ramos Almeida}, C., {Acosta Pulido}, J.~A., {et~al.} 2023, arXiv e-prints, arXiv:2309.10572, \dodoi{10.48550/arXiv.2309.10572}

\bibitem[{{Hummel} \& {Saikia}(1991)}]{Hum91}
{Hummel}, E., \& {Saikia}, D.~J. 1991, \aap, 249, 43

\bibitem[{{Ichikawa} {et~al.}(2017){Ichikawa}, {Ricci}, {Ueda}, {Matsuoka}, {Toba}, {Kawamuro}, {Trakhtenbrot}, \& {Koss}}]{2017ApJ...835...74I}
{Ichikawa}, K., {Ricci}, C., {Ueda}, Y., {et~al.} 2017, \apj, 835, 74, \dodoi{10.3847/1538-4357/835/1/74}

\bibitem[{{Immler} {et~al.}(1998){Immler}, {Pietsch}, \& {Aschenbach}}]{Imm98}
{Immler}, S., {Pietsch}, W., \& {Aschenbach}, B. 1998, \aap, 331, 601

\bibitem[{{Jones} {et~al.}(2011){Jones}, {McHardy}, {Moss}, {Seymour}, {Breedt}, {Uttley}, {K{\"o}rding}, \& {Tudose}}]{2011MNRAS.412.2641J}
{Jones}, S., {McHardy}, I., {Moss}, D., {et~al.} 2011, \mnras, 412, 2641, \dodoi{10.1111/j.1365-2966.2010.18105.x}

\bibitem[{Jones {et~al.}(2011)Jones, McHardy, Moss, Seymour, Breedt, Uttley, Körding, \& Tudose}]{Jon11}
Jones, S., McHardy, I., Moss, D., {et~al.} 2011, Monthly Notices of the Royal Astronomical Society, 412, 2641, \dodoi{10.1111/j.1365-2966.2010.18105.x}

\bibitem[{{Kennicutt} \& {Evans}(2012)}]{2012ARA&A..50..531K}
{Kennicutt}, R.~C., \& {Evans}, N.~J. 2012, \araa, 50, 531, \dodoi{10.1146/annurev-astro-081811-125610}

\bibitem[{{Khachikian} \& {Weedman}(1974)}]{Kha74}
{Khachikian}, E.~Y., \& {Weedman}, D.~W. 1974, \apj, 192, 581, \dodoi{10.1086/153093}

\bibitem[{{Knapen} {et~al.}(1993){Knapen}, {Arnth-Jensen}, {Cepa}, \& {Beckman}}]{Kna93}
{Knapen}, J.~H., {Arnth-Jensen}, N., {Cepa}, J., \& {Beckman}, J.~E. 1993, \aj, 106, 56, \dodoi{10.1086/116620}

\bibitem[{{Kumari} {et~al.}(2023){Kumari}, {Jana}, {Naik}, \& {Nandi}}]{Kum23}
{Kumari}, N., {Jana}, A., {Naik}, S., \& {Nandi}, P. 2023, \mnras, 521, 5440, \dodoi{10.1093/mnras/stad867}

\bibitem[{{Kuo} {et~al.}(2011){Kuo}, {Braatz}, {Condon}, {Impellizzeri}, {Lo}, {Zaw}, {Schenker}, {Henkel}, {Reid}, \& {Greene}}]{Kuo11}
{Kuo}, C.~Y., {Braatz}, J.~A., {Condon}, J.~J., {et~al.} 2011, \apj, 727, 20, \dodoi{10.1088/0004-637X/727/1/20}

\bibitem[{{Lammers} {et~al.}(2023){Lammers}, {Iyer}, {Ibarra-Medel}, {Pacifici}, {S{\'a}nchez}, {Tacchella}, \& {Woo}}]{2023ApJ...953...26L}
{Lammers}, C., {Iyer}, K.~G., {Ibarra-Medel}, H., {et~al.} 2023, \apj, 953, 26, \dodoi{10.3847/1538-4357/acdd57}

\bibitem[{{Liszt} \& {Dickey}(1995)}]{Lis95}
{Liszt}, H.~S., \& {Dickey}, J.~M. 1995, \aj, 110, 998, \dodoi{10.1086/117579}

\bibitem[{Lu {et~al.}(2021)Lu, Wang, Zhang, Huang, Xu, Xin, Yu, Ding, Wang, \& Feng}]{Lu21}
Lu, K.-X., Wang, J.-G., Zhang, Z.-X., {et~al.} 2021, The Astrophysical Journal, 918, 50, \dodoi{10.3847/1538-4357/ac0c78}

\bibitem[{{Lu} {et~al.}(1993){Lu}, {Hoffman}, {Groff}, {Roos}, \& {Lamphier}}]{Lu93}
{Lu}, N.~Y., {Hoffman}, G.~L., {Groff}, T., {Roos}, T., \& {Lamphier}, C. 1993, \apjs, 88, 383, \dodoi{10.1086/191826}

\bibitem[{{Maiolino} \& {Rieke}(1995)}]{Mai95}
{Maiolino}, R., \& {Rieke}, G.~H. 1995, \apj, 454, 95, \dodoi{10.1086/176468}

\bibitem[{{Maiolino} {et~al.}(2017){Maiolino}, {Russell}, {Fabian}, {Carniani}, {Gallagher}, {Cazzoli}, {Arribas}, {Belfiore}, {Bellocchi}, {Colina}, {Cresci}, {Ishibashi}, {Marconi}, {Mannucci}, {Oliva}, \& {Sturm}}]{2017Natur.544..202M}
{Maiolino}, R., {Russell}, H.~R., {Fabian}, A.~C., {et~al.} 2017, \nat, 544, 202, \dodoi{10.1038/nature21677}

\bibitem[{{Marconi} \& {Hunt}(2003)}]{2003ApJ...589L..21M}
{Marconi}, A., \& {Hunt}, L.~K. 2003, \apjl, 589, L21, \dodoi{10.1086/375804}

\bibitem[{{Marquez} \& {Moles}(1994)}]{Mar94}
{Marquez}, I., \& {Moles}, M. 1994, \aj, 108, 90, \dodoi{10.1086/117048}

\bibitem[{{Martin} {et~al.}(2005){Martin}, {Fanson}, {Schiminovich}, {Morrissey}, {Friedman}, {Barlow}, {Conrow}, {Grange}, {Jelinsky}, {Milliard}, {Siegmund}, {Bianchi}, {Byun}, {Donas}, {Forster}, {Heckman}, {Lee}, {Madore}, {Malina}, {Neff}, {Rich}, {Small}, {Surber}, {Szalay}, {Welsh}, \& {Wyder}}]{2005ApJ...619L...1M}
{Martin}, D.~C., {Fanson}, J., {Schiminovich}, D., {et~al.} 2005, \apjl, 619, L1, \dodoi{10.1086/426387}

\bibitem[{{McHardy} {et~al.}(2004){McHardy}, {Papadakis}, {Uttley}, {Page}, \& {Mason}}]{McH04}
{McHardy}, I.~M., {Papadakis}, I.~E., {Uttley}, P., {Page}, M.~J., \& {Mason}, K.~O. 2004, \mnras, 348, 783, \dodoi{10.1111/j.1365-2966.2004.07376.x}

\bibitem[{{McLure} \& {Dunlop}(2002)}]{2002MNRAS.331..795M}
{McLure}, R.~J., \& {Dunlop}, J.~S. 2002, \mnras, 331, 795, \dodoi{10.1046/j.1365-8711.2002.05236.x}

\bibitem[{{Mountrichas} \& {Buat}(2023)}]{2023A&A...679A.151M}
{Mountrichas}, G., \& {Buat}, V. 2023, \aap, 679, A151, \dodoi{10.1051/0004-6361/202347392}

\bibitem[{Mundell {et~al.}(1999)Mundell, Pedlar, Shone, \& Robinson}]{Mun99}
Mundell, C.~G., Pedlar, A., Shone, D.~L., \& Robinson, A. 1999, Monthly Notices of the Royal Astronomical Society, 304, 481, \dodoi{10.1046/j.1365-8711.1999.02331.x}

\bibitem[{{Nandi} {et~al.}(2023{\natexlab{a}}){Nandi}, {Stalin}, {Saikia}, {Muneer}, {Mountrichas}, {Wylezalek}, {Sagar}, \& {Kissler-Patig}}]{2023ApJ...950...81N}
{Nandi}, P., {Stalin}, C.~S., {Saikia}, D.~J., {et~al.} 2023{\natexlab{a}}, ApJ, 950, 81, \dodoi{10.3847/1538-4357/accf1e}

\bibitem[{{Nandi} {et~al.}(2023{\natexlab{b}}){Nandi}, {Stalin}, {Saikia}, {Riffel}, {Manna}, {Pal}, {Dors}, {Wylezalek}, {Paliya}, {Saikia}, {Dabhade}, {Patig}, \& {Sagar}}]{2023ApJ...959..116N}
---. 2023{\natexlab{b}}, \apj, 959, 116, \dodoi{10.3847/1538-4357/ad0c57}

\bibitem[{{Pan} \& {Kuno}(2017)}]{Pan17}
{Pan}, H.-A., \& {Kuno}, N. 2017, \apj, 839, 133, \dodoi{10.3847/1538-4357/aa60c2}

\bibitem[{{Papadakis} {et~al.}(2008){Papadakis}, {Ioannou}, {Brinkmann}, \& {Xilouris}}]{2008A&A...490..995P}
{Papadakis}, I.~E., {Ioannou}, Z., {Brinkmann}, W., \& {Xilouris}, E.~M. 2008, \aap, 490, 995, \dodoi{10.1051/0004-6361:200810298}

\bibitem[{{Parkash} {et~al.}(2018){Parkash}, {Brown}, {Jarrett}, \& {Bonne}}]{2018ApJ...864...40P}
{Parkash}, V., {Brown}, M. J.~I., {Jarrett}, T.~H., \& {Bonne}, N.~J. 2018, \apj, 864, 40, \dodoi{10.3847/1538-4357/aad3b9}

\bibitem[{{Paturel} {et~al.}(2003){Paturel}, {Petit}, {Prugniel}, {Theureau}, {Rousseau}, {Brouty}, {Dubois}, \& {Cambr{\'e}sy}}]{Pat03}
{Paturel}, G., {Petit}, C., {Prugniel}, P., {et~al.} 2003, \aap, 412, 45, \dodoi{10.1051/0004-6361:20031411}

\bibitem[{{Pedlar} {et~al.}(1992){Pedlar}, {Howley}, {Axon}, \& {Unger}}]{1992MNRAS.259..369P}
{Pedlar}, A., {Howley}, P., {Axon}, D.~J., \& {Unger}, S.~W. 1992, \mnras, 259, 369, \dodoi{10.1093/mnras/259.2.369}

\bibitem[{{P{\'e}rez-Torres} \& {Alberdi}(2007)}]{2007MNRAS.379..275P}
{P{\'e}rez-Torres}, M.~A., \& {Alberdi}, A. 2007, \mnras, 379, 275, \dodoi{10.1111/j.1365-2966.2007.11944.x}

\bibitem[{{Peterson} {et~al.}(2000){Peterson}, {McHardy}, {Wilkes}, {Berlind}, {Bertram}, {Calkins}, {Collier}, {Huchra}, {Mathur}, {Papadakis}, {Peters}, {Pogge}, {Romano}, {Tokarz}, {Uttley}, {Vestergaard}, \& {Wagner}}]{Pet00}
{Peterson}, B.~M., {McHardy}, I.~M., {Wilkes}, B.~J., {et~al.} 2000, \apj, 542, 161, \dodoi{10.1086/309518}

\bibitem[{{Peterson} {et~al.}(2014){Peterson}, {Grier}, {Horne}, {Pogge}, {Bentz}, {De Rosa}, {Denney}, {Martini}, {Sergeev}, {Kaspi}, {Minezaki}, {Zu}, {Kochanek}, {Siverd}, {Shappee}, {Araya Salvo}, {Beatty}, {Bird}, {Bord}, {Borman}, {Che}, {Chen}, {Cohen}, {Dietrich}, {Doroshenko}, {Drake}, {Efimov}, {Free}, {Ginsburg}, {Henderson}, {King}, {Koshida}, {Mogren}, {Molina}, {Mosquera}, {Motohara}, {Nazarov}, {Okhmat}, {Pejcha}, {Rafter}, {Shields}, {Skowron}, {Skowron}, {Valluri}, {van Saders}, \& {Yoshii}}]{Pet14}
{Peterson}, B.~M., {Grier}, C.~J., {Horne}, K., {et~al.} 2014, \apj, 795, 149, \dodoi{10.1088/0004-637X/795/2/149}

\bibitem[{{Pounds} {et~al.}(2004){Pounds}, {Reeves}, {King}, \& {Page}}]{Pou04}
{Pounds}, K.~A., {Reeves}, J.~N., {King}, A.~R., \& {Page}, K.~L. 2004, \mnras, 350, 10, \dodoi{10.1111/j.1365-2966.2004.07639.x}

\bibitem[{{Renzini} \& {Peng}(2015)}]{2015ApJ...801L..29R}
{Renzini}, A., \& {Peng}, Y.-j. 2015, \apjl, 801, L29, \dodoi{10.1088/2041-8205/801/2/L29}

\bibitem[{{Riffel} {et~al.}(2022){Riffel}, {Dahmer-Hahn}, {Riffel}, {Storchi-Bergmann}, {Dametto}, {Davies}, {Burtscher}, {Bianchin}, {Ruschel-Dutra}, {Ricci}, \& {Rosario}}]{2022MNRAS.512.3906R}
{Riffel}, R., {Dahmer-Hahn}, L.~G., {Riffel}, R.~A., {et~al.} 2022, \mnras, 512, 3906, \dodoi{10.1093/mnras/stac740}

\bibitem[{{Risaliti} {et~al.}(2005){Risaliti}, {Elvis}, {Fabbiano}, {Baldi}, \& {Zezas}}]{Ris05}
{Risaliti}, G., {Elvis}, M., {Fabbiano}, G., {Baldi}, A., \& {Zezas}, A. 2005, \apjl, 623, L93, \dodoi{10.1086/430252}

\bibitem[{{Risaliti} {et~al.}(2009){Risaliti}, {Miniutti}, {Elvis}, {Fabbiano}, {Salvati}, {Baldi}, {Braito}, {Bianchi}, {Matt}, {Reeves}, {Soria}, \& {Zezas}}]{Ris09}
{Risaliti}, G., {Miniutti}, G., {Elvis}, M., {et~al.} 2009, \apj, 696, 160, \dodoi{10.1088/0004-637X/696/1/160}

\bibitem[{{Rodriguez Espinosa} {et~al.}(1987){Rodriguez Espinosa}, {Rudy}, \& {Jones}}]{1987ApJ...312..555R}
{Rodriguez Espinosa}, J.~M., {Rudy}, R.~J., \& {Jones}, B. 1987, \apj, 312, 555, \dodoi{10.1086/164901}

\bibitem[{{Saikia} {et~al.}(1994){Saikia}, {Pedlar}, {Unger}, \& {Axon}}]{1994MNRAS.270...46S}
{Saikia}, D.~J., {Pedlar}, A., {Unger}, S.~W., \& {Axon}, D.~J. 1994, \mnras, 270, 46, \dodoi{10.1093/mnras/270.1.46}

\bibitem[{{Salim} {et~al.}(2007){Salim}, {Rich}, {Charlot}, {Brinchmann}, {Johnson}, {Schiminovich}, {Seibert}, {Mallery}, {Heckman}, {Forster}, {Friedman}, {Martin}, {Morrissey}, {Neff}, {Small}, {Wyder}, {Bianchi}, {Donas}, {Lee}, {Madore}, {Milliard}, {Szalay}, {Welsh}, \& {Yi}}]{2007ApJS..173..267S}
{Salim}, S., {Rich}, R.~M., {Charlot}, S., {et~al.} 2007, \apjs, 173, 267, \dodoi{10.1086/519218}

\bibitem[{{S{\'a}nchez-Portal} {et~al.}(2004){S{\'a}nchez-Portal}, {D{\'\i}az}, {Terlevich}, \& {Terlevich}}]{San04}
{S{\'a}nchez-Portal}, M., {D{\'\i}az}, {\'A}.~I., {Terlevich}, E., \& {Terlevich}, R. 2004, \mnras, 350, 1087, \dodoi{10.1111/j.1365-2966.2004.07720.x}

\bibitem[{{Sandage} \& {Bedke}(1994)}]{San94}
{Sandage}, A., \& {Bedke}, J. 1994, {The Carnegie atlas of galaxies}, Vol. 638

\bibitem[{{Sandqvist} {et~al.}(1995){Sandqvist}, {Joersaeter}, \& {Lindblad}}]{1995A&A...295..585S}
{Sandqvist}, A., {Joersaeter}, S., \& {Lindblad}, P.~O. 1995, \aap, 295, 585

\bibitem[{{Sargent} {et~al.}(2024){Sargent}, {Fischer}, {Johnson}, {van der Horst}, {Secrest}, {Shuvo}, {Cigan}, \& {Smith}}]{2024ApJ...961..230S}
{Sargent}, A.~J., {Fischer}, T.~C., {Johnson}, M.~C., {et~al.} 2024, \apj, 961, 230, \dodoi{10.3847/1538-4357/ad11d4}

\bibitem[{{Sarzi} {et~al.}(2002){Sarzi}, {Rix}, {Shields}, {McIntosh}, {Ho}, {Rudnick}, {Filippenko}, {Sargent}, \& {Barth}}]{Sar02}
{Sarzi}, M., {Rix}, H.-W., {Shields}, J.~C., {et~al.} 2002, \apj, 567, 237, \dodoi{10.1086/338351}

\bibitem[{{Seifina} {et~al.}(2018){Seifina}, {Chekhtman}, \& {Titarchuk}}]{Sei18}
{Seifina}, E., {Chekhtman}, A., \& {Titarchuk}, L. 2018, \aap, 613, A48, \dodoi{10.1051/0004-6361/201732235}

\bibitem[{{Shin} {et~al.}(2019){Shin}, {Woo}, {Chung}, {Baek}, {Cho}, {Kang}, \& {Bae}}]{2019ApJ...881..147S}
{Shin}, J., {Woo}, J.-H., {Chung}, A., {et~al.} 2019, \apj, 881, 147, \dodoi{10.3847/1538-4357/ab2e72}

\bibitem[{{Singh} {et~al.}(2019){Singh}, {Gulati}, \& {Bagla}}]{Sin19}
{Singh}, A., {Gulati}, M., \& {Bagla}, J.~S. 2019, \mnras, 489, 5582, \dodoi{10.1093/mnras/stz2523}

\bibitem[{{Slavcheva-Mihova} \& {Mihov}(2011)}]{Sla11}
{Slavcheva-Mihova}, L., \& {Mihov}, B. 2011, \aap, 526, A43, \dodoi{10.1051/0004-6361/200913243}

\bibitem[{{Springob} {et~al.}(2005){Springob}, {Haynes}, {Giovanelli}, \& {Kent}}]{sp05}
{Springob}, C.~M., {Haynes}, M.~P., {Giovanelli}, R., \& {Kent}, B.~R. 2005, \apjs, 160, 149, \dodoi{10.1086/431550}

\bibitem[{{Stevens} {et~al.}(1999){Stevens}, {Forbes}, \& {Norris}}]{1999MNRAS.306..479S}
{Stevens}, I.~R., {Forbes}, D.~A., \& {Norris}, R.~P. 1999, \mnras, 306, 479, \dodoi{10.1046/j.1365-8711.1999.02543.x}

\bibitem[{{Storchi-Bergmann} \& {Schnorr-M{\"u}ller}(2019)}]{2019NatAs...3...48S}
{Storchi-Bergmann}, T., \& {Schnorr-M{\"u}ller}, A. 2019, Nature Astronomy, 3, 48, \dodoi{10.1038/s41550-018-0611-0}

\bibitem[{{Sweet} {et~al.}(2018){Sweet}, {Fisher}, {Glazebrook}, {Obreschkow}, {Lagos}, \& {Wang}}]{2018ApJ...860...37S}
{Sweet}, S.~M., {Fisher}, D., {Glazebrook}, K., {et~al.} 2018, \apj, 860, 37, \dodoi{10.3847/1538-4357/aabfc4}

\bibitem[{{Tandon} {et~al.}(2020){Tandon}, {Postma}, {Joseph}, {Devaraj}, {Subramaniam}, {Barve}, {George}, {Ghosh}, {Girish}, {Hutchings}, {Kamath}, {Kathiravan}, {Kumar}, {Lancelot}, {Leahy}, {Mahesh}, {Mohan}, {Nagabhushana}, {Pati}, {Rao}, {Sankarasubramanian}, {Sriram}, \& {Stalin}}]{2020AJ....159..158T}
{Tandon}, S.~N., {Postma}, J., {Joseph}, P., {et~al.} 2020, \aj, 159, 158, \dodoi{10.3847/1538-3881/ab72a3}

\bibitem[{{Terashima} {et~al.}(1999){Terashima}, {Kunieda}, \& {Misaki}}]{Ter99}
{Terashima}, Y., {Kunieda}, H., \& {Misaki}, K. 1999, \pasj, 51, 277, \dodoi{10.1093/pasj/51.3.277}

\bibitem[{{Tody}(1986)}]{1986SPIE..627..733T}
{Tody}, D. 1986, in Society of Photo-Optical Instrumentation Engineers (SPIE) Conference Series, Vol. 627, Instrumentation in astronomy VI, ed. D.~L. {Crawford}, 733, \dodoi{10.1117/12.968154}

\bibitem[{{Tortosa} {et~al.}(2018){Tortosa}, {Bianchi}, {Marinucci}, {Matt}, {Middei}, {Piconcelli}, {Brenneman}, {Cappi}, {Dadina}, {De Rosa}, {Petrucci}, {Ursini}, \& {Walton}}]{Tor18}
{Tortosa}, A., {Bianchi}, S., {Marinucci}, A., {et~al.} 2018, \mnras, 473, 3104, \dodoi{10.1093/mnras/stx2457}

\bibitem[{{Troyer} {et~al.}(2016){Troyer}, {Starkey}, {Cackett}, {Bentz}, {Goad}, {Horne}, \& {Seals}}]{Tro16}
{Troyer}, J., {Starkey}, D., {Cackett}, E.~M., {et~al.} 2016, \mnras, 456, 4040, \dodoi{10.1093/mnras/stv2862}

\bibitem[{{Tsai} \& {Hwang}(2015)}]{2015AJ....150...43T}
{Tsai}, M., \& {Hwang}, C.-Y. 2015, \aj, 150, 43, \dodoi{10.1088/0004-6256/150/2/43}

\bibitem[{{Tueller} {et~al.}(2008){Tueller}, {Mushotzky}, {Barthelmy}, {Cannizzo}, {Gehrels}, {Markwardt}, {Skinner}, \& {Winter}}]{Tue08}
{Tueller}, J., {Mushotzky}, R.~F., {Barthelmy}, S., {et~al.} 2008, \apj, 681, 113, \dodoi{10.1086/588458}

\bibitem[{{Ulvestad} \& {Wilson}(1984)}]{1984ApJ...285..439U}
{Ulvestad}, J.~S., \& {Wilson}, A.~S. 1984, \apj, 285, 439, \dodoi{10.1086/162520}

\bibitem[{{Urbanik} {et~al.}(1986){Urbanik}, {Klein}, \& {Graeve}}]{Urb86}
{Urbanik}, M., {Klein}, U., \& {Graeve}, R. 1986, \aap, 166, 107

\bibitem[{{Veilleux} {et~al.}(1999){Veilleux}, {Bland-Hawthorn}, \& {Cecil}}]{Vei99}
{Veilleux}, S., {Bland-Hawthorn}, J., \& {Cecil}, G. 1999, \aj, 118, 2108, \dodoi{10.1086/301095}

\bibitem[{{V{\'e}ron-Cetty} \& {V{\'e}ron}(2006)}]{2006A&A...455..773V}
{V{\'e}ron-Cetty}, M.~P., \& {V{\'e}ron}, P. 2006, \aap, 455, 773, \dodoi{10.1051/0004-6361:20065177}

\bibitem[{Wang {et~al.}(2010)Wang, Risaliti, Fabbiano, Elvis, Zezas, \& Karovska}]{Wan10}
Wang, J., Risaliti, G., Fabbiano, G., {et~al.} 2010, The Astrophysical Journal, 714, 1497, \dodoi{10.1088/0004-637x/714/2/1497}

\bibitem[{{Ward} {et~al.}(2022){Ward}, {Harrison}, {Costa}, \& {Mainieri}}]{2022MNRAS.514.2936W}
{Ward}, S.~R., {Harrison}, C.~M., {Costa}, T., \& {Mainieri}, V. 2022, \mnras, 514, 2936, \dodoi{10.1093/mnras/stac1219}

\bibitem[{{Weiler} {et~al.}(1981){Weiler}, {van der Hulst}, {Sramek}, \& {Panagia}}]{1981ApJ...243L.151W}
{Weiler}, K.~W., {van der Hulst}, J.~M., {Sramek}, R.~A., \& {Panagia}, N. 1981, \apjl, 243, L151, \dodoi{10.1086/183463}

\bibitem[{{Whitmore} {et~al.}(2023){Whitmore}, {Chandar}, {Rodr{\'\i}guez}, {Lee}, {Emsellem}, {Floyd}, {Kim}, {Kruijssen}, {Mok}, {Sormani}, {Boquien}, {Dale}, {Faesi}, {Henny}, {Hannon}, {Thilker}, {White}, {Barnes}, {Bigiel}, {Chevance}, {Henshaw}, {Klessen}, {Leroy}, {Liu}, {Maschmann}, {Meidt}, {Rosolowsky}, {Schinnerer}, {Sun}, {Watkins}, \& {Williams}}]{Whit23}
{Whitmore}, B.~C., {Chandar}, R., {Rodr{\'\i}guez}, M.~J., {et~al.} 2023, \apjl, 944, L14, \dodoi{10.3847/2041-8213/acae94}

\bibitem[{{Williams} {et~al.}(2017){Williams}, {McHardy}, {Baldi}, {Beswick}, {Argo}, {Dullo}, {Knapen}, {Brinks}, {Fenech}, {Mundell}, {Muxlow}, {Panessa}, {Rampadarath}, \& {Westcott}}]{2017MNRAS.472.3842W}
{Williams}, D.~R.~A., {McHardy}, I.~M., {Baldi}, R.~D., {et~al.} 2017, \mnras, 472, 3842, \dodoi{10.1093/mnras/stx2205}

\bibitem[{{Wolfinger} {et~al.}(2013){Wolfinger}, {Kilborn}, {Koribalski}, {Minchin}, {Boyce}, {Disney}, {Lang}, \& {Jordan}}]{Wol13}
{Wolfinger}, K., {Kilborn}, V.~A., {Koribalski}, B.~S., {et~al.} 2013, \mnras, 428, 1790, \dodoi{10.1093/mnras/sts160}

\bibitem[{{Xu} {et~al.}(1999){Xu}, {Livio}, \& {Baum}}]{Xu99}
{Xu}, C., {Livio}, M., \& {Baum}, S. 1999, \aj, 118, 1169, \dodoi{10.1086/301007}

\bibitem[{{Xu} \& {Wang}(2022)}]{Xu22}
{Xu}, X., \& {Wang}, J. 2022, \apj, 933, 110, \dodoi{10.3847/1538-4357/ac7222}

\bibitem[{{Yang} {et~al.}(2024){Yang}, {Dav{\'e}}, {Cui}, {Cai}, {Peacock}, \& {Sorini}}]{2024MNRAS.527.1612Y}
{Yang}, T., {Dav{\'e}}, R., {Cui}, W., {et~al.} 2024, \mnras, 527, 1612, \dodoi{10.1093/mnras/stad3223}

\bibitem[{{Zhang} \& {Ho}(2023)}]{2023ApJ...953L...9Z}
{Zhang}, L., \& {Ho}, L.~C. 2023, \apjl, 953, L9, \dodoi{10.3847/2041-8213/acea73}

\bibitem[{{Zhuang} {et~al.}(2021){Zhuang}, {Ho}, \& {Shangguan}}]{2021ApJ...906...38Z}
{Zhuang}, M.-Y., {Ho}, L.~C., \& {Shangguan}, J. 2021, \apj, 906, 38, \dodoi{10.3847/1538-4357/abc94d}

\bibitem[{{Zinn} {et~al.}(2013){Zinn}, {Middelberg}, {Norris}, \& {Dettmar}}]{2013ApJ...774...66Z}
{Zinn}, P.~C., {Middelberg}, E., {Norris}, R.~P., \& {Dettmar}, R.~J. 2013, \apj, 774, 66, \dodoi{10.1088/0004-637X/774/1/66}

\end{thebibliography}
\bibliographystyle{aasjournal}



\end{document}